\documentclass[twocolumn]{aastex701}

\usepackage{amsmath}
\usepackage{subfigure,wrapfig}
\usepackage{booktabs}
\usepackage[normalem]{ulem}
\usepackage{soul}
\usepackage{hyperref}
\usepackage{cancel}

\usepackage{etoolbox}

\makeatletter
\patchcmd{\appendix}{\setcounter{table}{0}}{
  \setcounter{table}{0}
  \renewcommand{\thetable}{\thesection\arabic{table}}
}{}{}
\makeatother

\newcommand{\e}{\mathrm{e}}

\begin{document}

\title{Probing hadronic gamma-ray and high-energy neutrino emission from Gaia DR2 star clusters}

\author[orcid=0000-0003-3631-5648,sname='Mitchell']{Alison M.W. Mitchell}
\affiliation{Erlangen Centre for Astroparticle Physics, Friedrich-Alexander-Universit\"at Erlangen-N\"urnberg, Nikolaus-Fiebiger-Str. 2, 91058 Erlangen, Germany}
\email[show]{alison.mw.mitchell@fau.de, giovanni.morlino@inaf.it,\\  silvia.celli@roma1.infn.it, smenchiari@iaa.es}  

\author[orcid=0000-0002-5014-4817,sname='Morlino']{Giovanni Morlino}
\affiliation{Istituto Nazionale di Astrofisica, Osservatorio Astrofisico di Arcetri,
                L.go E. Fermi 5, Firenze, Italy}
\email{giovanni.morlino@inaf.it}

\author[orcid=0000-0002-7592-0851,sname='Celli']{Silvia Celli}
\affiliation{Sapienza Universit\`a di Roma, Physics Department,
                 P.le Aldo Moro 5, 00185, Rome, Italy}
\affiliation{ Istituto Nazionale di Fisica Nucleare, Sezione di Roma,
                 P.le Aldo Moro 5, 00185, Rome, Italy}
\email{silvia.celli@roma1.infn.it}

\author[orcid=0009-0006-6386-3702,sname='Menchiari']{Stefano Menchiari}
\affiliation{Instituto de Astrof\`isica de Andaluc\`ia, CSIC, 18080 Granada, Spain}
\affiliation{Istituto Nazionale di Astrofisica, Osservatorio Astrofisico di Arcetri,
                L.go E. Fermi 5, Firenze, Italy}
\email{smenchiari@iaa.es}

\author[sname='Specovius']{Andreas Specovius}
\affiliation{Erlangen Centre for Astroparticle Physics, Friedrich-Alexander-Universit\"at Erlangen-N\"urnberg, Nikolaus-Fiebiger-Str. 2, 91058 Erlangen, Germany}
\email{andreas.specovius@fau.de}

\begin{abstract}

Young and massive stellar clusters are a potential source of galactic cosmic rays due to at least two acceleration mechanisms. Collective stellar winds from massive stars form a wind-blown bubble with a termination shock at which particle acceleration to PeV energies may be achieved. Furthermore, shock acceleration may occur at supernova remnants (SNRs) expanding inside the bubble.  
We apply a model of cosmic ray acceleration at both the collective wind termination shock and SNR shocks to the catalog of known stellar clusters derived from the Gaia DR2. 
Predictions for the secondary fluxes of $\gamma$-ray and neutrino emission are derived based on hadro-nuclear interactions with the surrounding medium. We compare our modelling under baseline and optimistic scenarios to available data, finding consistent results. 
An anticipated flux range is provided for a shortlist of the most promising stellar clusters. Approximately 10 clusters may be detectable with future facilities, and $1-3$ could be currently operating as PeVatrons. Among these, data from three $\gamma$-ray detected clusters can be consistently described by our model.
Several further as-yet-undetected stellar clusters offer promising targets for future $\gamma$-ray observations, 
although the flux range allowed by our model can be broad ($\gtrsim$ factor 10). The large angular size of the wind-blown bubble may pose a challenge, leading to low surface brightness emission, thus exacerbating the problem of source confusion. Nevertheless, we discuss how further work will help to constrain stellar clusters as PeVatron candidates. 

\end{abstract}


\section{Introduction}
\label{sec:introduction}

The all particle cosmic ray (CR) spectrum spans many orders of magnitude in both flux and energy, with Galactic accelerators thought to provide the bulk of the CR flux at energies up to at least the so-called `knee' - a spectral softening feature at $\sim 10^{15}$\,eV (1\,PeV). Galactic sources capable of acceleration up to this energy are colloquially termed `PeVatrons'. 
Although the rate and energetics of supernova remnants (SNRs) can approximately account for the origins of Galactic CRs, to date clear evidence for the presence of particles at PeV energies in the vicinity of SNRs is still lacking. Gamma-ray emission at $\gtrsim 100$\,TeV is a signature of the presence of hadronic particles with PeV energies. The first experimental confirmation of the existence of such powerful Galactic sources occurred only in recent years \citep{hawc2020_56tev,Lhaaso2021_pevatrons}. However, so far none of the possible counterparts have been unambiguously associated with SNRs. 

Young and massive stellar clusters (SC) have recently been proposed as a suitable alternative source of CRs at PeV energies \citep{Aharonian2019_natast}. Indeed, several SCs are plausibly associated with known very-high-energy $\gamma$-ray sources, such as Westerlund\,1 and the Cygnus OB2 \citep{Westerlund1_2022}. Among the most recent results, LHAASO \citep{LHAASOcygnus:2024} observed the extended bubble surrounding the Cygnus OB2 association with 66 photon-like events above 400 TeV, of which 8 have reconstructed energies above PeV. This result suggests that this source class could contribute to the Galactic CR flux, although at high energy a contribution from past old SNR or the microquasar Cyg X-3 may be relevant. According to the model of \cite{Morlino2021}, diffusive shock acceleration at the collective wind termination shock (WTS) driven by the stellar winds of massive stars may result in the production of PeV protons in the case it proceeds in the Bohm or Kraichnan domain. In particular, in the case of the Cygnus bubble, a diffusion regime intermediate between Kraichnan and Bohm is favoured under this scenario to explain the entire $\gamma$-ray spectrum detected from GeV up to hundreds of TeV \citep{Menchiari+2024}.
For the case of the Cygnus cocoon, the formation of a collective termination shock is not obvious, given the loose nature of the stellar associations.

Energetic particles accelerated at the WTS of SC can interact with ambient target material to produce $\gamma$-ray emission; in the case of hadronic particles via proton interactions resulting in pion production and the subsequent decay of neutral pions into $\gamma$ rays. Similarly, signature neutrino emission may be produced via the decay of charged pions. 
We aim here at extending the investigation performed on the Cygnus bubble \citep{Menchiari+2024}, by accounting for the whole sample of young open clusters observed in our Galaxy by the Gaia satellite \citep{gaia2020}, to infer the presence of potential PeVatron candidates in this population and, more generally, to estimate which among those SCs may be detectable at TeV energies and beyond. Theoretical considerations indicate that the main source of SC energetics depends on its age, being dominated by winds of massive OB stars in the first  2-3 million years of evolution, then powered mainly by the winds of Wolf-Rayet (WR) stars and finally by the luminosity of their supernova (SN) explosions \citep{vieu2020, vieu2022}. In \cite{celli2023} we identified 387 clusters from the Gaia sample with age smaller than 30~Myr, corresponding to the characteristic main sequence lifetime of a $8\,M_\odot$ star, i.e. its time of explosion into an SN. For these clusters, we provided lower limits to their mass and mechanical wind luminosity, both crucial ingredients to describe the bubble structure and the particle acceleration as the energy converted into non-thermal particles is expected to scale linearly with the wind luminosity. 

In this study, we derive both the CR properties and the related hadronic interaction signatures, in terms of $\gamma$-ray and neutrino emission from the young cluster catalog. In Sec.~\ref{sec:model} we introduce the geometry of the system, describing in particular the expected target gas distribution. We then discuss in Sec.~\ref{sec:acceleration} the adopted particle acceleration model, including both the contribution at the WTS following \citep{Morlino2021}, as well as the CR production due to SNRs exploded inside the clusters, a calculation that we provide here for the first time. For the wind model, we additionally estimate the contribution of observed WR stars to the cluster kinetic luminosity, as in \cite{celli2023}.
In Sec.~\ref{sec:secondaries} we focus on the hadronic interactions of such particles, and obtain spectra for the emerging secondaries from a benchmark cluster, while in  Sec.~\ref{sec:DR2results} we present the results from the entire cluster sample discussing which clusters are potentially observable by next-generation instruments, such as LHAASO and CTA for $\gamma$-rays, and KM3NeT for neutrinos. 
In Sec.~\ref{sec:gammadata_compare} our results are compared to currently available $\gamma$-ray data from three specific clusters, namely  Westerlund~1 (Wd1),  Westerlund~2 (Wd2) and NGC~3603, while in Sec.~\ref{sec:catalogues} we compare the positions of clusters in our sample with existing $\gamma$-ray catalogues. Finally, we conclude in Sec.~\ref{sec:conclusions} with an outlook towards future observations.

\begin{figure}
    \includegraphics[width=0.6\columnwidth]{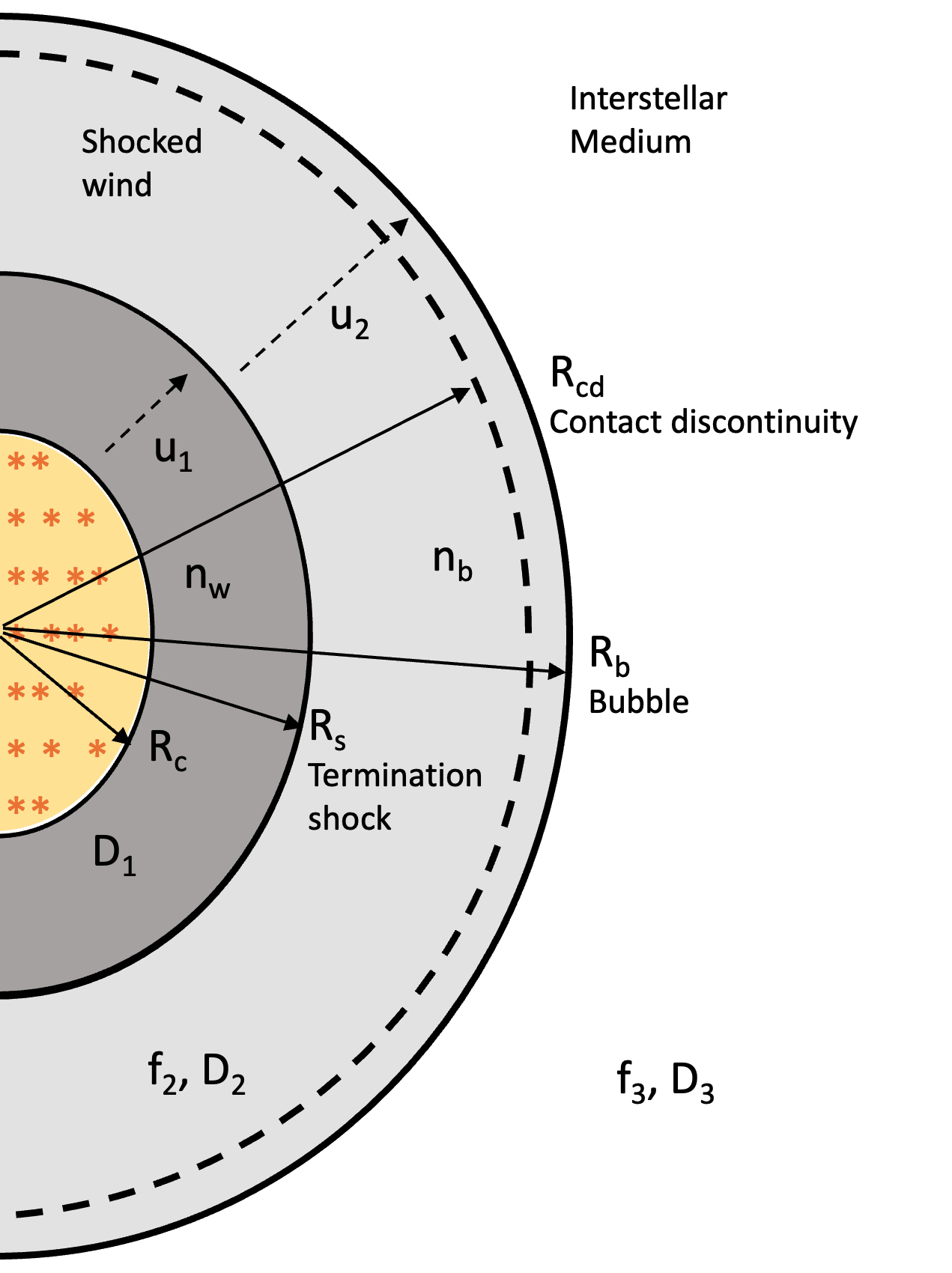}
    \caption{Schematic model for a stellar cluster. CR acceleration occurs at the WTS with radius $R_s$, the wind blown bubble extends to a radius $R_b$ whilst the member stars of the cluster itself are confined to a radius $R_c$. The shocked wind region is indicated by the subscript $2$ whilst the ISM is denoted by the subscript $3$.}
    \label{fig:schematic}
\end{figure}

\section{The wind-blown bubble structure}
\label{sec:model}
During the first few Myrs, when no SN has exploded yet, the circumstellar medium around SCs is shaped by stellar winds, which inflate a low density bubble producing a forward shock, expanding into the interstellar medium (ISM), and a reverse shock, i.e. the WTS, expanding into the cold fast wind. Among the two, only the reverse shock can have a Mach number $\gg 1$ and hence represents a possible location for particle acceleration \citep{Lancaster+2021a}. We describe the structure of the wind-blown bubble in analogy to the stellar model by \citep{Weaver+1977}. In particular, the location of the WTS is affected by the surrounding density, as well as by the cluster age and properties of  member stars, e.g. their mass loss rate and wind velocity. A schematic representation of the wind-driven bubble is provided in Fig.~\ref{fig:schematic}, showing both the forward shock of the wind blown bubble at distance $R_b$ and the WTS at distance $R_{s}$, that for a pure adiabatic bubble are equal to
\begin{equation}
\label{eqn:Rs}
\begin{aligned}
    R_{\rm s} (t) =& 26\, \eta_m^{-1/5}
        \, \left(\frac{\dot{M}_{\rm c}}{10^{-4}\mathrm{M}_\odot\mathrm{yr}^{-1}}\right)^{3/10}
        \left(\frac{v_{\rm w}}{2000\,\mathrm{km\,s}^{-1}}\right)^{1/10}  \\
        &\times \left(\frac{n_0}{10\,\mathrm{cm}^{-3}}\right)^{-3/10}
        \left(\frac{t}{10\,\mathrm{Myr}}\right)^{2/5}
        \,\mathrm{pc}
\end{aligned}
\end{equation}
and
\begin{equation}\label{eqn:Rb}
\begin{aligned}
        R_{\rm b} (t) =& 112
        \,\left( \frac{\eta_m \, L_{\rm w,c}}{10^{37}\,\mathrm{erg\,s}^{-1}}\right)^{1/5}
        \left(\frac{n_0}{10\,\mathrm{cm}^{-3}}\right)^{-1/5} \\
        & \times \left(\frac{t}{10\,\mathrm{Myr}}\right)^{3/5}
        \mathrm{pc}
\end{aligned}
\end{equation}
where $t$ is the cluster age, $n_0$ the number density of the ISM outside of the wind bubble, $\dot{M}_{\rm c}$ the SC mass loss rate, $v_{\rm w}$ its wind velocity, $L_{\rm w,c}=\frac{1}{2}\dot{M}_{\rm c} v_{\rm w} ^2$ its wind luminosity, and $\eta_m$ its mechanical efficiency, namely the fraction of wind energy converted into kinetic energy. The latter parameter accounts for the cooling of the system, resulting into a fraction of the wind luminosity lost into radiation. The impact of radiative cooling is not straightforward to estimate because it has a strong dependence on the level of fragmentation of the bubble shell and on the ISM density, which enhance the cooling due to radiative losses \cite[see, e.g.][]{Yadav+2017}. The effect of $\eta_m$ is to reduce the internal pressure of the bubble, which can be written as \citep{Morlino2021}
\begin{equation} \label{eqn:Pb}
  P_{\rm b} = \frac{7}{(3850 \pi)^{2/5}} \, \left(\eta_{m} L_{\rm w,c}\right)^{2/5} \, \rho_0^{3/5} \, t^{-4/5} 
\end{equation}
with $\rho_0=n_0 m_p$ being the mass ambient density in which the wind bubble expands, that will be assumed as homogeneous, and $m_p$ the proton mass. The impact of $\eta_m$ is hence such that the bubble size is reduced while the size of the termination shock increases. Typical values of $\eta_m$ estimated with numerical simulations are a few tens of percent \citep{Yadav+2017}: in this work, we simply set $\eta_m=0.3$.

In addition to stellar winds, clusters older than $\sim 3$\,Myr will host supernova (SN) explosions that inject further energy inside the bubble. In such a case, we evaluate the bubble size and internal pressure by substituting the wind luminosity in Eqs.~\eqref{eqn:Rb} and \eqref{eqn:Pb} with the total luminosity given by 
\begin{equation} \label{eq:L_tot}
    L_{\rm tot} = L_{\rm w,c}\ + \frac{N_{\rm SN}(t) E_{\rm SN}}{t}
\end{equation}
where $N_{\rm SN}$ is the number of SNe exploded during the SC lifetime $t$ and $E_{\rm SN}= 10^{51}$\,erg is the kinetic energy released by each explosion.

The different parameters determining the energy output of a star cluster can be obtained from the characterization of its stellar population, as described in \cite{celli2023}, where we developed a method to infer the wind luminosity of each cluster via a stellar mass distribution that reproduces the observed number of member stars per cluster by Gaia \citep{gaia2020}. Our procedure accounted for both the bolometric correction and the light extinction from each individual cluster direction to derive the correct normalization of the stellar mass function, assumed in the form of a Kroupa distribution \citep{WeidnerKroupa2004}. As such, we could extend the mass domain of our investigation beyond Gaia observations by including all stellar masses between $0.08 \, M_\odot$ and the maximum stellar mass expected to still exist in the cluster in its main sequence phase. The latter was determined by theoretical considerations as limited by either the age of the parent cluster, $M^*_{\rm max}(t_{\rm age})$, or by its mass, $M^*_{\rm max,0}$. We obtained reasonable cluster mass values (namely within a factor 2 of literature estimations), which we find acceptable given the intrinsic limitations of our procedure that strongly relies on individual star detection inside SCs. 
Once the stellar population is determined for each SC, we determine the SC wind mass loss rate and velocity starting from the properties of each stellar wind, from which we could derive the cluster wind luminosity required to model the SC particle acceleration. 

\begin{figure}
    \centering
    \includegraphics[width=\columnwidth]{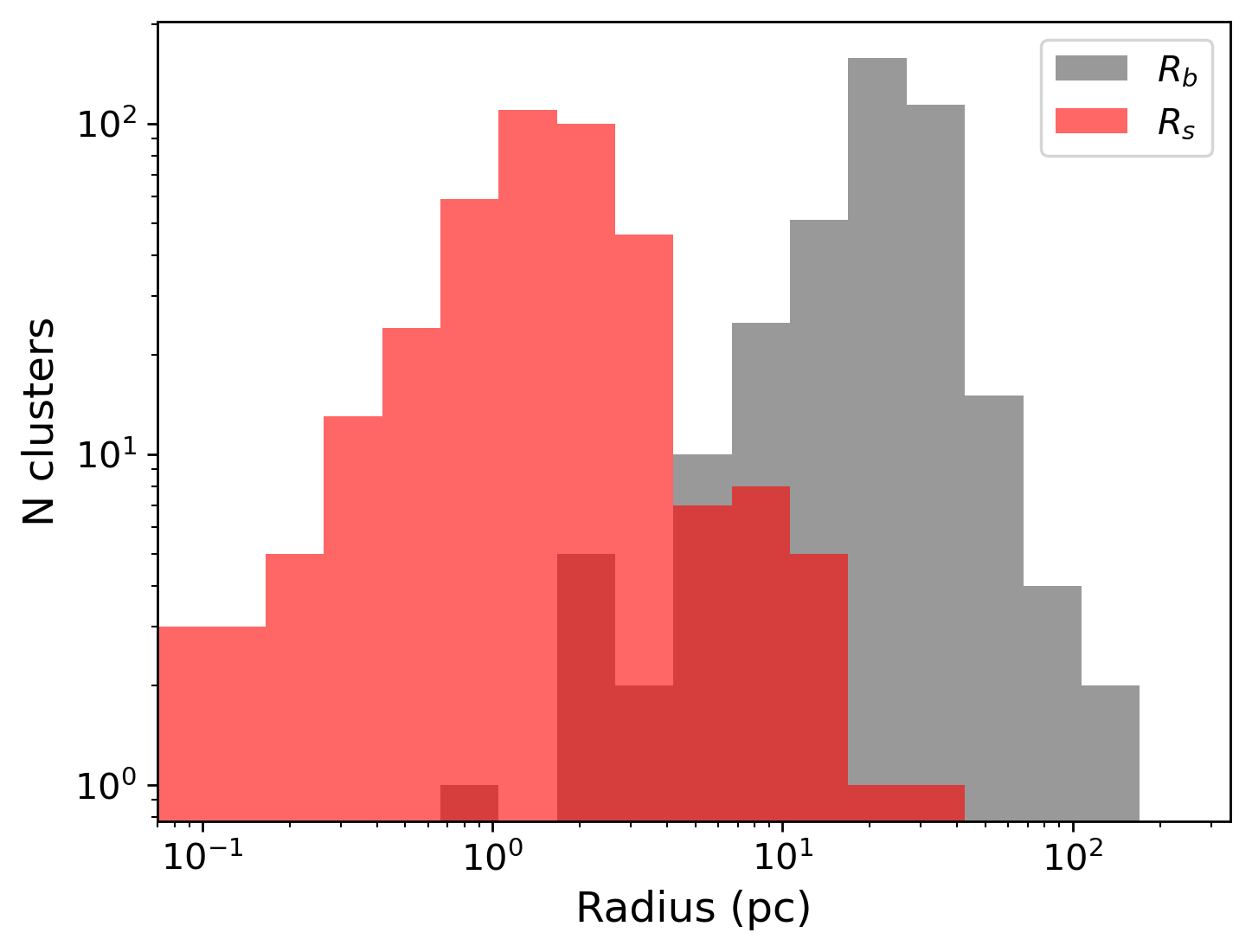}
    \caption{Distribution of wind termination shock $R_s$ and bubble radii $R_b$ evaluated from Eqs.~\eqref{eqn:Rs} and \eqref{eqn:Rb} for the selected sample of stellar clusters from Gaia DR2.}
    \label{fig:sc_radii_hist}
\end{figure}

Regarding the ambient density $\rho_0$, because young SCs evolve while being still surrounded by the dense material of the parent Giant Molecular Cloud (GMC) from which they formed, one can estimate the average density of the parent GMC after assuming a star formation efficiency and a mass-radius relation for GMCs. The former can be inferred from other galaxies \citep{Kruijssen2019}, amounting to approximately 1\%. For the latter, several relations are provided in the literature, thus introducing an uncertainty in the calculated $\rho_0$. We account for this uncertainty by considering two mass-radius relations which return two bounding values for $\rho_0$ (see Appendix~\ref{sec:AppDens} for more detail). 
The resulting distributions of $R_{\rm s}$ and $R_{\rm b}$ in our sample of SCs younger than 30\,Myr are shown in Figure~\ref{fig:sc_radii_hist}, with $R_{\rm s}$ and $R_{\rm b}$ peaking at $\sim 2$\,pc and $\sim 24$\,pc respectively.

The density profile of target gas inside of the bubble follows the description of \citep{Weaver+1977}. Within the cluster core, we assume for simplicity a constant value, fixed by a continuity condition with the cold wind profile at the cluster core $R_{\rm c}$, namely
\begin{equation}
n_{\rm c}=\frac{\dot{M_{\rm c}}}{4\pi R^2_{\rm c} v_{w} \, m_p} \qquad r<R_{\rm c}
\label{eq:nc}
\end{equation}
where for $R_{\rm c}$ we adopt the $R50$ value from \cite{gaia2020}, which is the radius that contains half of the cluster members. This angular size is converted into a physical size using the cluster distance, either as reported in the same catalogue, or from the updated Gaia DR3 where available \citep{2023A&A...673A.114Hunt}. 
The model scenario for a wind-blown bubble around a stellar cluster is only valid for compact clusters, namely when $R_{\rm c} \leq R_{\rm s}$. For the cases where this condition is not satisfied, it remains unclear whether a WTS can actually be formed, therefore those clusters are removed from our sample, leaving $\sim 79$ young, compact stellar clusters.

Within the cold wind region, the ambient density profile is expected to drop as $\propto r^{-2}$, and the density of material is determined by the mass loss rate:
\begin{equation}
n_{\rm w}(r)=\frac{\dot{M_{\rm c}}}{4\pi r^2 v_{\rm w}\, m_p} \qquad R_{\rm c} \leq r < R_{\rm s} \,.
\label{eq:nw}
\end{equation}
In the hot wind region, namely between $R_{\rm s}$ and the contact  discontinuity located at $r=R_{\rm cd}$, the density profile is constant, determined by both the mass injected by the wind and by the mass evaporated from the outer shell, $M_{\rm sh}$, whose loss rate evolves as \citep{Weaver+1977}:
\begin{equation}
\begin{aligned}
 \dot{M}_{\rm sh} =& 2 \times 10^{-4} \left(\frac{L_{\rm w,c}}{10^{37} \, {\rm erg/s}} \right)^{\frac{27}{35}}  \left(\frac{n_0}{10\, {\rm cm}^{-3}} \right)^{-\frac{2}{35}} \\
 & \times \left(\frac{t}{1\, {\rm Myr}} \right)^{\frac{6}{35}} \, M_\odot \, {\rm yr}^{-1}\,,
\end{aligned}
\label{eq:mdotS}
\end{equation}
such that the total mass present within the bubble depends on the cluster wind luminosity. Given its weak dependence on time, we treated the mass evaporation rate as constant and computed the amount of evaporated mass from the shell as $M_{\rm sh}=\dot{M}_{\rm sh}(t) t$. This mass will contribute to the mass contained in the shocked wind region, in addition to the mass flow from the wind, that can be estimated as $\dot{M}_{\rm w}t-\dot{M}_{\rm w}/v_{\rm w}$, the two terms representing respectively the total mass ejected during the cluster lifetime and the mass going into the cold wind, which guarantees mass conservation. An additional contribution to the bubble mass comes from the ejecta of all SN explosions occurred during the cluster lifetime, $N_{\rm sn}$, that we compute from the stellar mass function $\xi(M)$ as:
\begin{equation} \label{eq:N_sn}
 N_{\rm sn}(t)=\int_{M^*_{\rm min}}^{M^*_{\rm max,0}} \xi(M) dM
\end{equation}
where $M^*_{\rm min}=\min[8\,M_\odot, M^*_{\rm max}(t)]$. In other words, the total mass in the hot wind reads as
\begin{equation}
 M_{\rm b}=(\dot{M}_{\rm sh}+\dot{M}_{\rm w})\,t -\frac{\dot{M}_{\rm w}(R_{\rm s}-R_{\rm c})}{v_{\rm w}} + N_{\rm sn} (t) M_{\rm ej}
\label{eq:mb}
\end{equation}
where the latter term embeds the contribution of SNe to the bubble mass via their ejecta, whose mass $M_{\rm ej}$ we take equal to 5\,$M_\odot$ for each SN. We hence obtain the number density $n_{\rm b}$ in the bubble as:
\begin{equation}
 n_{\rm b} = \frac{M_{\rm b}}{\frac{4}{3} \pi \left(R^3_{\rm cd}-R^3_{\rm s} \right) m_{\rm p}}~. 
\label{eq:nb}
\end{equation}
Finally, the density in the shell is simply determined by the compressed ISM density, corresponding to the entire amount of swept-up mass, depleted by the evaporated shell mass \eqref{eq:mdotS} as:
\begin{equation}
 n_{\rm sh} = \frac{n_0}{1- R^3_{\rm cd}/R^3_{\rm b}} - \frac{\dot{M}_{\rm sh} t}{m_{\rm p} V_{\rm sh}}~, 
\label{eq:n_sh}
\end{equation}
where $V_{\rm sh}$ is the volume of the shell embedded between the bubble radius $R_{\rm b}$ and the contact discontinuity radius $R_{\rm cd}\simeq 0.95 R_{\rm b}$\citep{gupta2016}.  

\section{Non-thermal particle production in star clusters}
\label{sec:acceleration}
Particles in SCs are accelerated by two distinct processes, i.e. continuous acceleration occurring at the WTS and impulsive acceleration from SNR shocks. The former dominates before the first SN explosions start taking place around $\sim 3$ Myrs. Following the explosion, each SNR will dissolve inside the bubble after a typical time of few $10^4$\,yr (see Eq.~\eqref{eq:t_to} below). Compared to the average time between two SN explosions, which is usually $\gtrsim 10^5$\,yr, we  consider the repeated formation of the WTS and its continuous activity of particle acceleration between different SN events. Here we neglect a possible contribution due to turbulent acceleration \citep{Vink2024}. 

\subsection{Contribution from the wind termination shock}
\label{sec:WTS} 

With regards to acceleration at the WTS, we use the model developed by \cite{Morlino2021}. 
The authors solved the spherically symmetric transport equation of particles within the environment of a SC to obtain the spectrum of accelerated particles $f(r,p)$ as a function of the particle momentum $p$ and the distance to the SC centre $r$.
In our work, we will refer to quantities in the cold wind region ($r<R_s$) with subscript 1, inside the shocked wind ($R_s<r<R_b$) with subscript 2 and outside of the wind blown bubble ($r>R_b$) with subscript 3, see Figure \ref{fig:schematic}. With this nomenclature, the particle distribution function in the shocked wind region $f_2(r,p)$ can be written as: 
\begin{equation}
\label{eqn:f2}
    \begin{aligned}
    f_2(r,p) =& f_s(p) \e^{\alpha(r)}
        \frac{1 + \beta(\e^{\alpha(R_b)}\e^{-\alpha(r)}-1)}
        {1+\beta(\e^{\alpha(R_b)} -1)} + \\
        &+ f_\mathrm{Gal}(p) \frac{\beta\left(\e^{\alpha(r)} - 1\right)}
        {1+\beta(\e^{\alpha(R_b)} -1)}
    \end{aligned}
\end{equation}
while that outside the wind blown bubble $f_3(r,p)$ is
\begin{equation}
\label{eqn:f3}
    f_3(r,p) = f_b(p) \frac{R_b}{r} + f_\mathrm{Gal}(p)\left(1-\frac{R_b}{r}\right)
\end{equation}
where $f_b(p) = f_2(R_b,p)$ is the particle distribution at the bubble boundary, and $f_s(p)$ is the one at the WTS:
\begin{equation}\label{eqn:fs}
    f_s(p) = 
        \frac{3 \, \xi_\mathrm{CR} \, n_1 u_1^2}{4 \pi \Lambda (m_p c)^3 c^2}
        \left( \frac{p}{m_p c} \right)^{-s_{\rm wts}}
        \e^{-\Gamma(p)} \,.
\end{equation}
In the latter expression, $c$ is the in-vacuum speed of light and $\xi_{\mathrm{cr}}$ is the acceleration efficiency expressed as the fraction of incoming momentum flux converted into CR pressure\footnote{Note that the efficiency can  be defined with respect to the fraction of wind luminosity converted into CR luminosity, i.e. $4\pi R_{\rm TS}^2u_2 \int d^3 p \, \epsilon(p)f_{s}(p)= \xi_{\rm cr}' L_w$, with $\epsilon(p)$ being the particle kinetic energy. It is easy to show that $\xi_{\rm cr}'/\xi_{\rm cr} = 6 \Lambda'/(\sigma\Lambda)$ where $\Lambda'= \int_{x_{\rm inj}}^{\infty} x^{2-s_{\rm wts}} [(1+x^2)^{1/2}-1] \e^{-\Gamma(x)} dx$. For $s_{\rm wts}=4$ the ratio is 1.38 and decreases for $s_{\rm wts}>4$.}, i.e. $P_{\rm cr} = \xi_{\rm cr}\, \rho_1 u_1^2$ , $\rho_1$ and $u_1=v_w$ being respectively the plasma density and velocity upstream of the WTS. The slope is $s_{\rm wts}=3u_1/(u_1-u_2)$, $u_2$ being the plasma downstream velocity, whilst the normalisation factor is $\Lambda = \int_{x_{\rm inj}}^{\infty} x^{4-s_{\rm wts}} \left( 1+x^2\right)^{-1/2} \e^{-\Gamma(x)} dx$, with $x=p/m_{\rm p}c$. In general for strong shocks $s_{\rm wts}\simeq 4.0$, such that $\Lambda \simeq \ln\frac{p_{\max,1}}{m_p c}$. We assume that the sonic and Alfv\'enic Mach number are equal and, because the latter is close to 6 (see discussion below Eq.\eqref{eq:B_r}) we have $s_{\rm wts} \simeq 2.1$ 
The functions $\alpha$ and $\beta$ are:
\begin{eqnarray}
  \alpha(r,p) &=& \frac{u_2 R_s}{D_2(p)} \left(1-\frac{R_s}{r}\right) \\
  \beta(p) &=& \frac{D_3(p) R_b}{u_2 R_s^2}
\end{eqnarray}
\noindent
where the diffusion coefficients are assumed to be spatially constant within each of the regions: $D_2$ in the shocked wind region and $D_3$ in the ISM. 
Finally, the cutoff term, $e^{-\Gamma(p)}$, depends on the maximum energy and has an involved expression that depends on the particle distribution function itself. However, it can be approximated by:
\begin{equation} \label{eq:cutoff}
    \e^{-\Gamma(p)} \simeq \left[1+a_1\left(\frac{p}{p_\mathrm{max,1}}\right)^{a_2}\right] \,\exp\left[-a_3\left(\frac{p}{p_\mathrm{max,1}}\right)^{a_4}\right]~,
\end{equation}
where the values of the $a_i$ parameters (provided below) are obtained from a fitting procedure to the full solution and depend on the diffusion scenario adopted.
Far away from the bubble, the CR spectrum reduces to the average sea of Galactic CR $f_\mathrm{Gal}(p)$, which is assumed to be the same as that measured close to Earth \citep{AMS2015}.

In the model by \cite{Morlino2021}, the cluster wind velocity is assumed to be constant in the cold wind region $u_1$, compressed when passing the TS to $u_2=u_1/\sigma$ with compression factor $\sigma$ (typically $\sigma=4$) and decreasing with radial distance to the cluster centre as $\propto r^{-2}$ in the shocked wind region, hence:
\begin{equation}
    u(r) = 
    \begin{cases}
        u_1 & r < R_s \\
        u_2 \left(\frac{R_s}{r}\right)^2 & R_s < r < R_b \\
        0 & r > R_b \,.
    \end{cases} 
\end{equation}
The magnetic field inside the bubble is likely dominated by magneto-hydrodynamic (MHD) instabilities triggered by the non-uniform plasma motion. We assume that the turbulence is preferentially generated at a typical length scale, $L_{\rm coh}$, comparable to the size of the SC core, and hence that the magnetic field can be treated as isotropic on the bubble scale. Motivated by the modelling of diffusion properties inside the wind blown bubble around Cygnus OB2 \citep{Blasi-Morlino:2023, Menchiari+2024}, we assume magnetic turbulence inside SC wind bubbles to follow a Kraichnan scaling with diffusion coefficient 
\begin{equation}\label{eqn:Kra_diffusion}
    D_{\rm Kra}(p,B) = \frac{c}{3} \sqrt{r_L(p,B) \, L_\mathrm{coh}}
\end{equation}
with magnetic field strength $B$ and particle Larmor radius
\begin{equation}
    r_L(p,B)=\frac{p c}{e B}
\end{equation}
$e$ being the elementary charge. The coherence scale of the magnetic field in the shocked wind region is fixed to $L_\mathrm{coh} = 2\, \rm pc$ for all SCs because this is the peak value of the SC core in our sample. 
However, we emphasise that detailed knowledge about the diffusion properties inside SC wind bubbles is limited: a preferential direction of the magnetic field - orthogonal or parallel to the wind direction - will significantly change the propagation of accelerated particles. We will show below that in some cases the Bohm diffusion seems to be more consistent with some observed $\gamma$-ray spectra. In such a case the diffusion coefficient is $D_{\rm Bohm}= r_L(p,B)\, c/3$.
In the scenario of Kraichnan diffusion, the parameters used in Eq.\eqref{eq:cutoff} are $a_1=5$, $a_2=0.449$, $a_3=12.52$, $a_4=0.643$, while for Bohm $a_1=8.94$, $a_2=1.296$, $a_3=5.310$, $a_4=1.132$ \citep{Menchiari+2024}.
The assumption of Bohm diffusion may sound extreme, because it requires an equal power over all scales of the turbulence. However, at lower spatial scales such a flat spectrum can be explained by self-generated turbulence \citep{Morlino2021}, while at the highest scale there is the possibility that the turbulence is injected at several scales with a similar power such that the final convolution could resemble an effective Bohm scaling. Such hypotheses should be verified by means of numerical simulations.

The assumption of spatially uniform diffusion in the bubble may also be violated. In particular, the diffusion coefficient may increase if the magnetic field is damped while the plasma is advected towards the bubble boundary. Such a situation may occur, e.g. due to ion-neutral damping by neutral material left from the shell fragmentation. However, HD instabilities that develop at the contact discontinuity may enhance the local magnetic turbulence resulting into a suppressed diffusion \citep{vieu2020}. In the former scenario we expect that the size of the gamma-ray/neutrino emitting region will be reduced [cfr. \S~\ref{sec:Wd1}], while the latter should  enhance the emission at the bubble boundary. Additionally, on length scales $\lesssim L_\mathrm{coh}$ there may be a locally preferred direction to the diffusion; yet, as $L_\mathrm{coh}$ is typically $\ll R_b$, this is not relevant on the scale of the full SC bubble. We therefore neglect those effects here, retaining the assumption of spatially uniform diffusion throughout.
The local direction of the magnetic field also affects the particle injection in the accelerator, thus determining the overall acceleration efficiency \citep{Caprioli-Spitkovsky:2014}. For this reason, the acceleration efficiency is not determined a priori, but regarded as a free parameter with a fiducial value of a few percent.

In Eq.~\eqref{eq:cutoff}, $p_{\max,1}$ is the maximum momentum obtained from the confinement condition in the upstream region, i.e. 
\begin{equation} \label{eq:pmax1}
    D_1(p_{\max,1})/u_1 = R_s \,,
\end{equation}
which gives, for the Bohm and Kraichnan cases:
\begin{align}
    c p_{\max,1}^{\rm Bohm} &= 3 u/c \, eB \, R_s \,\\
    c p_{\max,1}^{\rm Kra} &=  \left(3 u/c \, R_s \right)^2  eB/L_{\rm coh} \,.     
\end{align}
Whilst in the Bohm case the maximum energy does not depend on the turbulence coherence scale, in the Kraichnan case it is inversely proportional to it.

We stress that $p_{\max,1}$ represents the \emph{nominal} maximum momentum of accelerated particles and not the \emph{effective} maximum momentum $p_{\max,\it{eff}}$, which is instead further affected by the escaping condition in the downstream and accounted for in the shape of the cut-off in Eq.~\eqref{eq:cutoff} \cite[see the discussion in][]{Morlino2021}. The latter can be defined, in analogy to the purely exponential cut-off spectral shape, through the condition $\Gamma(p_{\max,\it{eff}}) = 1$. Using Eq.~\eqref{eq:cutoff} one obtains $p_{\max,\it{eff}}/p_{\max, 1} \simeq 0.53\,(0.05) $ for the Bohm (Kraichnan) case, which shows that $p_{\max,1}$ should be regarded as an upper limit for the maximum momentum. 
Moreover the approach used by \cite{Morlino2021} does not account for the adiabatic losses due to the expansion of the TS.
The energy where those losses become relevant can be estimated by comparing the expansion time, $R_b/u_1$, with the diffusion time, $R_b^2/4 D_1$. Not surprisingly such a condition is similar to Eq.~\eqref{eq:pmax1}.
In the rest of this work we will always refer to $p_{\max,1}$ as the maximum momentum but the caveat discussed above should be kept in mind.

The magnetic field in the unshocked wind region is computed assuming that a fraction $\xi_{B}$ of the wind kinetic luminosity is converted into magnetic energy flux:
\begin{equation} \label{eq:B_r}
    B(r) = \frac{1}{r} \sqrt{\frac{\xi_B}{2} \dot{M}_c u_1}
\end{equation}
while in the shocked wind region the magnetic field is assumed to be uniform and stronger by a factor $\sqrt{11}$ due to compression at the WTS of a randomly oriented magnetic field.
Benchmark values for $\xi_\mathrm{CR}$, $\xi_\mathrm{B}$, $L_\mathrm{coh}$ and $\sigma$ assumed in this work are listed in Table~\ref{tab:default}, unless differently specified. Typical values of $B$ obtained in the shocked wind region are $\approx (1-12)\, \mu$G for wind luminosity in the range $\sim 10^{34}-10^{38}$\,erg\,s$^{-1}$. Interestingly, Eq.~\eqref{eq:B_r} allows us to write the Alfv\'en Mach number in a compact form as $M_A = \sqrt{2/\xi_B}$. For the adopted benchmark value of $\xi_B = 0.05$ we have $M_A=6.3$, independent of the wind luminosity. Assuming equipartition between thermal and magnetic energy,  the sonic Mach number has the same value.

Lastly, the stationary solution assumed here requires that the acceleration time to achieve the maximum energy should be smaller than the elapsed time between two SN explosions. Estimating the acceleration time for plane parallel shocks as $t_{\rm acc} \simeq 8 D_1(p)/u_1^2$, together with Eq.\eqref{eq:pmax1} we obtain: 
\begin{eqnarray}
  t_{\rm acc}(p_{\max,1}) &\simeq& 8 \, \frac{R_{\rm s}}{u_1} = \\
  &=& 3.9 \times 10^4 \left( \frac{R_s}{10\,\rm pc} \right) \left(\frac{u_1}{2000 \, \rm km/s} \right)^{-1}\, \rm yr \nonumber
\end{eqnarray}
which is a factor of a few smaller than the average time occurring between two subsequent SN explosions. This assumption hence applies in general, whilst for individual SCs this stationary solution may not always be established between every SN explosion.

\subsection{Contribution from SNRs inside SCs}
\label{sec:SNR}
In addition to CRs accelerated at the WTS, SCs older than $\approx 3$\,Myr are expected to be populated by non-thermal particles produced at SNR shocks.
A full and consistent treatment of this scenario is not easy to achieve for several reasons.
Firstly, SN explosions inject energy on a time scale much smaller than the cluster age, hence the bubble structure is probably modified in a non-stationary way, producing transient phases \citep{vieu2020}. Moreover, the evolution of an SNR shock inside the bubble itself is complicated by the inhomogeneity of the medium: the presence of the SC core, the stellar winds and the termination shocks results in a non-smooth evolution of the shock and may produce multiple reflected shocks \citep{Dwarkadas:2005, Dwarkadas:2007}. Consequently, a correct description of particle acceleration requires a time-dependent calculation. Particle acceleration is also strongly affected by the conditions of the plasma inside the bubble where the shock expands, especially by the uncertain magnetic field configuration.

Despite these complications, here we give an approximate estimation of the SNR contribution to the SC $\gamma$-ray emission, making several simplifying assumptions. 
Firstly, we remark that when SNe have exploded, the size of the bubble is determined using the total luminosity given by Eq.~\eqref{eq:L_tot}, as anticipated in \S\ref{sec:model}. Secondly, when an SN explodes, it initially expands inside the SC core, then in the cold fast wind until it reaches the WTS, and finally proceeds through the hot bubble. Because the mass enclosed in the SC core and in the cold wind is usually much smaller than the mass of the ejecta, the  kinetic energy of SN is released mainly in the hot bubble\footnote{This condition has been verified for all the SCs in our sample with one exception, namely Wd1. For this cluster, the wind mass enclosed within the TS radius is $\sim 5.5\, \rm M_{\odot}$ hence of the same order of the typical ejecta mass of a SN.}. Hence we approximate the SNR evolution as if it were expanding only inside the hot and uniform bubble. Despite the fact that we are thus neglecting the very early evolutionary stage of SNRs, when the shock expands inside the SC core and particle acceleration probably reaches the highest energies \citep{2010MNRAS.406.2633Schure}, we note that the total swept-up mass at this stage remains small enough that the amount of accelerated particles is negligible with respect to the subsequent phase and contributes only marginally to the final spectrum \citep{Cristofari+2020}. Our model, therefore, underestimates the CR spectrum at the highest energies.

We can approximate the SNR evolution mainly in two phases: the ejecta dominated (ED) phase, when the swept-up mass is smaller than the ejecta mass, and the Sedov-Taylor (ST) phase, when $M_{\rm swept-up}>M_{\rm ej}$. During the ED phase, the shock speed is roughly constant and is given by $u_{\rm ed} = (2 E_{\rm sn}/M_{\rm ej})^{1/2}$. The ED phase lasts a time $t_{\rm st} = R_{\rm st}/u_{\rm ed}$ where the radius $R_{\rm st}$ is determined by the condition that the swept-up mass is equal to the ejecta mass, i.e. $M_{\rm ej} = \int_{0}^{R_{\rm st}} \rho(r) \, 4\pi r^2 dr$ which gives
\begin{equation}
    R_{\rm st}= R_{\rm s} + \left[\frac{3}{4 \pi n_{\rm b} m_{\rm p}} \left( M_{\rm ej} - \frac{\dot{M}_c R_{\rm s}}{v_w} \right) \right]^{1/3}
\end{equation}
In contrast, during the ST phase, the shock's radius and velocity scale as $R_{\rm sh}(t) = R_{\rm st} (t/t_{\rm st})^{2/5}$ and $u_{\rm sh}(t) = dR_{\rm sh}/dt = (2/5)\, R_{\rm sh}(t) /t$. As we show below, the amount of accelerated particles during the two phases is roughly the same.

For a constant shock speed, the CR distribution at the shock is given by the usual power-law with exponential cut-off
\begin{equation}    
\label{eq:f_snr}
    f_{\rm snr}(p) = \frac{3 \, \xi_{\rm cr} n_{b} u_{\rm sh}^2}{4\pi\, \Lambda (m_p c)^3 c^2} 
    \left( \frac{p}{m_p c}\right)^{-s_{\rm sn}} \e^{-p/p_{\max}} \,.
\end{equation}
Eq.~\eqref{eq:f_snr} is identical to Eq.~\eqref{eqn:fs}, with the exception of the exponential term which differs in maximum momentum (see below). Also, note that the numerical value of the slope $s_{\rm sn}$ can be different from the $s_{\rm wts}$ that appears in Eq.~\eqref{eqn:fs}. 
For SNR we expect Mach numbers larger than the WTS one. Typical shock speeds in the ejecta dominated phase are $\gtrsim 5000$ km s$^{-1}$ while the temperature in the bubble ranges between $10^6$ and $10^7$ K \citep{El-Badry+2019MNRAS,Rodriguez+2025}, implying a sound speed of a few hundreds km/s. Hence Mach numbers between 10 and 50 are expected. This allows for efficient magnetic field amplification which typically results in spectra steeper than $p^{-4}$. Results from simulations \citep{Haggerty-Caprioli:2020} as well as direct observations of isolated SNRs suggest spectral slopes between 4.2 and 4.4 \cite[see, e.g.,][]{Morlino-Caprioli:2012, Caprioli:2012}.
Finally, we assume that the acceleration efficiency remains constant during the ED and ST phases as suggested by results from particle-in-cell simulations \cite[e.g.][]{Caprioli-Spitkovsky:2014}.
The amount of CRs produced by a single SNR during the ED phase is
\begin{equation}
    F_{\rm snr}(t<t_{\rm st}) = \int_{0}^{t_{\rm st}} f_{\rm snr}(p) \frac{u_{\rm sh}}{4} \, 4\pi R_{\rm sh}^2 dt
    =  \frac{\pi}{3} \, R_{\rm st}^3 \, f_{\rm snr}(p)
\end{equation}
where $u_{\rm sh}/4=u_{\rm ed}/4$ is the downstream speed and we have used $u_{\rm ed} dt= dR_{\rm sh}$. 
For later times, usually it is assumed that the shock keeps accelerating particles at least up to the beginning of the radiative phase. However, a SN exploding inside hot bubbles usually becomes subsonic before becoming radiative \citep{Parizot+2004}; hence, for the ST phase we integrate up to the {\it turn-off} time $t_{\rm to}$, namely the moment when the shock speed becomes equal to the sound speed.
We obtain:
\begin{equation}
    F_{\rm snr}(t>t_{\rm st}) = 3 \left( \frac{2}{5} \right)^{3} \ln\left( \frac{t_{\rm to}}{t_{\rm st}} \right) \, F_{\rm snr}(t<t_{\rm st}) \,.
\end{equation}
We can estimate $t_{\rm to}$ using the sound speed, $c_{s,b}= \sqrt{\gamma P_b/\rho_b}$, where the bubble pressure is given by Eq.~\eqref{eqn:Pb}, the density is approximated as $\rho_{\rm b} \simeq \dot{M}_c t /(4\pi/3 R_{\rm b}^3)$, and $\gamma=5/3$ is the adiabatic index. The sound speed can thus be written as
\begin{equation} \label{eq:c_s}
  c_{s,b} \simeq 0.5 \,\eta_{m}^{1/2} v_w \,.
\end{equation}
By equating it with the SNR shock speed, we can write the {\it turn-off} time as
\begin{equation} \label{eq:t_to}
  t_{\rm to} \simeq 3.1 \, \eta_m^{-5/6} \left(u_{\rm ed}/v_w \right)^{5/3} t_{\rm st} \,.
\end{equation}
For typical SNR parameters and for mechanical efficiencies $\eta_m$ in the interval [0.1,1], the ratio $t_{\rm to}/t_{\rm st}$ ranges from a few to a few tens. Using 10 as a fiducial value, the ratio between the two contributions is $F_{\rm snr}(t>t_{\rm st})/F_{\rm snr}(t<t_{\rm st})\approx 0.44$. In addition, the maximum energy decreases over time during the ST phase, such that we can neglect the contribution of accelerated particles produced during it\footnote{We neglect all complications due to adiabatic losses and particle escape from the shock upstream.}.
Eq.~\eqref{eq:t_to} also tells us that the typical turn-off time of a SNR is $\lesssim 10^5$\,yr. Hence, we can assume, as anticipated at the beginning of  \S\ref{sec:acceleration} that the TS is re-formed between two SN explosions. 

On the million year long timescale of evolution of clusters, the particles produced by a single SNR will escape the bubble because of both diffusion and advection. Hence the CR distribution is contributed to by all the SNe exploded during an escape time, $N_{\rm sn}(t_{\rm esc})$. We set this escape time to be equal to the advection time in the bubble, namely $t_{\rm esc} \approx t_{\rm adv} = \int_{R_s}^{R_b} dr/u(r)$, providing us with a value equal to about 60\% of the cluster age. The average distribution over the bubble volume will then be
\begin{equation} \label{eq:f_snr_1}
    \langle f_{\rm snr} \rangle =  N_{\rm sn}(t_{\rm esc}) 
    \frac{F_{\rm snr}}{V_{\rm b}} 
    = \frac{N_{\rm sn}(t_{\rm esc})}{4} \, \frac{R_{\rm st}^3}{R_b^3} \, f_{\rm snr}(p)\,.\end{equation}
The ratio between CRs produced by SNRs and those produced by the WTS at a specific particle momentum is:
\begin{flalign} \label{eq:f_ratio}
    \mathcal{R} &\equiv \left.\frac{\langle f_{\rm snr} \rangle}{f_{\rm ts}}\right|_{p=m_p c} =
    \frac{n_b u_{\rm ed}^2}{n_{w}(R_s) v_w^2}
    \frac{R_{\rm st}^3}{R_{\rm b}^3} 
    \frac{N_{\rm sn}(t_{\rm esc})}{4} =  \nonumber \\
    &= 1.65 \, \frac{N_{\rm sn} E_{\rm sn}}{L_{\rm w,c} t_{\rm adv}}  
    = 1.74 \, N_{\rm sn}(t_{\rm esc}) \times \nonumber\\
    & \times \left( \frac{E_{\rm sn}}{10^{51}\, \rm erg}\right)
    \left( \frac{L_{\rm w,c}}{10^{37}\rm erg \, s^{-1}} \right)^{-1}
    \left( \frac{t}{\rm 3 \,Myr} \right)^{-1}  
\end{flalign}
such that Eq.~\eqref{eq:f_snr_1} can be rewritten as 
\begin{equation} \label{eq:f_snr_2}
    \langle f_{\rm snr}(p) \rangle =  \mathcal{R}\, f_{\rm ts}(m_p c) \, \left( \frac{p}{m_p c}\right)^{-s_{\rm snr}} \e^{-p/p_{\max}} \,.
\end{equation}
\noindent
Note that in Eq.~\eqref{eq:f_ratio} we used $R_{\rm b}$ from Eq.~\eqref{eqn:Rb} and assumed that the acceleration efficiency is the same for both types of shocks. Not surprisingly, the ratio $\mathcal{R}$ depends solely on the energy input of SNe and stellar winds.

Now, we evaluate the maximum momentum $p_{\max}$ of particles accelerated at SNRs, which represents the most uncertain part of the calculation due to the lack of knowledge about the magnetic turbulence development inside the bubble. We adopt the time-limited condition of SNRs by equating the acceleration time, $t_{\rm acc}\simeq 8 D/u_{\rm ed}^2$, with the beginning of the ST age, $t_{\rm st}= R_{\rm st}/u_{\rm ed}$. 
Writing the diffusion coefficient as $D= r_L c/(3 \mathcal{F})$, where $\mathcal{F}$ is the logarithmic power spectrum of the magnetic turbulence containing information about the turbulence injection scale, we get a maximum energy $E_{\max}=cp_{\max}$ equal to
\begin{equation}
\label{eq:EmaxSNR}
\begin{aligned}
    E_{\max} =& \frac{3}{8} \frac{u_{\rm ed} R_{\rm st}}{c} e\,  B \, \mathcal{F} = \\
    =& 580 \, \mathcal{F} \left( \frac{B}{10 \, \rm \mu G} \right) 
    \left( \frac{u_{\rm ed}}{5000 \, \rm km/s} \right)
    \left( \frac{R_{\rm st}}{10 \, \rm pc} \right) \, \rm TeV
    \end{aligned}
\end{equation}

If we use solely the diffusion determined by the wind turbulence, i.e. Eq.~\eqref{eqn:Kra_diffusion}, such that $\mathcal{F} = (r_L/L_{\rm coh})^{1/2}$, the maximum energy is: 
\begin{eqnarray}
\label{eq:EmaxSNR_Kra}
    E_{\max} &=& \frac{e B}{L_{\rm coh}} \left(\frac{3}{8} \frac{u_{\rm ed} R_{\rm st}}{c} \right)^2 = \\
    &=& 48 \left( \frac{B}{10 \, \rm \mu G} \right)
    \left( \frac{L_{\rm coh}}{2 \rm pc} \right)^{-1}
    \left( \frac{u_{\rm ed}}{5000 \, \rm km/s} \right)^{\frac{1}{2}}
    \left( \frac{R_{\rm st}}{10 \, \rm pc} \right)^{\frac{1}{2}} \, \rm TeV \nonumber
\end{eqnarray}

On top of the pre-existing turbulence, the magnetic field may further be enhanced by the CR streaming instability (SI) through either resonant or non-resonant modes. The latter dominates over the former only when the shock speed and the upstream density are both very large ($u_{\rm sh} \gtrsim 5000 \, \rm km/s$ and $n \gtrsim 10\, \rm cm^{-3}$), conditions that are typically realized during the first hundred years of the SNR evolution \citep{Bell+2013}. Because such a phase is not described here with accuracy, we will only consider the resonant mode.
During the ED phase, the resonant SI is strong enough that $\mathcal{F}\approx 1$, hence we should use the expression for the saturation in the non linear regime \citep{Blasi:2013}, i.e.
\begin{equation} \label{eq:F_res}
\begin{aligned}
    \mathcal{F}_{\rm res} =& \left( \frac{\pi \, \xi_{\rm cr} \, c}{6\, \Lambda \, u_{\rm sh}} \right)^{1/2} = \\
    =& 0.5\, \left( \frac{\xi_{\rm cr}}{0.1} \right)^{\frac{1}{2}}
    \left( \frac{E_{\rm sn}}{10^{51} \rm erg} \right)^{\frac{1}{4}}
    \left( \frac{M_{\rm ej}}{5 M_{\odot}} \right)^{-\frac{1}{4}}
\end{aligned}
\end{equation}
With such a value of $\mathcal{F}$ and assuming $B \simeq 10\,\mu$G, the maximum energy in Eq.~\eqref{eq:EmaxSNR} can achieve $\sim 200$\,TeV.
In this approach, the maximum energies achieved at SNR shock and at the WTS are not independent quantities but are connected through the value of the magnetic field in the bubble (see Fig.~\ref{fig:emax} and related discussion in Sec.~\ref{sec:SC_emax}).

\cite{Vieu-Reville:2023} performed a similar calculation for the SNR maximum energy, but distinguished between compact and loose clusters. In the former case, where a collective WTS is not formed, they estimated a maximum energy that is compatible, within a factor two, to Eq.~\eqref{eq:EmaxSNR}. In the latter case, they instead estimated values roughly one order of magnitude larger than ours. This is due to the different scenario adopted, in which they only consider SNe exploding at the edge of a SC core, with the shock expanding inside the free wind region where the magnetic field is highly amplified due to MHD instabilities. Even if such a phase is probably more favourable in achieving higher energies, as discussed above, the swept-up mass inside the free wind region is usually much smaller than the ejecta mass, such that the majority of the power in CRs is released when the shock expands inside the hot bubble. To properly address this issue, one would need to correctly describe the shock evolution through the entire bubble structure, which is beyond the scope of this work.

\section{Secondary particle production in hadronuclear collisions}
\label{sec:secondaries}
Proton-proton collisions result in the production of secondary particles, including neutral and stable messengers such as $\gamma$ rays and neutrinos, that can hence be used as astronomical probes for the occurrence of such interactions. Given the very mild energy dependence of the cross-section, the energy spectrum of final products closely resembles that of primaries \citep{celli2020}. A formal computation of the emissivity $\psi_{\gamma,\nu}(E_\gamma,r)$ of secondary particles as a function of space is performed here following the treatment of \cite{Kelner06}: 
\begin{equation}
  \psi_{\gamma,\nu} (E_{\gamma,\nu},r) = c \, n(r) \int_{E_{\gamma,\nu}}^{\infty}  \sigma_{\mathrm{pp}}(E) \, f(E,r) \, K_{\gamma,\nu} \left(\frac{E_{\gamma,\nu}}{E},E\right)\frac{dE}{E}~,
\label{eq:gflux}
\end{equation}
where $n(r)$ is the number density profile of the target medium, discussed in Sec.~\ref{sec:model}, $\sigma_{\mathrm{pp}}$ the total inelastic proton cross-section, while $f(E,r)$ is the particle distribution function in energy (i.e. $f(E) dE = f(p) d^3p$). Kernel functions describing the production of $\gamma$ rays and neutrinos $K_{\gamma,\nu} (E_{\gamma,\nu}/E,E)$ are taken from \cite{Kelner06}. The particle fluxes on Earth are obtained by integrating the emissivities (in ph~cm$^{-3}$~s$^{-1}$~TeV$^{-1}$) over the bubble size and accounting for the cluster distance $d$, such that these can finally be expressed as: 
\begin{equation}
    F(E_{\gamma,\nu}) = \frac{1}{4\pi d^2} \int_{0}^{R_b} 4\pi r^2 \psi_{\gamma,\nu} (E_{\gamma,\nu},r) \, dr .
    \label{eq:earthflux}
\end{equation}
Although cluster distance estimates are provided in \cite{gaia2020}, where available we instead adopt distances estimated from the third Gaia data release (DR3) \citep{2023A&A...673A.114Hunt}.

\begin{figure}
    \centering
    \includegraphics[width=\columnwidth]{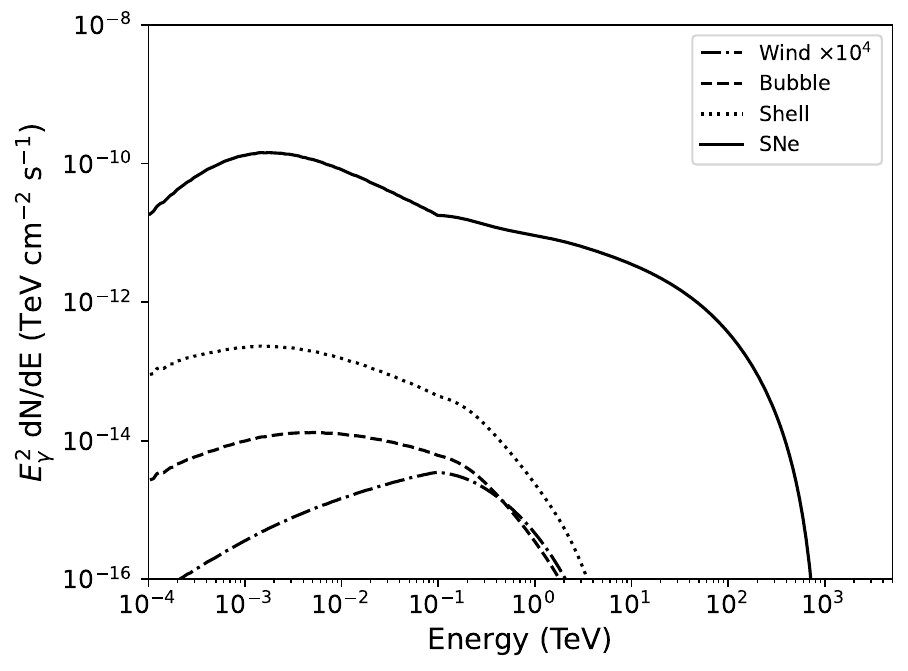}
    \caption{Predicted $\gamma$-ray emission arising from the different regions of the SC bubble, also showing the influence of SNe. The contribution from Galactic CRs is not included (i.e. $f_{\rm Gal}=0$ in Eq.~\eqref{eqn:f2}).}
    \label{fig:gamma_dummy}
\end{figure}

Before applying the model to the sample of Gaia SCs, we first discuss some general properties of $\gamma$-ray emission from a typical cluster located at a distance of 1~kpc. When estimating the $\gamma$-ray flux, we set $f_{\rm Gal}$ to zero, to probe the contribution from the SC only. In $\gamma$-ray data analysis, the contribution due to $f_{\rm Gal}$ is often not relevant due to the background subtraction methods applied\footnote{Note that the background subtraction could remove some gamma-ray emission due to particles escaping from the clusters. However this effect should be minor.}.
We adopt representative values for the various cluster-dependent properties as listed in Table~\ref{tab:default}. 
Figure~\ref{fig:gamma_dummy} shows the $\gamma$-ray emission due to accelerated particles emerging from different parts of the system: cold wind, hot bubble and thin shell. 
The $\gamma$-ray emission depends primarily on the density of the medium and therefore also on the mass-loss rate of the cluster, as in Eqs.~\eqref{eq:nb} and \eqref{eq:n_sh}.
The average density of the bubble for typical cluster properties is $\sim 0.01$\,cm$^{-3}$, whilst the density of the surrounding shell of swept up ISM material is $\sim 100$\,cm$^{-3}$, considerably higher than that of the cold wind region from Eq.~\eqref{eq:nw}. Unsurprisingly, the total $\gamma$-ray flux is dominated by the contributions from the shell.
Figure~\ref{fig:gamma_dummy} also shows that the contribution from the bubble is harder than the one from the shell, reflecting the fact that the particle distribution function is harder at distances closer to the termination shock.

Figure \ref{fig:gamma_dummy} also shows the emission due to SNRs (solid red line) assuming that a representative number of 3 SNe occurred along the last advection time. Note that the spectrum due to SNRs is steeper in that we assumed $s_{\rm snr}= 4.3$, while for the WTS $s_{\rm wts}=4.1$. The acceleration efficiency is set to 5\% for both WTS and SNRs.

\begin{figure}
    \centering
   \includegraphics[width=\columnwidth]{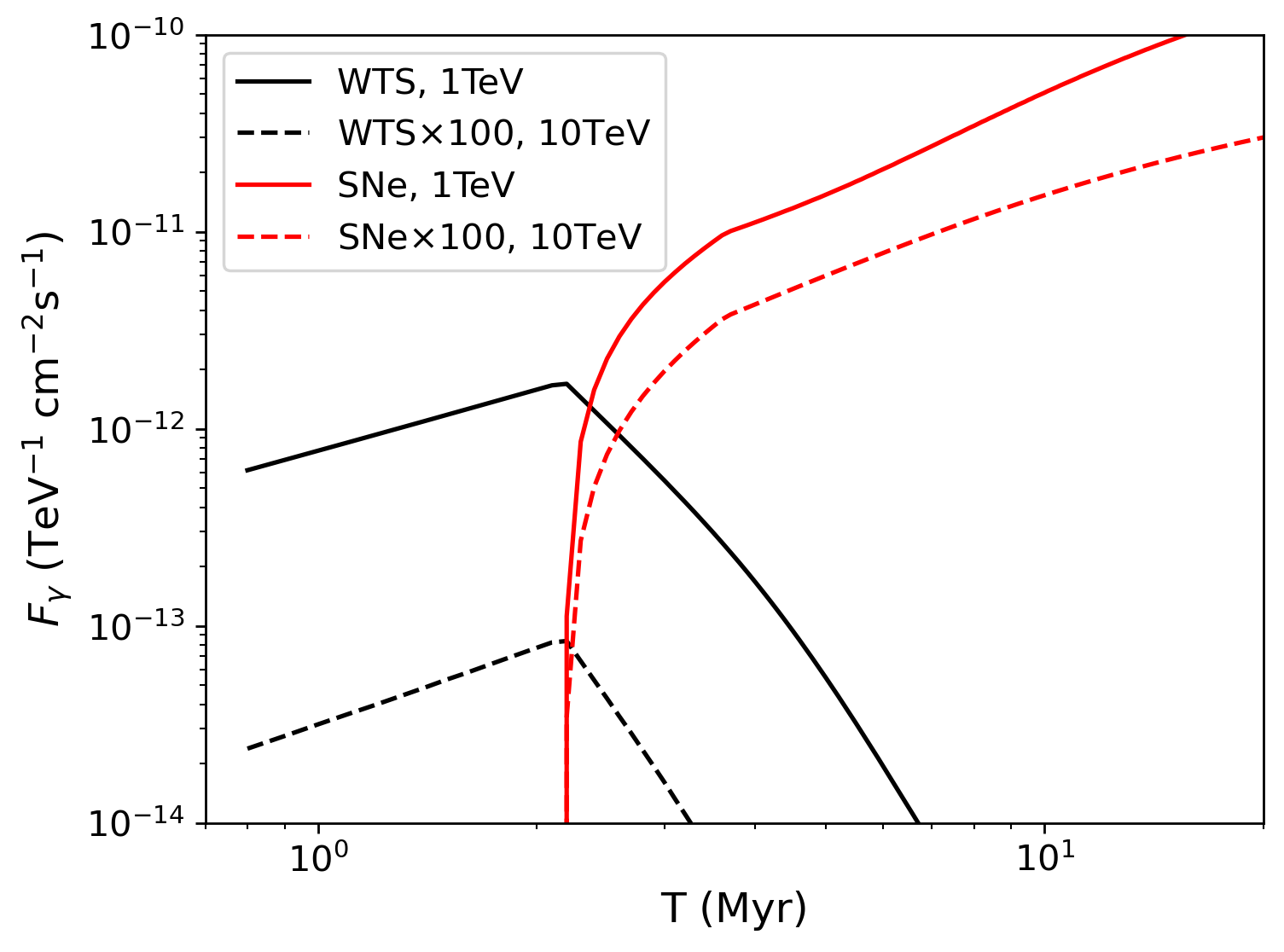}
    \caption{Expected time variation of gamma-ray flux due to particles accelerated at the WTS and at SNR shocks at 1 TeV and 10 TeV. Note that the latter is multiplied by a factor 100 for ease of visibility. }
    \label{fig:cases}
\end{figure}

The relative contribution between the WTS and SNRs is better explored in Figure~\ref{fig:cases}, showing the total differential $\gamma$-ray flux at 1\,TeV and 10\,TeV as a function of the cluster age: the WTS contribution dominates up to 3\,Myr, after which it is overtaken by the SNR one. 
Note that the latter is a smooth function because the number of SNe is approximated by the continuous expression provided in Eq.~\eqref{eq:N_sn}, while in reality it will be an integer number for each cluster.

\begin{table}
    \centering
    \begin{tabular}{ll|l}
        \toprule\toprule
        {} & Parameter & Value \\ 
        \midrule
        CR & $\xi_\mathrm{CR}$ & $5\%$ \\
        {} & $\xi_\mathrm{B}$  & $5\%$ \\
        {} & $L_\mathrm{coh}$  & $2\,\mathrm{pc}$ \\
        {} & $s_{\rm wts}$     & $4.1$ \\
        {} & $\delta$          & 0.5 (Kraichnan) \\
        {} & $D_3$ (10\,GeV)   & $3.0\times10^{28}$\,cm$^2$/s \\        
        \midrule
        Cluster & $t$& 3\,Myr \\
        {} & $M_{\rm sc}$      & $10^4\, M_{\odot}$  \\ 
        {} & $\dot{M}_{\rm c}$         & $10^{-5}$\,M$_\odot$ /yr \\
        {} & $L_{\rm w,c}$         & $10^{37}$\,erg s$^{-1}$ \\
        {} & $n_{0}$     & 10\,cm$^{-3}$ \\
        {} & $d$               & 1\,kpc \\
        {} & $\eta_{\rm m}$   & 0.3 \\
        \midrule
        SNR & $N_{\rm sn}$     & 3   \\
        {} & $M_{\rm ej}$      & 5\,$M_\odot$ \\
        {} & $E_{\rm sn}$      & $10^{51}$\,erg \\
        {} & $s_{\rm sn}$      & $4.3$ \\
        \bottomrule\bottomrule
    \end{tabular}
    \caption{Default parameter values used for the prototype SC model discussed in Sec.~\ref{sec:secondaries}. }
    \label{tab:default}
\end{table}

\section{High-energy emission from Young Star Clusters from Gaia DR2}
\label{sec:DR2results}


We now apply our model to the sample of young stellar clusters from Gaia DR2. In addition to the winds from main sequence stars, we include a contribution from the associated WR stars as reported in the Galactic catalogue \citep{2015MNRAS.447.2322R_WRCat}. Our method allows us to characterise the stellar population of each cluster as a function of time and hence to compute the number of stars that have already undergone SN explosions at the SC age. For simplicity, we model this additional contribution to the kinetic luminosity of the system as an injection term at the collective cluster wind, following Sec.~\ref{sec:SNR}.

\subsection{Maximum energy of accelerated particles}
\label{sec:SC_emax}
Before computing hadronic interactions, we first {investigate which clusters among the selected sample may generate PeV emission, thereby providing an indication for how many clusters are likely to be PeVatrons. 
Figure~\ref{fig:emax} shows the distribution of maximum energies to which CR protons can be accelerated for each SC in our sample as due to the WTS and SNRs. In the case of Kraichnan diffusion, $E_{\max}^{\rm SNR}$ is always larger than $E_{\max}^{\rm WTS}$ except for one cluster, Wd1. The maximum energy $E_{\max}^{\rm WTS}$ is always smaller than 1\,PeV while $E_{\max}^{\rm SNR} \sim1$\,PeV only for one SC, Danks\,1. For the case of Bohm diffusion six SCs have $E_{\max}^{\rm WTS} \gtrsim E_{\max}^{\rm SNR}$ and, among these, only Wd1 is found to have $E_{\max}^{\rm WTS} \geq 1$\,PeV (see Table~\ref{tab:cluster_top}).

The technique applied for the computation of cluster masses relies on the number of stars detected by the Gaia satellite and associated to a cluster using machine-learning techniques. As discussed in \cite{celli2023}, this procedure provides a lower limit to the total SC mass. 
In \cite{celli2023} we quantified an underestimation of SC masses by a factor $\sim 3$ on average. This translates into an equivalent underestimation of the cluster wind luminosity, as well as an underestimation of the maximum energy achieved at the WTS, $E_{\max}^{\rm WTS}$. The latter scales roughly as $\sim L_{\rm w,c}^{4/5}$ for Kraichnan diffusion and $\sim L_{\rm w,c}^{1/2}$ for Bohm diffusion \citep{Menchiari+2024}, such that we might expect an increase in the estimated WTS maximum energy up to a factor 2.4 and 1.7, respectively.
On the other hand, the uncertainty in the SC masses has a milder effect on the maximum energy reached at SNRs, which depends mainly on the magnetic field strength in the bubble (see Eq.~\eqref{eq:EmaxSNR}): considering that $B\sim L_{\rm w,c}^{1/5}$, a factor 3 uncertainty in $L_{\rm w,c}$ translates into a 25\% uncertainty in $E_{\rm max}^{\rm SNR}$. 
We account for this uncertainty by providing lower and upper bound estimations for the $\gamma$-ray and neutrino emission expected from each SC.

\begin{figure}
    \centering
    \includegraphics[width=\columnwidth]{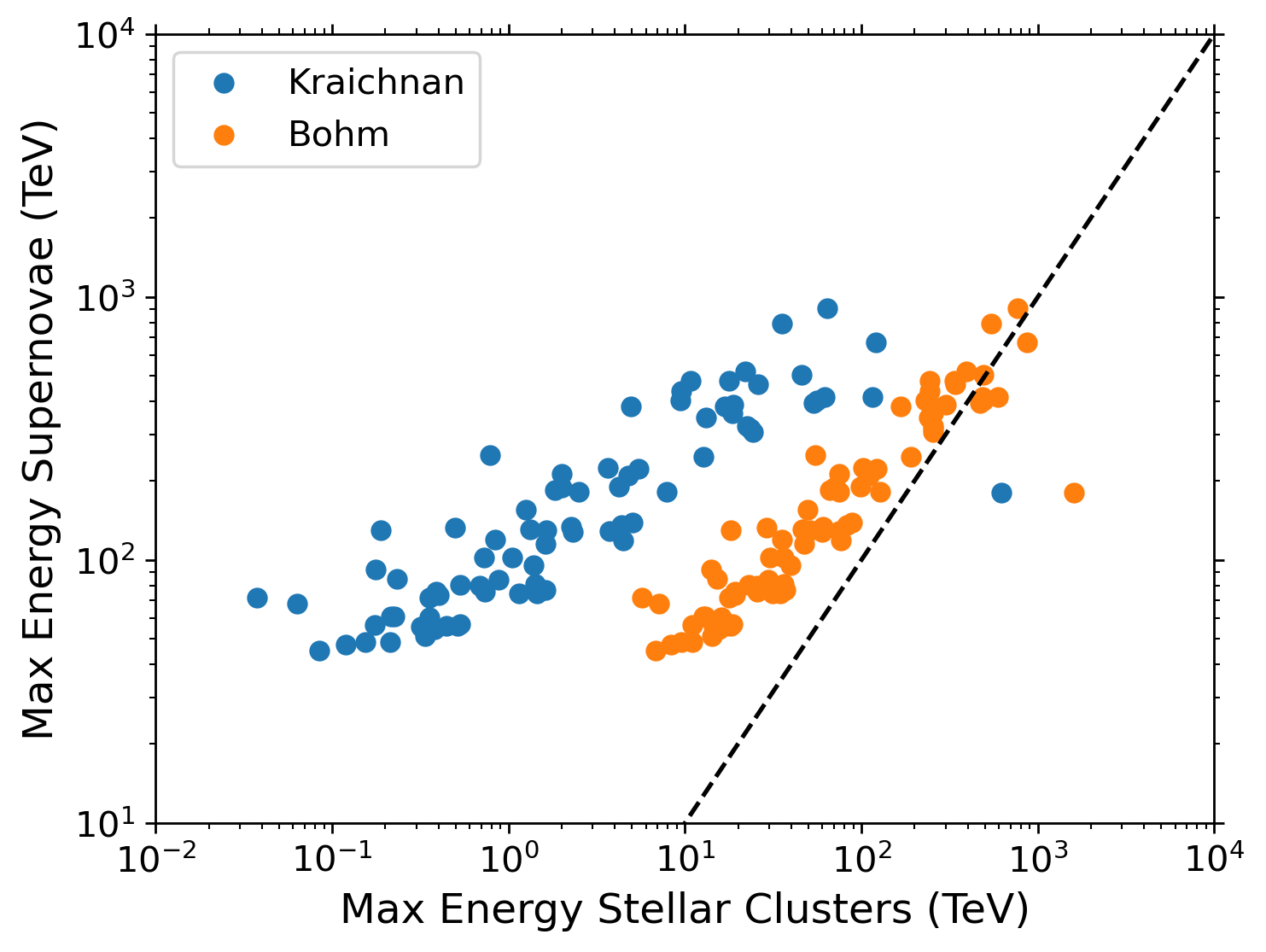} 
    \caption{The predicted proton maximum energy from acceleration in the wind for stellar clusters in our sample (x-axis), compared to the maximum energy due to SNe (y-axis), assuming Kraichnan or Bohm diffusion in the bubble. The dashed line shows $E_{\max}^{\rm SNR} = E_{\max}^{\rm WTS}$.}
    \label{fig:emax}
\end{figure}

\begin{figure*}
    \centering
    \includegraphics[trim=5mm 1cm 1cm 1.6cm,clip,width=0.95\textwidth]{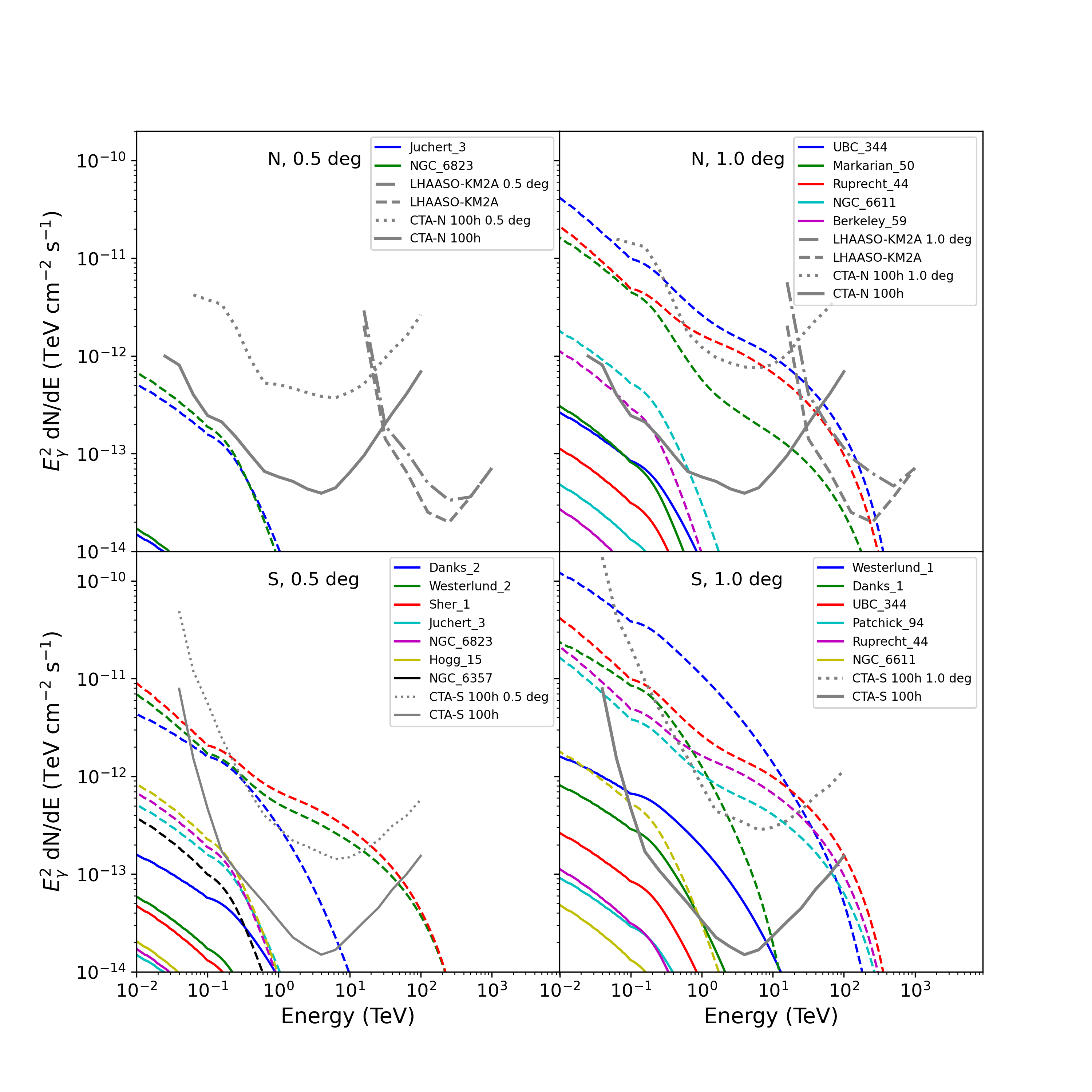}
    \caption{ The predicted spectral energy distribution (SED) for $\gamma$-ray emission integrated over the wind-blown bubble, with angular size in the \emph{baseline} scenario comparable to (i.e. within $0.25^\circ$ of) $0.5^\circ$ (left panels) and $1.0^\circ$ (right panels). Top and bottom panels show the sources visible in the Northern and Southern sky, respectively. Each source is shown with two coloured lines representing the \emph{baseline} (solid) and \emph{maximal} (dashed) models for the case of Kraichnan diffusion.  Sensitivity curves for point source and extended sources for CTA-N and CTA-S (at 100 hours exposure) and LHAASO-KM2A (at 1 year exposure) are also shown for reference (grey lines) \citep{Celli-Peron_Extended:2024}. } 
    \label{fig:mwsc_gamma}
\end{figure*}

\subsection{Gamma-ray emission}
\label{sec:mwscgamma}

For all of the selected SCs in our sample, we evaluate the expected hadronic $\gamma$-ray emission from the wind-blown bubble. To account for model uncertainties, we define \emph{minimal} and \emph{maximal} models as follows. 
In the \emph{minimal} model, corresponding to our \emph{baseline} scenario, we only account for particles accelerated at the WTS in the Kraichnan diffusion regime, using the wind luminosity as estimated in \citep{celli2023}.
For the \emph{maximal} model, we instead assume a wind luminosity 3 times larger, to account for the SC mass uncertainty discussed above. In addition, for the \emph{maximal} model we also include the contribution due to SNe. 
For all models, the acceleration efficiency is fixed to $\xi_{\rm cr}= 0.05$. The ISM density, relevant for the expansion of the wind during the early history of the SC, is derived individually for each cluster from mass-radius relations of the parent GMCs \citep{Menchiari_PhD:2023}. We use two such relations to account for uncertainty in the ISM density, with \cite{Chen2020} for the \emph{baseline} and \cite{Larson1981} for the \emph{maximal} scenarios, typically resulting in values from few to tens of particles per cm$^3$ (see appendix \ref{sec:AppDens}). 

Figure~\ref{fig:mwsc_gamma} compares the predicted $\gamma$-ray emission from the brightest clusters according to our model to sensitivity curves for current and next-generation ground-based instruments, split according to their visibility from either the Northern or the Southern hemisphere.
As shown in Fig. \ref{fig:sc_radii_hist}, the typical radii of cluster bubbles are of order $\sim 10-100$\,pc, which yields a $\sim$degree scale angular size on the sky, with a median bubble radius of $0.7^\circ$. Consequently, detection prospects for these objects are more realistically evaluated through the comparison with instrument sensitivities with regards to extended sources. Hence we report both the point-like and the degraded sensitivity curves of either LHAASO-KM2A and CTA-N (in the North) or CTA-S (in the South), appropriate for bubbles with angular sizes of 0.5$^\circ$ or 1.0$^\circ$, as calculated in \cite{Celli-Peron_Extended:2024} for on-axis observations.
Figure \ref{fig:mwsc_gamma} therefore only shows clusters whose angular size (in the \emph{baseline} scenario) is comparable to the extended sensitivity curves shown, namely $0.25^\circ <R_b<0.75^\circ$ for the left hand panels and $0.75^\circ <R_b<1.5^\circ$ for the right hand ones. 
The predicted $\gamma$-ray emission for wind-blown bubbles of other angular sizes is not shown in Figure \ref{fig:mwsc_gamma}. Despite many predicted fluxes being above the point-like sensitivity of current detectors, in several cases the degraded sensitivity curve for the corresponding angular size is no longer adequate to suggest that a detection is possible within a comparable observation time. 
Also in Figure~\ref{fig:mwsc_gamma}, the emission is shown for the \emph{baseline} (solid lines) and \emph{maximal} models (dashed lines). 

Given the strong dependence of instrument sensitivity on angular size, in Figure~\ref{fig:intfluxRb} we show the predicted integral $\gamma$-ray flux from the SC bubbles in comparison to the expected CTA-S and CTA-N 100\,hour sensitivity at the corresponding angular bubble size. Integral sensitivity lines are obtained following \cite{Celli-Peron_Extended:2024} that is based on the official response functions from each detector specifically for on-axis observations of a Crab-like spectrum source. However, for sources with radial extension beyond $2^\circ$, a further degradation of the telescope performance should be considered due to off-axis observations. Such a degradation is also energy dependent, e.g. for a source radius of $4^\circ$ the worsening with respect to the point-like on-axis case is expected to amount to a factor 4 above 1 TeV  \cite[see figure~3 from][]{Celli-Peron_Extended:2024}. We neglect this complication here, and simply provide in Fig.~\ref{fig:intfluxRb} the extended source sensitivity without accounting for any off-axis worsening, indicating with a dashed line the angular sizes where we expect this extrapolation to be less reliable. 
The error bars represent the range between the \emph{baseline} and \emph{maximal} model predictions, affecting both the absolute flux as well as the expected size of the emitting bubble. 
To prevent overcrowding of the plot, we show in Figure \ref{fig:intfluxRb} only those clusters with a predicted bubble size $< 5^\circ$ and either a mean predicted flux $F_\gamma > 10^{-14}$\,$ \mathrm{TeV cm^{-2} s^{-1}}$ or a bubble size $< 0.5^\circ$, given that the sensitivity improves to flux levels $\lesssim 10^{-13}\,\mathrm{TeV cm^{-2} s^{-1}}$ for point-like sources. Figure \ref{fig:intfluxRbBohm} shows the case of Bohm diffusion, which is the most promising in terms of detection prospects. 
Correspondingly, the flux limits for Figure \ref{fig:intfluxRbBohm} are one order of magnitude larger than for Figure \ref{fig:intfluxRb}.

\begin{figure*}
    \centering
    \includegraphics[width=\columnwidth]{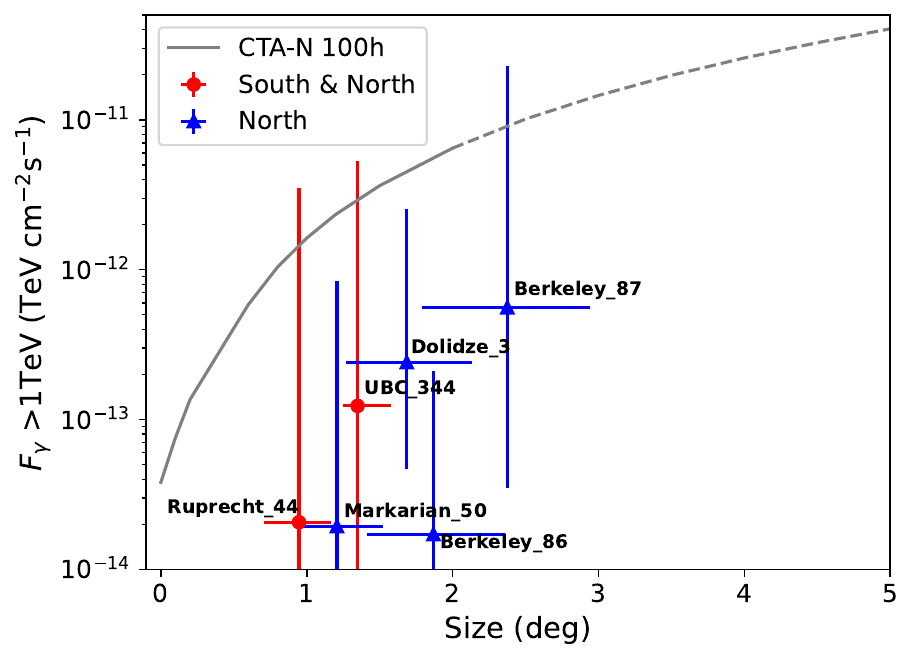} 
    \includegraphics[width=\columnwidth]{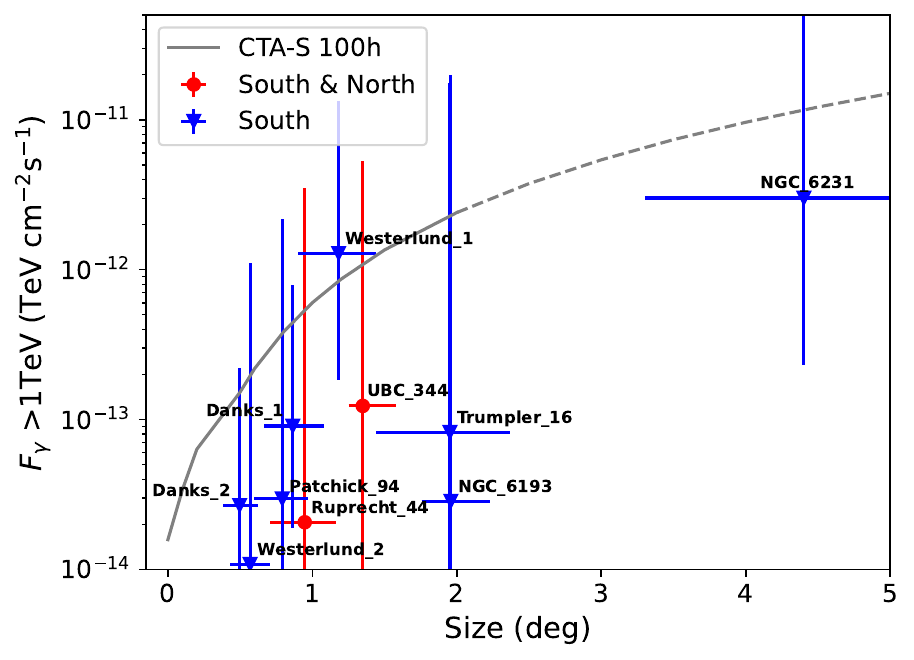}
    \caption{Integral $\gamma$-ray flux above 1\,TeV from the cluster bubble, plotted as a function of the bubble size. Left: North and Right: South. The CTA sensitivity of the corresponding observation site in 100\,hr is shown for comparison \citep{Celli-Peron_Extended:2024}, as obtained for an extended source of Crab-like spectrum. The uncertainties show, for each cluster, the \emph{baseline} and \emph{maximal} predictions from the model for the case of Kraichnan diffusion as described in the text.  } 
    \label{fig:intfluxRb}
\end{figure*}

Figure \ref{fig:intfluxRb} is obtained with \emph{baseline} values of the model as listed in Table \ref{tab:default}, e.g. in terms of acceleration efficiency and target density. Therefore, dedicated studies of individual clusters should be performed to specifically constrain the model parameters, as we do for three clusters in Section \ref{sec:gammadata_compare}.

\begin{figure*}
    \centering
    \includegraphics[width=\columnwidth]{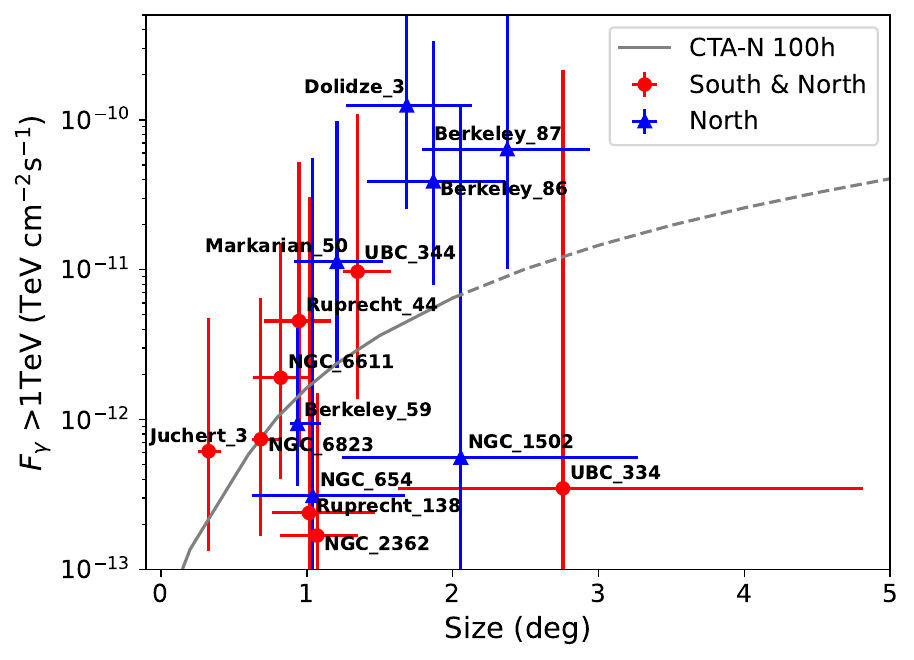} 
    \includegraphics[width=\columnwidth]{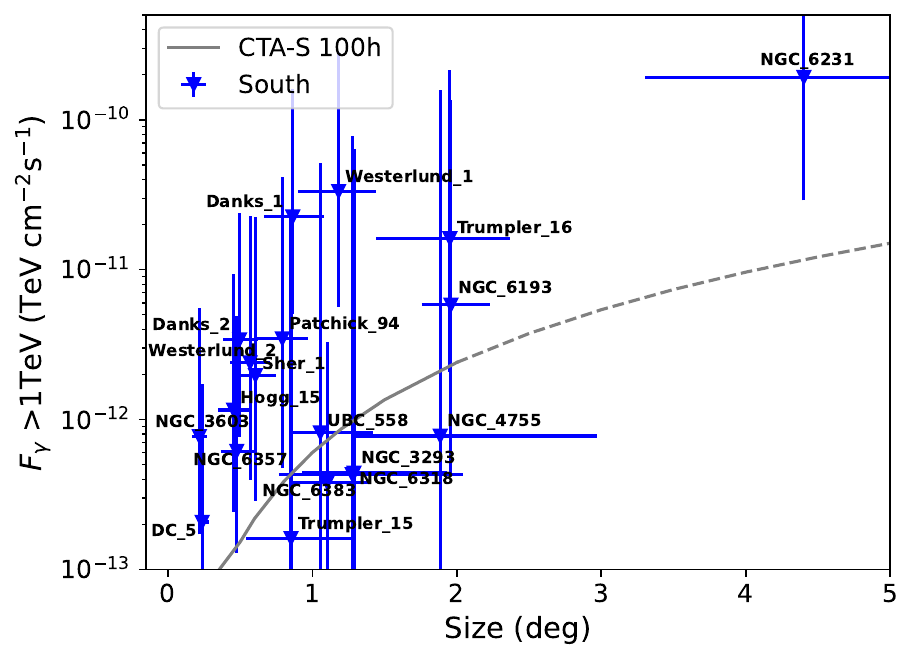}
    \caption{As in Figure \ref{fig:intfluxRb} except shown here for the case of Bohm diffusion. To improve visibility, the right hand panel depicts only those clusters that are uniquely observable in the South. }
    \label{fig:intfluxRbBohm}
\end{figure*}

Results for the 25 stellar clusters with the highest predicted $\gamma$-ray emission (integrated over the bubble) are listed in ranked order in Tables~\ref{tab:clusterflux_min} and \ref{tab:clusterflux_max} for the \emph{baseline} and \emph{maximal} models respectively, while Table~\ref{tab:cluster_top} summarises the physical properties of the same clusters, including the expected WTS and bubble radii, together with the resulting expected angular size of the bubble radius under the \emph{baseline} scenario. One should keep in mind that the ranked order in terms of $\gamma$-ray surface brightness will be different due to the variation in angular size of the wind-blown bubbles. 
Tables \ref{tab:clusterflux_min} and \ref{tab:clusterflux_max} present the predicted integral $\gamma$-ray energy fluxes above 1\,TeV and 10\,TeV compared to upper limits from the H.E.S.S. Galactic Plane Survey (HGPS) \citep{2018HGPS} where available. 
Whilst several of the clusters have predicted fluxes higher than the quoted upper limit, it should be noted that upper limits from the HGPS correspond to the angular size of the bubble only up to a maximum angular size of $0.6^\circ$, the largest size to which the upper limit map provided by \cite{2018HGPS} can be considered valid. By contrast, the majority of the bubble radii are larger than this value ($\gtrsim 0.6^\circ$), implying that these upper limits are not constraining over the cluster angular scales.
Tables~\ref{tab:clusterflux_min} and \ref{tab:clusterflux_max} also provide the predicted differential $\gamma$-ray flux at 7\,TeV for comparison to upper limits reported from the 3HWC survey where available \citep{3hwc_2020ApJ...905...76A}, which are evaluated from the sky maps provided for the nearest fixed angular size, of $0.5^\circ$ $1^\circ$ or $2^\circ$ respectively. None of these latter upper limits are violated by the predicted fluxes in our \emph{baseline} scenario.

\subsection{Neutrino Emission }
\label{sec:mwscneutrino}
For neutrino experiments, the muon neutrino flux is the most relevant astronomical channel, affording good reconstruction prospects via the track-like channel. Due to neutrino oscillations, the muon neutrino flux at Earth is roughly a third of the all-flavour flux, whilst the $\gamma$-rays have a similar normalisation. 
Therefore, the ranked order of stellar clusters according to the amount of emission produced is unchanged when considering the $\gamma$-ray or the neutrino flux above the same energy threshold. Figure \ref{fig:mwsc_neutrino} shows the predicted single-flavour muon neutrino SEDs for the brightest clusters, whilst the integral muon neutrino flux values $>10$\,TeV are provided in Tables \ref{tab:clusterflux_min} and \ref{tab:clusterflux_max}. 

\begin{figure*}
    \centering
    \includegraphics[width=0.9\textwidth]{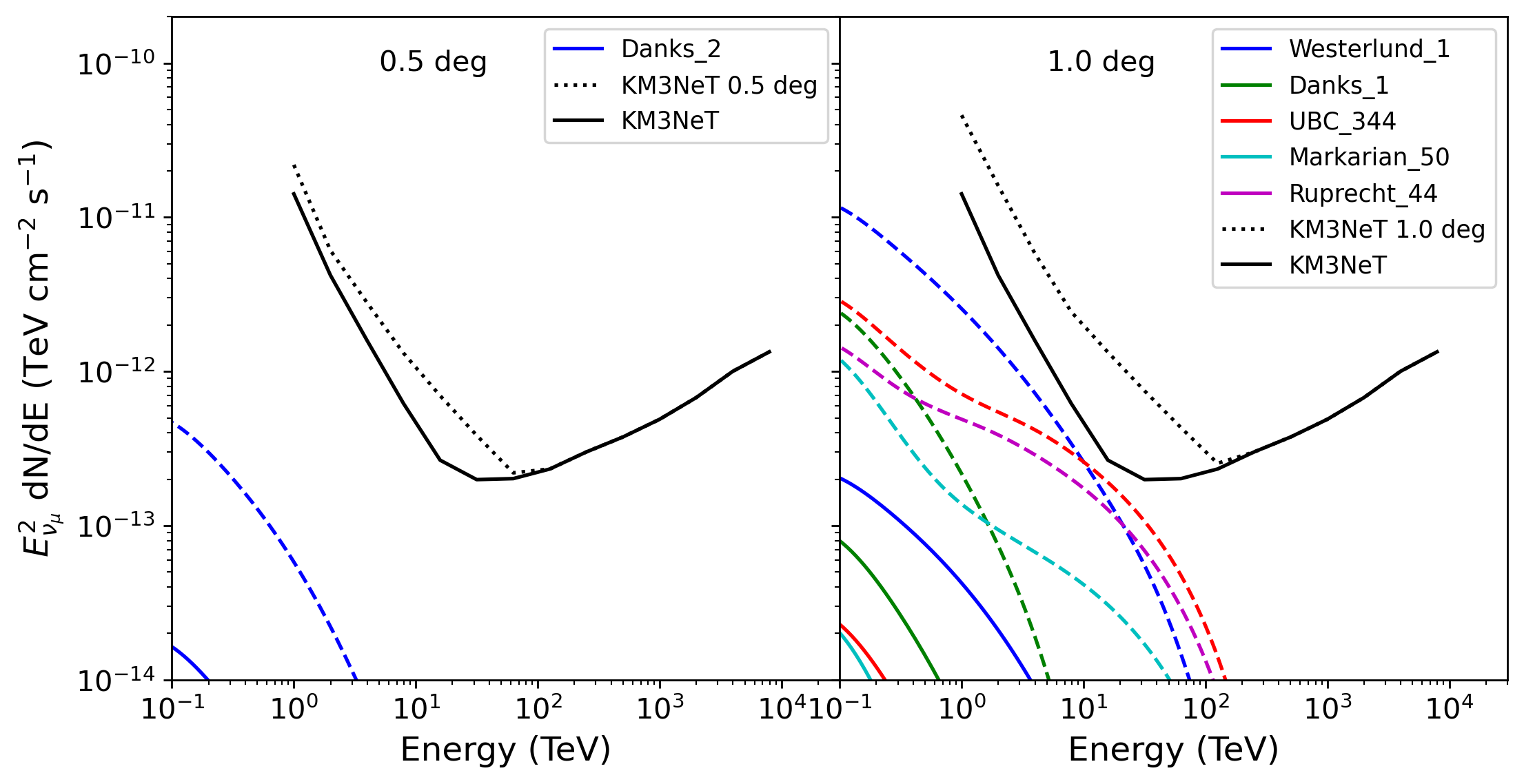}
    \caption{Predicted single-flavour muon neutrino flux from stellar cluster bubbles compared to the KM3NeT sensitivity (at 10 years exposure), under \emph{baseline} (solid) and \emph{maximal} (dashed) cases,  for angular sizes in the \emph{baseline} scenario of 0.5$^\circ$ (left) and 1.0$^\circ$ (right). }
    \label{fig:mwsc_neutrino}
\end{figure*}

Similarly to Figure~\ref{fig:mwsc_gamma}, Figure~\ref{fig:mwsc_neutrino} shows predicted muon neutrino spectral energy distributions (SEDs) compared to the KM3NeT sensitivity curves (at 10 years exposure) for both a point-like source and for a source with an angular extension comparable to 0.5$^\circ$ or 1$^\circ$ respectively, as computed in \cite{2018APh...100...69A}. Westerlund~1 (Wd~1) has the highest expected flux, coming closest to the KM3NeT sensitivity in the \emph{maximal} model: our detection prospects are found to be consistent with the expectations obtained by the KM3NeT Collaboration in the hypothesis of hadronic $\gamma$-ray emission \citep{tim2024}. As such, either a clear detection or constraining upper limits are expected to emerge in the near future for this source, that will allow us to constrain its hadronic acceleration efficiency. A large source sample might be accessible by including events from the cascade channel in order to access the entirety of the expected neutrino flux. Shower reconstruction algorithms for KM3NeT have been shown to provide an angular resolution below $2^\circ$ at energies above 100~TeV \citep{thijs}, compatible with and even smaller than the size of several of the SCs investigated here. Proper evaluations of the combined track and shower sensitivities are currently being performed by the KM3NeT Collaboration, so that the improved performances will demand further future investigations in SC searches. Along similar lines, a stacking analysis could be a more suitable approach in searching for neutrino emission from SC wind-blown bubbles, rather than relying on individual systems.

\section{Comparison to $\gamma$-ray detected stellar clusters}
\label{sec:gammadata_compare}
To date, three stellar clusters in the Gaia DR2 sample have already been detected in GeV and TeV $\gamma$ rays, i.e. Wd1, Wd2 and NGC~3603 (GeV only). For these specific clusters we can hence compare our flux predictions with the corresponding $\gamma$-ray data, as shown in Figure \ref{fig:data}. The grey bands show, for each plot, the flux range described by reasonable variation in model parameters. This is adjusted case-by-case, accounting for known literature values regarding cluster properties, as explained in the following subsections. 
We emphasise that no fitting was performed: the shaded bands in Figure~\ref{fig:data} serve to simply illustrate the behaviour of the model and how the range of reasonable parameters compares to the data. A summary of the model parameters corresponding to the by-eye match to data, as indicated by the red curves in Figure~\ref{fig:data}, is given in Table \ref{tab:indivSCs}  while the values in square brackets correspond to the lower and upper bounds of the shaded area. All curves shown assume Bohm diffusion, providing a better agreement with the shape of the gamma-ray flux. A detailed modelling of each source is required to better address the validity of the hadronic model by also using other information, such as the spatial morphology which is not considered here. 

In all three cases, the uncertainty is dominated by the variation in the mass estimation. Our \emph{baseline} scenario using mass estimates from \cite{celli2023} is known to underestimate the total cluster mass compared to literature estimations. When accounting for this, several physical parameters are correspondingly scaled, namely $\dot{M}_c$, $L_{\rm w,c}$, $n_0$ and $N_{\rm sn}$ (adv), which yields the model curve and upper bound in each case. The specific details of the modelling for each of the three clusters is described below. 

\begin{table*}
 \centering
    \begin{tabular}{l|ccc}
    \toprule \toprule
         Cluster & Westerlund\,1 & Westerlund\,2 & NGC\,3603 \\
    \midrule
     $t$ (Myr)   & 7.94 & 2.5 & 1 \\
     $d$ (kpc)   & 3.7 & 4.5 & 6.8 \\
     $M_{\rm sc}$ ($10^3\,\rm M_\odot$)   & 49 [22-49] & 3.6 [2.2-3.6] & 13 [4.8-13] \\
     $\dot{M}_c$ ($10^{-5} \rm M_\odot$ yr$^{-1}$)  & 66.4 [30.1-66.4] & 4.3 [2.6-4.3] & 19.0 [7.2-19]\\
     $L_{\rm w,c}$ ($10^{37}\,$erg\,s$^{-1}$)  & 69 [32-69] & 2.9 [1.8-2.9] & 12 [4.6-12] \\
     $n_0$ (cm$^{-3}$)   & 6.3 [5.9-19] & 25 [25-88] & 79 [41-210]\\     
     $N_{\rm sn}$ (adv)  & 28 & 0 [0-1] & 0 \\
     $\xi_{\rm CR}$   & 5\% [5\%-10\%] & 10\% [5\% - 10\%] & 10\% [5\%-10\%] \\
     $\delta$ & 1 & 1 & 1    \\
     \bottomrule
    \end{tabular}
    \caption{Parameter values corresponding to the model curves indicated by a red solid line in Figure \ref{fig:data}, which represent a by-eye match to observational data. The values in the square brackets correspond to the lower and upper bounds of the grey shaded area in Fig~\ref{fig:data}. }
    \label{tab:indivSCs}
\end{table*}

\subsection{Westerlund~1} \label{sec:Wd1}
Wd1 has an estimated age and mass of $\sim 8$\,Myr and $2.2\times 10^4\, \rm M_{\odot}$ \citep{celli2023} respectively, and as such a considerable number of SNe have already occurred within it. We estimate a total number of 51 SNe, contributing to the average density within the cluster, among which 28 SNe have occurred within one advection time, thereby contributing to the non thermal particles present inside the bubble. 
The estimated age in the literature ranges from $\sim 4$\,Myr \citep{Gennaro+2011} to 10\,Myr \citep{Navarete+2022-Wd1}. This uncertainty is probably the result of an intrinsic temporal spread of several Myrs for the star-forming process instead of a single starburst episode scenario, which consequently may yield different ages when using a different sample of stars or different modelling assumptions.  
If we consider the smaller age value, the number of SNe would be reduced to $\sim 17$ (14 of which in one advection time).
We adopt a distance of 4\,kpc, as it is the preferred distance estimated in the most recent literature based on Gaia DR3 \citep{Navarete+2022-Wd1, Rocha+2022-Wd1}, although a distance of $\sim 7$\, kpc has also been adopted in past works \citep{gaiadr2_2018}.

The shaded band in the top left panel of Figure~\ref{fig:data} shows that the observational data from H.E.S.S. \citep{Westerlund1_2022} and Fermi-LAT \citep{2013MNRAS.434.2289OhmWest1} can be well bracketed by our model. Whilst the lower bound corresponds to the cluster wind only using the estimated mass of $2.2\times 10^4\, \rm M_{\odot}$, the upper bound is determined by 
accounting for the contribution from 62 SNe and for an approximate factor 2 underestimation of the total cluster mass \citep{celli2023}, the latter being obtained by comparing our mass estimate to literature values available for this cluster that estimate a total mass of $49000 M_\odot$ \citep{Gennaro+2011}.
We assume Bohm diffusion, and the ISM density is in the range $5.9-19\,{\rm cm}^{-3}$, following the approach defined in Appendix~\ref{sec:AppDens}, and the CR efficiency varies from a lower bound  of 5\% to a maximum of 10\% to cover the data. 

The highest energy data point measured by H.E.S.S. lies above the shaded band and the H.E.S.S. spectrum above $\sim3$\,TeV seems slightly harder than our curved models. Such a discrepancy may point towards a harder mass function in the stellar distribution. 
The \emph{nominal} maximum energy in Wd1 is $E_{\max}^{\rm WTS} \simeq $\,PeV, however, the shape of the predicted curves do not show a sharp cut-off at such an energy, but rather curve at lower energies corresponding to the \emph{effective} maximum energy (see section \ref{sec:WTS}). This is also the consequence of the spherical geometry of the system as discussed in detail by \cite{Morlino2021}.

The results for the grey band are calculated accounting for the predicted gamma-ray emission from the entire bubble, whose estimated radius ranges between $\sim 110$ and 130\,pc, depending on the assumed cluster mass. Such a radius corresponds to an angular size of $(1.6-1.8)^\circ$ at 4~kpc distance. However, the size of the gamma-ray emission detected by H.E.S.S. is $\sim 1^{\circ}$ \citep{Westerlund1_2022}, corresponding to a radius of $\sim 60$\,pc. Such a discrepancy poses a challenge to our hadronic model. In fact, this argument was already raised by \cite{Harer+2023}, who claimed a dominant leptonic contribution because the size of the emitting region is smaller than the estimated bubble size. However, this model requires a very large electron acceleration efficiency, of the order or 1\%, which is at odds with the typical values of 0.1\%-0.01\%  inferred from SNR shocks.
Although we estimate a rather larger bubble size, the shell fragmentation can be responsible for an apparent reduced size of the bubble, especially if CR are allowed to easily escape from the region where the fragmented shell is located, a scenario supported by the recent finding by \cite{Lemoine-Goumard+2025} who detected gamma-ray emission immediately outside the Wd1 bubble compatible with an outflow of high energy particles from the bubble itself. 
In fact, the magnetic turbulence may be damped while the plasma is advected away from the termination shock, implying a faster escape of particles and a correspondingly reduced $\gamma$-ray emission from the bubble interior. Our model does not account for such an effect, which we consider important to assess the full validity of the hadronic model. 
If we account only for the gamma-ray emission from a region of $\sim 60$\,pc, by simply rescaling for the volume ratio between the 60~pc bubble and the one we derived, the required efficiency increases to $\gtrsim30\%$ in the Kraichnan diffusion scenario, while considering the case of Bohm diffusion and keeping the external density of 6.3 cm$^{-3}$ the efficiency required is $\sim5\%$, as shown by the red line (see also Table \ref{tab:indivSCs}). 
As a consequence, we find that the hadronic model is a viable explanation of the gamma-ray emission of Wd1, provided that the external density is of the order of ten particles per cm$^{3}$. A more detailed model than the one presented here is needed to properly account for the spatial morphology.

\begin{figure}
    \centering
    \includegraphics[width=\columnwidth]{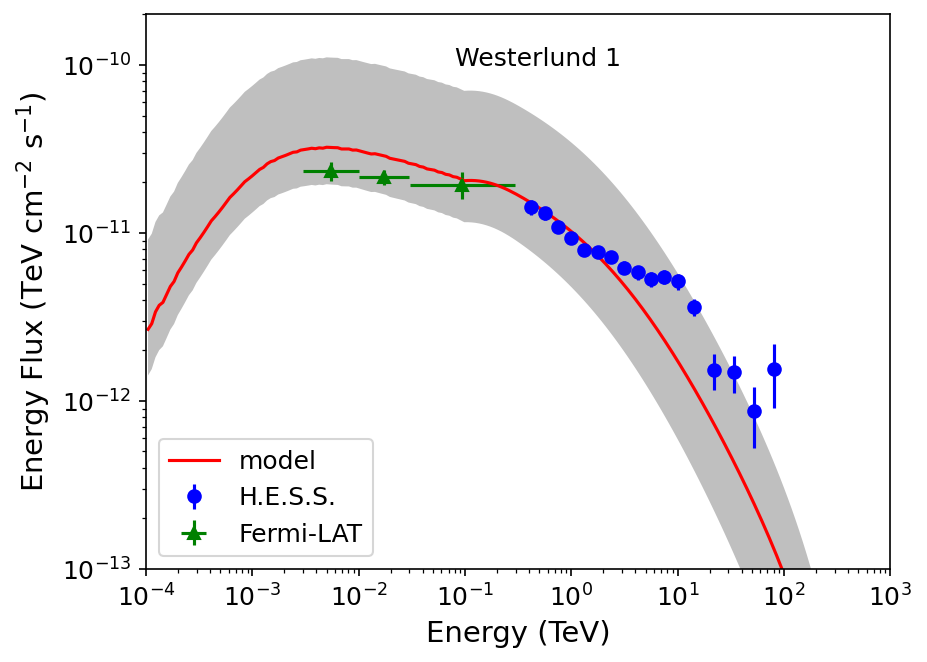}
    \includegraphics[width=\columnwidth]{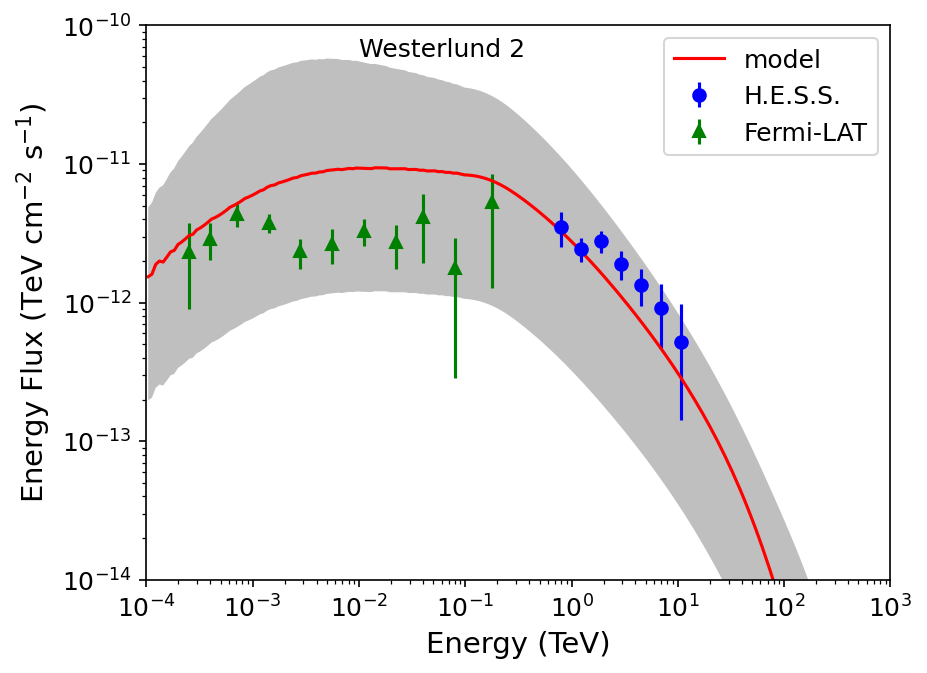}
    \includegraphics[width=\columnwidth]{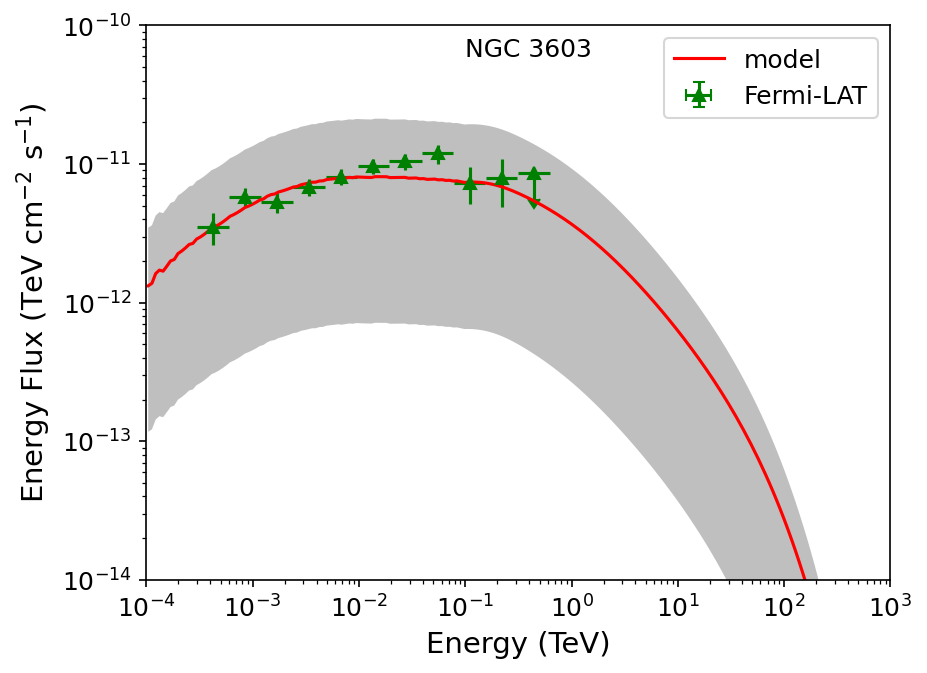}
    \caption{Comparison of measured $\gamma$-ray emission from three stellar clusters listed in the Gaia DR2 catalogue to predictions from this model. Shaded bands indicate reasonable model variations, see the text in \S~\ref{sec:gammadata_compare} for the detailed explanation of each curve. Top left: Wd1, data from \citep{2013MNRAS.434.2289OhmWest1,Westerlund1_2022} Top right: Wd2, data from \citep{2021MNRAS.505.2731Mestre,2011A&A...525A..46H_West2} Bottom: NGC~3603, data from \citep{2020ApJ...897..131Saha}. }
    \label{fig:data}
\end{figure}

\subsection{Westerlund 2}
\label{sec:Wd2} 
In the case of Wd2, which has an estimated age of 2.5\,Myr and a mass of $2200\,\rm M_{\odot}$ \citep{celli2023}, there is only a 50\% chance of one SNe having already occurred there, hence we expect a small influence of SNe on the predicted flux over the \emph{baseline} wind scenario. We show in Figure~\ref{fig:data} a shaded band that incorporates the effects on the flux of parameter variation as indicated in Table \ref{tab:indivSCs}, 
including when correcting for an underestimation of the total cluster mass 
compared to the literature value of $3600\, M_\odot$. The ISM density is set to $25\,{\rm cm}^{-3}$ minimum and ranges up to a maximum of $88\,{\rm cm}^{-3}$, following the approach defined in Appendix~\ref{sec:AppDens}, and a CR efficiency ranging from 5\% to 10\%, where 10\% is required to explain the data as shown by the red curve. 
To simultaneously account for the Fermi-LAT and H.E.S.S. data \citep{2011A&A...525A..46H_West2} the emission spectrum should be quite flat, forcing us to adopt a Bohm diffusion coefficient for all the cases shown. 
In fact, the Bohm case results in a flatter distribution of energy flux below $\sim 1$\,TeV, more consistent with the shape of the Fermi-LAT data \citep{2021MNRAS.505.2731Mestre}.

\subsection{NGC~3603}
\label{sec:ngc3603}
NGC~3603 is a particularly young star cluster with an age of 1\,Myr hence no SN explosions are expected yet. A CR efficiency of 10\% is used in this case, consistent with our \emph{maximal} model, however our density estimations were found to be insufficient in this case for comparison to Fermi-LAT data \citep{2020ApJ...897..131Saha}. Note that we only consider the data detected from the extended source FGES J1109.4-6115e, neglecting the (lesser) contribution from a point like source also present in the region. As observational data suggests an enhanced gas density in this region, we adopt an average gas density obtained from \cite{Larson1981} ranging from $\sim 41$ to $\sim 79\,\mathrm{cm}^{-3}$ for the surrounding medium, (when also accounting for the uncertainty in cluster mass) and up to $210\,\mathrm{cm}^{-3}$ based on literature values which enables our model to approximately match the data in an optimistic scenario \citep{2025RocamoraNGC3603,2017Y}. 

The shaded region in Figure \ref{fig:data} shows the influence of accounting for a factor $\sim 3$ underestimation in the cluster mass, 
CR efficiency between 5\% and 10\% and using Bohm diffusion, which produces a shape more consistent with data than the Kraichnan case. As no TeV observation has been reported yet, our modelling provides a clear prediction of the expected gamma-ray slope in this domain within the termination shock model for particle acceleration, to be probed in the near future.

\section{Comparison with $\gamma$-ray catalogues}
\label{sec:catalogues}

We search for known $\gamma$-ray emitters in existing data from H.E.S.S., LHAASO, and Fermi-LAT as potential counterparts to the predicted emission from SC bubbles. Due to the limited expected fluxes for many of the systems in our sample, we constrain our discussion to the brightest objects listed in Table~\ref{tab:cluster_top}. The expected integral $\gamma$-ray flux above 1\,TeV of these sources are shown in Figure~\ref{fig:galplaneupper}, where we adopt the flux corresponding to the \emph{baseline} scenario. For visual comparison of the spatial coincidences as described above, we indicate the locations of sources from the first LHAASO catalogue (1LHAASO) \citep{Cao_2024_1lhaaso} with red circles and of unidentified 4FGL point sources \cite{2020ApJS..247...33A_4fgl} as cyan markers. The two brightest clusters, appearing yellow in Figure \ref{fig:galplaneupper}, correspond to the top two clusters listed in Table~\ref{tab:cluster_top}, namely NGC\,6231 and Wd1. 

\subsection{Comparison to the H.E.S.S. Galactic Plane Survey}

Within the HGPS \citep{2018HGPS}, there are three sources which coincide with SCs listed in Table~\ref{tab:cluster_top} (extended). These are: HESS\,J1646-458 which coincides with Wd1; HESS\,J1023-575 with Wd2; and HESS\,J1908+063 with Juchert\,3. 
The angular separation must lie within the region used for spectral extraction in the HGPS (RSpec) for the SC to be considered coincident with the $\gamma$-ray source. 
Wd1 and Wd2 are already the preferred associations for the aforementioned sources, where the $\gamma$-ray emission is indeed thought to be physically associated to the stellar cluster. For Wd1 the angular radius of the $\gamma$-ray emission $\sim1^\circ$ is comparable to the predicted size of the wind-blown bubble of $\sim1.15^\circ$, corresponding to a physical size of $\sim155$\,pc, whilst for Wd2 the physical size of $\sim45$\,pc corresponds to a predicted angular size of $\sim0.6^\circ$ at the adopted distance of 4.53\,kpc, slightly larger than the measured $\sim0.3^\circ$ \citep{Westerlund1_2022,2011A&A...525A..46H_West2}. 
HESS\,J1908+063 is an unidentified $\gamma$-ray source with multiple possible counterparts: to date the source has been most often attributed to either the energetic pulsar PSR\,J1907+0602, the SNR G40.5-0.5 or a combination thereof \citep{2021ApJ...913L..33Li_J1908,2021MNRAS.505.2309Crestan_J1908}. The coincidence of Juchert\,3 with HESS\,J1908+063, suggests another possible counterpart for this unidentified $\gamma$-ray source.

\subsection{Comparison to first LHAASO catalogue}
The first LHAASO catalogue lists 43 sources of ultra-high-energy (UHE, $>$100\,TeV) $\gamma$-ray emission, significantly increasing upon the 12 sources known as of 2021 \citep{Cao_2024_1lhaaso}. As such,  1LHAASO is a suitable catalogue to cross-reference against the predictions of our model. 
The angular separation must be less than the 39\% containment radius of the 2D Gaussian for the 1LHAASO morphology (R39) for the stellar cluster to be considered coincident with the $\gamma$-ray source. 
Although Wd1 and Wd2 are not visible to LHAASO, $\gamma$-ray emission coincident with Juchert\,3 is detected and given the identifier 1LHAASO\,J1908+0615u.

While both Dolidze\,3 and Berkeley\,87 are coincident with the source 1LHAASO\,J2020+3638 within $\sim1^\circ$, an association seems unlikely as the extent of both is $\sim1.7-2.4^\circ$ and 1LHAASO\,J2020+3638 is potentially associated with a known energetic pulsar in the vicinity. 

A more promising association is the coincidence of NGC\,6823 with the unidentified source 1LHAASO\,J1945+2424*, detected by HAWC as 2HWC\,J1949+244 \citep{Cao_2024_1lhaaso}. Although the angular separation between the stellar cluster and 1LHAASO\,J1945+2424* is $\sim1.2^\circ$, larger than the $1\,\sigma$ (39\% containment) angular size of $\sim0.4^\circ$ reported by LHAASO, the lack of alternative physical counterparts for this unidentified source makes this a strong candidate for further follow-up studies to establish a physical association. 

\subsection{Comparison to 4FGL catalogue} 

Sources from the 4FGL catalogue are here considered to be coincident with a SC if they lie within the bubble radius predicted by our \emph{baseline} model. 
NGC\,3603 (see section \ref{sec:ngc3603}) is already provisionally associated with 4FGL\,J1115.1-6118 by \cite{2020ApJ...897..131Saha}, which is the only 4FGL source coincident within $R_s$. 
(Note however that the comparison in section \ref{sec:ngc3603} is made to the extended source FGES\,J1109.4-6115e.)
Wd2 (see section \ref{sec:Wd2}) is also already associated with a 4FGL source, namely 4FGL\,J1023.3-5747e which is extended and best fit by a radial disk model \citep{2020ApJS..247...33A_4fgl}. Nevertheless, we find that there is at least one further unidentified 4FGL source within one bubble radius of Wd2 that hence could be associated. 

For all other SCs listed in Table~\ref{tab:cluster_top}, we simply state the number of coincident unidentified 4FGL sources, i.e. within an angular separation $<R_{\rm b}$. 
NGC\,6231 -- 34; Westerlund\,1 -- 4; Dolidze\,3 -- 3; Berkeley\,87 -- 5; Danks\,1 -- 6; Danks\,2 -- 4; UBC\,344 -- 7; Berkeley\,86 -- 2; Trumpler\,16 -- 6; Markarian\,50 -- 3;  Westerlund\,2 -- 1; NGC\,3603 -- 1; NGC\,6611 -- 1; NGC\,6193 -- 8; Berkeley\,59 -- 1;  NGC\,6357 -- 2; UBC\,558 -- 2; and Berkeley\,90 -- 1.

Assuming that all unidentified 4FGL sources are evenly distributed in the Galactic plane, $|b|<5^\circ$, then at any position there should be on average one unidentified source within a radius of $\sim1.2^\circ$. For most of the aforementioned clusters, the numbers are consistent with chance association, whilst for seven clusters, there are at least a factor 4 times more coincident unidentified sources than would be expected by chance. These are Westerlund\,1, Danks\,1, Danks\,2, UBC\,344, Westerlund\,2, NGC\,3603, and NGC\,6357, for each of which further analysis is warranted to confirm or refute a potential association with the stellar cluster.
For potential emission on angular scales $\gtrsim1^\circ$, it is challenging to unambiguously associate potential sources without performing a dedicated analysis.

If, however, we apply a tighter restriction by evaluating coincidences occurring within the termination shock radius, then the following clusters listed in table \ref{tab:cluster_top} (beyond Wd2 and NGC\,3603) have plausible associations: Berkeley\,87 with 4FGL\,J2022.6+3716c, Danks\,1 with 4FGL J1312.3-6257 \& 4FGL J1312.3-6231c, and Berkeley\,59 with 4FGL\,J0002.1+6721c. 
A dedicated study for each of these SCs in turn would be required to conclusively establish an association with these as yet unidentified Fermi-LAT sources. 

\begin{figure*}
    \centering
    \includegraphics[width=\textwidth]{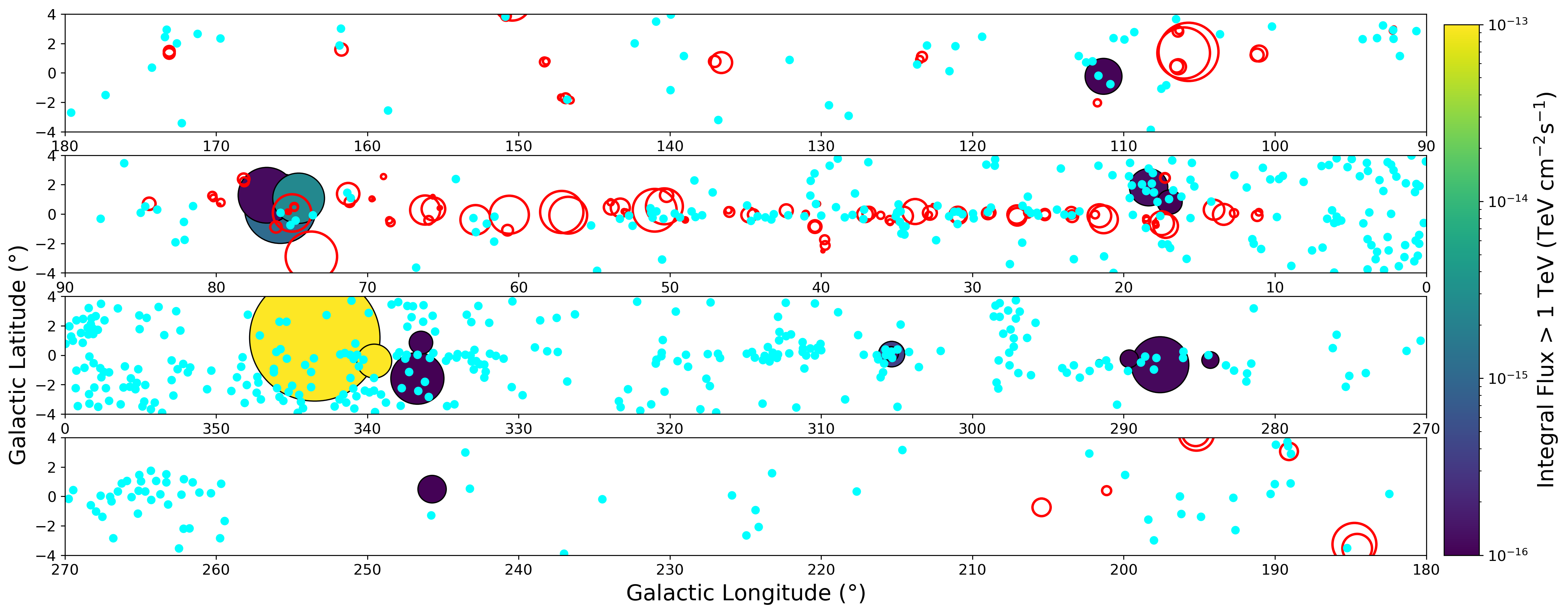} 
    \caption{Predicted $\gamma$-ray flux from star cluster bubbles along the Galactic Plane under the \emph{baseline} scenario. Also indicated are sources from the first LHAASO catalogue (red circles) and unidentified sources from the 4FGL (cyan markers). }
    \label{fig:galplaneupper}
\end{figure*}

\section{Conclusion and outlook}
\label{sec:conclusions}
Massive SCs are promising candidate PeVatron accelerators, potentially responsible for CR acceleration up to the ``knee'' region. Young SCs ($\lesssim 3$\,Myr), in particular, form a collective wind from member stars, that may terminate into a strong shock at which particle acceleration can occur, with a large, wind-blown bubble expanding into the surrounding medium. Beyond $\sim 3$\,Myr and up to $\sim 30$\,Myr particle acceleration in a SC will likely be dominated, instead, by SNRs occurring inside of the bubble. Therefore, clusters younger than $\sim 30$\,Myr may behave as non-thermal particle emitters.

In this study, we apply the SC acceleration and transport model of \cite{Morlino2021}, improved by incorporating the presence of SNRs, to known young and compact clusters catalogued by the Gaia satellite \citep{gaia2020}, reducing our sample to 79 out of 387 clusters in GaiaDR2. For these clusters, we first characterised the crucial physical parameters required in the application of the wind model, namely wind luminosities and surrounding density, thus defining the system geometry (i.e. the expected location of the termination shock radius as well as of the wind-blown bubble radius) and the maximum particle energy achieved at the WTS. Additionally, we included the anticipated contribution from the SNe exploded within each cluster lifetime, in terms of both energetics and maximum energy achieved. We found that the maximum energy achieved at the SNR shocks is usually larger than the one reached at the WTS and that there is a positive correlation between these two quantities due to the larger magnetic turbulence produced by powerful stellar winds.
Depending on the particle diffusion regime, we found that energies of $\gtrsim 1$\,PeV are reached at the WTS in only one SC with Bohm diffusion, whilst the maximum energy reached at SNR shocks can approach $\sim$\,PeV energy in a couple of cases.

We then predicted $\gamma$-ray and neutrino fluxes arising from hadronic collisions of the accelerated protons within the cluster bubble, thus evaluating which of the SCs among our sample could be detectable by CTA, LHAASO, and KM3NeT. Because the predicted emission is integrated over the entire wind-blown bubble of typically large angular size, it appears necessary to take the degradation of experimental sensitivity due to extended source observations into account. As such, the number of systems for which emission from the bubble would be detectable drastically reduces to a few. Dedicated follow-up analyses accounting for the large angular size would be needed to either verify or refute the predictions made in this work. 
Accompanying data containing the results of this work, with flux predictions, is described in Table~\ref{tab:fitsfile}.

Among all SCs in our sample, three of them, namely Wd1, Wd2 and NGC~3603, have been already detected in $\gamma$-rays. We show that the reported spectra lie within the prediction of our model using a CR acceleration efficiency ranging between 5\% and 10\%. In all three cases we found that the Bohm diffusion predicts spectra that are in better agreement with the detected flux, especially Wd2 and NGC~3603 which have a flatter spectrum below $\sim 1$\,TeV.
Our conclusions for these three SCs should be taken as preliminary, as more detailed analyses are required, where the sources morphology should be also taken into account 

Although the wind-blown bubbles around SCs may yield comparatively low surface brightness $\gamma$-ray (and neutrino) emission, the $\gamma$-ray flux could be locally enhanced by the presence of interstellar clouds. The availability of dense molecular material, causing increased hadronic interactions and subsequent emission on smaller angular scales, could highlight regions where the CR flux is higher than that of the Galactic CR sea, verifying that star clusters act as CR accelerators. The typically smaller angular size of interstellar clouds compared to that of cluster wind-blown bubbles may render such a search more easily feasible experimentally. This scenario will be considered in a forthcoming study complementing this work. 

As a final comment we note that in this work we have only considered hadronic production of high-energy photons, as the leptonic one is generally disfavoured, especially at the highest energies, in amplified magnetic fields \cite[as assumed in this work; see discussion in][Sec.~5.3]{Menchiari+2024}. However, we cannot rule out that inverse Compton scattering may be relevant for some of the SCs analysed here; additional work is required for a proper assessment of this issue.

\begin{acknowledgments}
The authors thank E. de O\~na Wilhelmi for providing data points for Wd2 and G.~Peron for useful feedback about the topics discussed in the text. SC gratefully acknowledges the financial support from Sapienza Universit\`a di Roma through the grant ID RM1221816813FFA3. AM is supported by the Deutsche Forschungsgemeinschaft (DFG, German Research Foundation) Project Number 452934793. 
SM and GM are partially supported by the INAF Theory Grant 2024 {\it Star Clusters As Cosmic Ray Factories II} and INAF Mini Grant 2023 {\it Probing Young Massive Stellar Cluster as Cosmic Ray Factories''}.
Author SM acknowledges financial support from the Severo Ochoa grant CEX2021-001131-S funded by MCIN/AEI/ 10.13039/501100011033.
The authors are also grateful for the organization of the TOSCA workshop 2024 that gave the opportunity to discuss several topics presented in this paper.
This research has made use of the CTA instrument response functions provided by the CTA Consortium and Observatory, see \url{https://www.ctao-observatory.org/science/cta-performance/} (version prod5 v0.1, \cite{zenodoCTA}) for more details.

\end{acknowledgments}

\begin{contribution}

AMWM was the main responsible for writing and preparing the manuscript. She also developed the numerical code to estimate the gamma-ray and neutrino fluxes and performed the model computations.
GM developed the analytical model for particle acceleration (content of Sec. \ref{sec:model} and \ref{sec:acceleration}).
SC has conceptualized the research activity, characterised the properties of star clusters by Gaia, and calculated the gamma-ray and neutrino instrument sensitivity.
SM estimated the gas density around star clusters. AS assisted with the numerical code and preparing the Gaia SC dataset. All authors contributed to the paper writing.


\end{contribution}

\software{ \texttt{astropy} \citep{2013A&A...558A..33A,2018AJ....156..123A,2022ApJ...935..167A},  
\texttt{numpy} \citep{numpy_harris2020array}, \texttt{scipy} \citep{2020SciPy-NMeth}.
}


\appendix
\renewcommand{\thetable}{\thesection.\arabic{table}}
\renewcommand{\theHtable}{\thesection.\arabic{table}}
\setcounter{table}{0}

\section{Ambient density around star clusters}
\label{sec:AppDens}
Young SCs are surrounded by the dense environment of their parent GMC. Assuming a spherical geometry for the GMC, the ambient density can be calculated as:
\begin{equation}
    \rho_0=\frac{M_{\rm gmc}}{\frac{4 \pi}{3} R_
    {\rm gmc}^3}\, ,
\end{equation}
where $M_{\rm gmc}$ and $R_{\rm gmc}$ are the mass and radius of the GMC respectively. The GMC mass is bound to the SC mass by the star formation efficiency ($\varepsilon_{\rm sfe}$), that is the fraction of the cloud's initial mass effectively transformed into stars:
\begin{equation}
    M_{\rm gmc}= M_{\rm sc} \left (\frac{1}{\varepsilon_{\rm sfe}} -1 \right )\, .
\end{equation}
In this work, we considered $\varepsilon_{\rm sfe}=0.01$, which is consistent with what is statistically inferred from other galaxies \citep{Kruijssen2019}. Empirical evidence shows that the size of a GMC correlates with its mass. Several mass-radius relations have been proposed of the form \citep[see e.g.][]{Larson1981, Miville2017, Chen2020, Sun2024}:
\begin{equation}
    R_{\rm gmc}= \left (\frac{M_{\rm gmc}}{M_0} \right )^\alpha \, \rm pc .
\end{equation}
Figure~\ref{fig:GMCdens} shows the GMC particle density distribution ($n_{\rm gmc}=\rho_0/2 m_p$) for our sample of SCs using different mass-radius relations. The values of $M_0$ and $\alpha$ for each of these relations are reported in Tab.~\ref{tab:GMC_MRR}. To account for the uncertainty associated with the mass-radius relation, we consider a maximum and minimum value for $\rho_0$: the minimum density is calculated using the relation reported in \cite{Chen2020}, and it is employed in our \emph{baseline} scenario, while the maximum density is obtained using the relation described by \cite{Larson1981}, which is used in our \emph{maximal} scenario. For NGC~3603 we instead use adopt a density of $\sim 210 $~cm$^{-3}$ for the \emph{maximal} scenario based on literature values from \citep{2025RocamoraNGC3603}, with the mass-radius relation described in \cite{Larson1981} as the lower bound, which approximately  reproduces values from observational data of $\sim 60 $~cm$^{-3}$ \citep{2017Y}.

\begin{table}[]
\begin{center}    
\begin{tabular}{ccc}
\toprule \toprule
Mass-Radius relation  & \begin{tabular}[c]{@{}c@{}}$M_0$\\ {[}M$_\odot${]}\end{tabular} & $\alpha$ \\
\midrule
Larson (1981) \citep{Larson1981} & 460 & 0.53 \\
Miville et al. (2017) \citep{Miville2017} & 36.7 & 0.45 \\
Chen et al. (2020) \citep{Chen2020} & 156.6 & 0.51 \\
Sun et al. (2024) \citep{Sun2024} & 12 & 0.41 \\ 
\bottomrule
\end{tabular}
\caption{Parameters for the mass-radius relation of GMCs.}
\label{tab:GMC_MRR}
\end{center}
\end{table}

\begin{figure*}
    \centering
    \includegraphics[width=\textwidth]{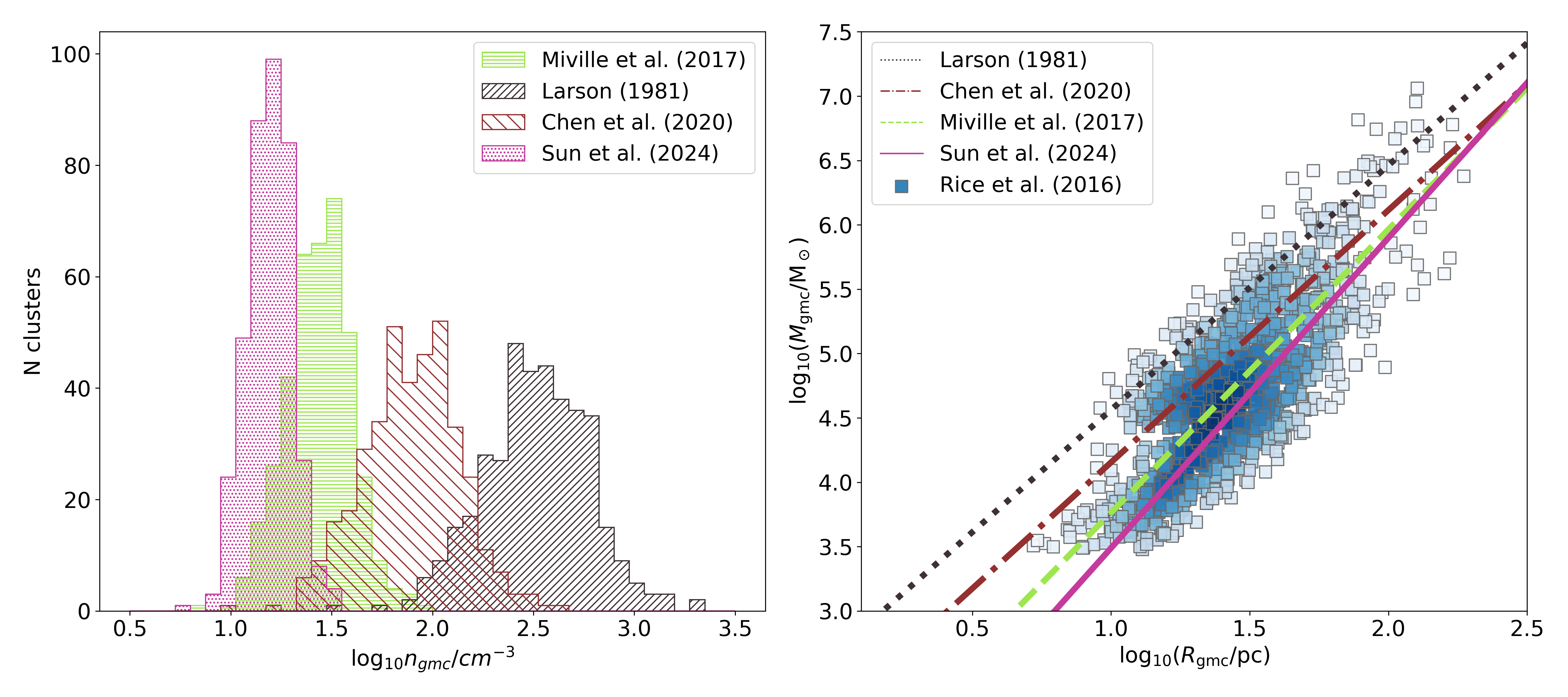}
    \caption{\textit{Left panel}: normalised particle density distributions for the sample of SCs considered using different mass radius relations. \textit{Right panel}: mass-radius relations compared to measured mass and radius from the catalogue provided by \cite{Rice16}. The colour code return the probability density distribution of the dataset calculated using a Gaussian kernel density estimation.}
    \label{fig:GMCdens}
\end{figure*}

\section{Star Cluster Physical Properties}
\label{sec:appA}

\renewcommand{\thetable}{\thesection.\arabic{table}}
\renewcommand{\theHtable}{\thesection.\arabic{table}}
\setcounter{table}{0}

Table~\ref{tab:cluster_top} provides the physical properties of the clusters with the brightest predicted gamma-ray flux according to our model. 

\begin{table*}
\begin{rotatetable*}
\begin{center}
    \begin{small}
    \caption{Properties of the stellar clusters with the brightest predicted gamma-ray emission according to our baseline model, ranked according to integral flux above 1\,TeV over the wind-blown bubble. The last three columns show the maximum energies obtained at the wind termination shock for Kraichnan and Bohm diffusion and the one achieved at SNR shocks. }
    \label{tab:cluster_top}
    \begin{tabular}{clccccccccccccccc}
    \toprule \toprule
& SC & GLON & GLAT & Age & Distance & $n_0$ & $R_{\rm c}$ & $R_{\rm s}$ & $R_{\rm b}$ &  $R_{\rm b}$ & $\dot{M}$ & $v_w$  & $L_{\rm w,c}$ & $E_{\rm max, Kra}^{\rm WTS}$ & $E_{\rm max, Bohm}^{\rm WTS}$ & $E_{\rm max}^{\rm SNe}$\\ 
    &  & $^\circ$ & $^\circ$ & Myr & kpc & cm$^{-3}$ & pc & pc & pc & $^\circ$ & M$_\odot$ / yr & cm / s & erg / s & PeV & PeV & PeV \\
    \midrule
    \midrule
1 & NGC\,6231 & 343.476 & 1.190 & 13.80 & 1.55 & 26.2 & 4.0 & 20.9 & 117.05 & 4.31 & 3.75e-05 & 1.87e+03 & 4.14e+37 & 0.12 & 0.59 & 0.42\\
2 & Westerlund\,1 & 339.546 & -0.401 & 7.94 & 7.69 & 9.4 & 3.1 & 42.4 & 154.85 & 1.15 & 0.000301 & 1.82e+03 & 3.15e+38 & 0.62 & 1.61 & 0.18\\
3 & Dolidze\,3 & 74.545 & 1.072 & 8.91 & 1.91 & 142.9 & 3.1 & 7.9 & 57.07 & 1.71 & 1.34e-05 & 2.34e+03 & 2.31e+37 & 0.05 & 0.49 & 0.51\\
4 & Berkeley\,87 & 75.756 & 0.361 & 8.32 & 1.69 & 42.7 & 2.7 & 11.1 & 69.53 & 2.36 & 1.36e-05 & 2.3e+03 & 2.28e+37 & 0.06 & 0.49 & 0.41\\
5 & Danks\,1 & 305.342 & 0.074 & 1.00 & 1.87 & 22.3 & 0.7 & 9.7 & 28.39 & 0.87 & 8.53e-05 & 1.7e+03 & 7.76e+37 & 0.06 & 0.77 & 0.90\\
6 & Danks\,2 & 305.390 & 0.089 & 2.00 & 5.15 & 16.3 & 1.2 & 11.5 & 44.79 & 0.50 & 3.86e-05 & 2.4e+03 & 7.01e+37 & 0.12 & 0.87 & 0.67\\
7 & UBC\,344 & 18.354 & 1.820 & 3.47 & 1.95 & 24.2 & 5.5 & 7.1 & 43.15 & 1.27 & 5.02e-06 & 3.22e+03 & 1.65e+37 & 0.06 & 0.49 & 0.40\\
8 & Berkeley\,86 & 76.650 & 1.276 & 10.96 & 1.68 & 120.8 & 2.1 & 8.4 & 55.65 & 1.89 & 1.2e-05 & 1.57e+03 & 9.25e+36 & 0.02 & 0.26 & 0.38\\
9 & Trumpler\,16 & 287.599 & -0.646 & 13.49 & 2.30 & 46.1 & 2.4 & 12.2 & 76.35 & 1.90 & 1.2e-05 & 1.56e+03 & 9.21e+36 & 0.02 & 0.25 & 0.31\\
10 & Patchick\,94 & 336.458 & 0.855 & 3.55 & 3.18 & 24.3 & 0.9 & 7.0 & 43.17 & 0.78 & 4.75e-06 & 3.21e+03 & 1.54e+37 & 0.05 & 0.47 & 0.40\\
11 & Markarian\,50 & 111.335 & -0.239 & 11.48 & 2.84 & 91.0 & 1.7 & 9.3 & 60.53 & 1.22 & 1.2e-05 & 1.56e+03 & 9.24e+36 & 0.02 & 0.25 & 0.36\\
12 & Ruprecht\,44 & 245.722 & 0.492 & 14.45 & 4.66 & 57.4 & 3.1 & 11.7 & 76.18 & 0.94 & 1.2e-05 & 1.56e+03 & 9.23e+36 & 0.02 & 0.25 & 0.31\\
13 & Westerlund\,2 & 284.272 & -0.328 & 3.98 & 4.53 & 32.2 & 1.6 & 10.4 & 44.85 & 0.57 & 2.61e-05 & 1.46e+03 & 1.75e+37 & 0.03 & 0.34 & 0.47\\
14 & NGC\,3603 & 291.624 & -0.518 & 1.00 & 6.78 & 21.1 & 4.6 & 9.2 & 25.83 & 0.22 & 7.21e-05 & 1.42e+03 & 4.6e+37 & 0.04 & 0.54 & 0.79\\
15 & Sher\,1 & 289.635 & -0.242 & 13.18 & 6.85 & 58.7 & 1.6 & 11.2 & 71.77 & 0.60 & 1.2e-05 & 1.56e+03 & 9.22e+36 & 0.02 & 0.25 & 0.32\\
16 & NGC\,6611 & 16.962 & 0.811 & 2.14 & 1.70 & 36.7 & 1.9 & 3.9 & 24.55 & 0.83 & 1.98e-06 & 3.19e+03 & 6.34e+36 & 0.02 & 0.30 & 0.39\\
17 & NGC\,6193 & 336.694 & -1.576 & 5.13 & 1.14 & 36.3 & 3.1 & 4.4 & 34.94 & 1.76 & 9.16e-07 & 3.03e+03 & 2.65e+36 & 0.01 & 0.19 & 0.25\\
18 & Juchert\,3 & 40.354 & -0.701 & 1.00 & 3.06 & 32.3 & 1.5 & 3.4 & 17.65 & 0.33 & 3.16e-06 & 3.24e+03 & 1.05e+37 & 0.02 & 0.39 & 0.52\\
19 & NGC\,6823 & 59.423 & -0.139 & 2.40 & 2.04 & 41.2 & 2.6 & 3.5 & 23.52 & 0.66 & 1.3e-06 & 3.15e+03 & 4.06e+36 & 0.01 & 0.24 & 0.35\\
20 & Hogg\,15 & 302.048 & -0.242 & 2.19 & 3.17 & 47.1 & 3.5 & 5.9 & 25.32 & 0.46 & 1.27e-05 & 1.49e+03 & 8.84e+36 & 0.01 & 0.24 & 0.48\\
21 & Berkeley\,59 & 118.230 & 5.019 & 1.26 & 1.02 & 41.9 & 2.4 & 2.6 & 15.67 & 0.88 & 1.21e-06 & 3.14e+03 & 3.75e+36 & 0.01 & 0.23 & 0.40\\
22 & DC\,5 & 286.795 & -0.502 & 1.12 & 4.42 & 34.7 & 2.4 & 3.3 & 17.61 & 0.23 & 2.42e-06 & 3.21e+03 & 7.87e+36 & 0.02 & 0.34 & 0.48\\
23 & NGC\,6357 & 353.166 & 0.890 & 1.00 & 1.67 & 40.9 & 1.5 & 2.5 & 14.02 & 0.48 & 1.34e-06 & 3.15e+03 & 4.18e+36 & 0.01 & 0.24 & 0.44\\
24 & UBC\,558 & 347.350 & -1.375 & 7.24 & 2.58 & 31.7 & 3.1 & 4.3 & 38.06 & 0.85 & 4.83e-07 & 2.88e+03 & 1.26e+36 & 0.01 & 0.13 & 0.18\\
25 & Berkeley\,90 & 84.883 & 3.794 & 5.13 & 2.73 & 54.1 & 2.6 & 3.0 & 27.00 & 0.57 & 3.82e-07 & 3.01e+03 & 1.09e+36 & 0.01 & 0.12 & 0.22\\
\bottomrule 
    \end{tabular}
    \end{small} 
    \end{center}
\end{rotatetable*}
\end{table*}

\section{Star Cluster Flux Predictions}
\label{sec:appB}

\renewcommand{\thetable}{\thesection.\arabic{table}}
\renewcommand{\theHtable}{\thesection.\arabic{table}}
\setcounter{table}{0}

Tables \ref{tab:clusterflux_min} and \ref{tab:clusterflux_max} provide the predicted gamma-ray and neutrino emission from our model under the \emph{baseline} and \emph{maximal} scenarios respectively, for the case of Kraichnan diffusion. 

\begin{table*}
     \begin{rotatetable*}
    \begin{center}
    \caption{Properties of the stellar clusters with the brightest predicted integral ($>1$~TeV) gamma-ray emission in ranked order according to the \emph{baseline} model. Integral flux values are given in units of TeV cm$^{-2}$ s$^{-1}$ and differential flux values in units of TeV$^{-1}$ cm$^{-2}$ s$^{-1}$. Upper limits from the HGPS (corresponding to the angular size of the bubble or $0.6^\circ$, whichever is smaller) and 3HWC (corresponding to the nearest available angular size) are provided where available \citep{2018HGPS,3hwc_2020ApJ...905...76A}. }
    \label{tab:clusterflux_min}
    \begin{tabular}{clccccccc}
\toprule \toprule
 & SC & $F_\gamma$ ($>$1\,TeV) & $F_\gamma^{\rm ul}$ ($>$1\,TeV) & $F_\gamma$ ($>$10\,TeV) & $F_\gamma^{\rm ul}$ ($>$10\,TeV) &  $F_\gamma$ (7\,TeV) & $F_\gamma^{\rm ul}$ (7\,TeV) & $F_{\nu_\mu}$ ($>$ 10\,TeV) \\
\midrule
\midrule
1 & NGC\,6231 & 2.3e-13 & 2e-11 & 1.5e-15 & 2.7e-11 & 2.7e-16 & -- & 9.1e-17 \\
2 & Westerlund\,1 & 1.8e-13 & 8.1e-12 & 7.9e-15 & 4.1e-11 & 5e-16 & -- & 1e-15 \\
3 & Dolidze\,3 & 4.7e-14 & 0 & 3.1e-17 & 0 & 1.8e-17 & 1.2e-14 & 1.1e-18 \\
4 & Berkeley\,87 & 3.5e-14 & -- & 5.7e-17 & -- & 2.3e-17 & 1.2e-14 & 2.4e-18 \\
5 & Danks\,1 & 1.9e-14 & 1.8e-12 & 3.1e-17 & 1.2e-11 & 1.2e-17 & -- & 1.3e-18 \\
6 & Danks\,2 & 5.7e-15 & 1e-12 & 4e-17 & 9.6e-12  & 7.2e-18 & -- & 2.4e-18 \\
7 & UBC\,344 & 4e-15 & 4.7e-12 & 4.9e-18 & 2.2e-11 & 2.2e-18 & 9.4e-15 & 1.9e-19 \\
8 & Berkeley\,86 & 3.2e-15 & -- & 3e-20 & -- & 8.7e-20 & 2.7e-15 & 5.5e-22 \\
9 & Trumpler\,16 & 2.1e-15 & 6.8e-12 & 1.3e-19 & 2.1e-11 & 1.9e-19 & -- & 3e-21 \\
10 & Patchick\,94 & 1.3e-15 & 7.3e-12 & 1.4e-18 & 5.4e-11 & 6.8e-19 & -- & 5.4e-20 \\
11 & Markarian\,50 & 1.2e-15 & -- & 1.9e-20 & -- & 4.6e-20 & 2.5e-14 & 3.8e-22 \\
12 & Ruprecht\,44 & 5.9e-16 & -- & 3.1e-20 & -- & 4.9e-20 & 4e-14 & 7e-22 \\
13 & Westerlund\,2 & 4.1e-16 & 1.6e-11 & 3.1e-20 & 1.5e-10  & 4.2e-20 & -- & 7.7e-22 \\
14 & NGC\,3603 & 3.2e-16 & 1.1e-12 & 8.1e-20 & 1.1e-11  & 6.6e-20 & -- & 2.4e-21 \\
15 & Sher\,1 & 2.3e-16 & 2.4e-12 & 9.9e-21 & 2.2e-11  & 1.7e-20 & -- & 2.2e-22 \\
16 & NGC\,6611 & 1.9e-16 & 3.2e-12 & 3.2e-21 & 2.2e-11 & 7.6e-21 & 1.9e-14 & 6.4e-23 \\
17 & NGC\,6193 & 1.1e-16 & 4.4e-12 & 1.9e-22 & 1.5e-11 & 1e-21 & -- & 2.9e-24 \\
18 & Juchert\,3 & 7.8e-17 & 5.1e-12 & 2.8e-21 & 4.7e-11 & 4.9e-21 & 4.4e-14 & 6.1e-23 \\
19 & NGC\,6823 & 3.7e-17 & 2.7e-12 & 8e-23 & 2.3e-11 & 3.9e-22 & 3.7e-15 & 1.3e-24 \\
20 & Hogg\,15 & 3.4e-17 & 2.5e-12 & 1.8e-23 & 2.3e-11  & 1.4e-22 & -- & 2.4e-25 \\
21 & Berkeley\,59 & 3.2e-17 & -- & 6.8e-24 & -- & 7.3e-23 & 1.4e-14 & 8.4e-26 \\
22 & DC\,5 & 1.8e-17 & 1.7e-12 & 2.2e-22 & 1.6e-11  & 5.9e-22 & -- & 4.2e-24 \\
23 & NGC\,6357 & 1.1e-17 & 3.1e-12 & 2.5e-24 & 2.9e-11  & 2.7e-23 & -- & 3.1e-26 \\
24 & UBC\,558 & 2.5e-18 & 1.8e-12 & 1.3e-25 & 1.2e-11 & 2.3e-24 & -- & 1.5e-27 \\
25 & Berkeley\,90 & 4.8e-19 & -- & 8.7e-28 & -- & 4.8e-26 & 1.7e-15 & 7.3e-30 \\
\bottomrule 
\end{tabular}
    \end{center}
\end{rotatetable*}
\end{table*}

\begin{table*}
    \begin{rotatetable*}
    \begin{center}
    \caption{Same as for table \ref{tab:clusterflux_min}, except providing the flux values for the \emph{maximal} scenario with Kraichnan diffusion. }
    \label{tab:clusterflux_max}
    \begin{tabular}{clccccccc}
\toprule \toprule
 & SC & $F_\gamma$ ($>$1\,TeV) & $F_\gamma^{\rm ul}$ ($>$1\,TeV) & $F_\gamma$ ($>$10\,TeV) & $F_\gamma^{\rm ul}$ ($>$10\,TeV) &  $F_\gamma$ (7\,TeV) & $F_\gamma^{\rm ul}$ (7\,TeV) & $F_{\nu_\mu}$ ($>$ 10\,TeV) \\
\midrule
\midrule
1 & NGC\,2264 & 3.4e-10 & -- & 3.7e-11 & -- & 1.5e-12 & 1.1e-15 & 5.5e-12 \\
2 & NGC\,1960 & 2.6e-10 & -- & 2.9e-11 & -- & 1.2e-12 & 1.6e-15 & 4.4e-12 \\
3 & UBC\,334 & 1.6e-10 & 9.3e-12 & 2.2e-11 & 1.5e-11 & 7.2e-13 & -- & 3.7e-12 \\
4 & NGC\,6249 & 8.8e-11 & 5e-12 & 1.1e-11 & 1.2e-11 & 4e-13 & -- & 1.8e-12 \\
5 & NGC\,6231 & 8.2e-11 & 2e-11 & 1.8e-11 & 2.7e-11 & 3.5e-13 & -- & 4e-12 \\
6 & NGC\,1502 & 7.3e-11 & -- & 1.3e-11 & -- & 3.5e-13 & 4.5e-15 & 2.4e-12 \\
7 & UBC\,355 & 5.7e-11 & 3.3e-12 & 6.8e-12 & 9.5e-12 & 2.6e-13 & 1.2e-14 & 1.1e-12 \\
8 & NGC\,4755 & 5.5e-11 & 9.4e-11 & 9.6e-12 & 2.4e-10 & 2.6e-13 & -- & 1.8e-12 \\
9 & NGC\,3590 & 5.2e-11 & 1.3e-11 & 6.1e-12 & 3.9e-11 & 2.3e-13 & -- & 9.5e-13 \\
10 & Teutsch\,30 & 4.9e-11 & -- & 5.1e-12 & -- & 2.1e-13 & 3.6e-15 & 7.5e-13 \\
11 & NGC\,366 & 4.9e-11 & -- & 5.8e-12 & -- & 2.2e-13 & 1.5e-14 & 9.1e-13 \\
12 & NGC\,7510 & 4.1e-11 & -- & 5.8e-12 & -- & 1.9e-13 & 1.8e-14 & 9.9e-13 \\
13 & Trumpler\,1 & 3.9e-11 & -- & 4.3e-12 & -- & 1.7e-13 & 2.2e-14 & 6.5e-13 \\
14 & Berkeley\,47 & 3.6e-11 & 7.2e-12 & 4.5e-12 & 2.6e-11 & 1.7e-13 & 7e-15 & 7.3e-13 \\
15 & Dolidze\,11 & 3.4e-11 & -- & 4.2e-12 & -- & 1.6e-13 & 9.3e-15 & 6.6e-13 \\
16 & Haffner\,15 & 3.1e-11 & 0 & 3.7e-12 & 0 & 1.4e-13 & -- & 5.8e-13 \\
17 & NGC\,7128 & 3.1e-11 & -- & 3.6e-12 & -- & 1.4e-13 & 8.6e-15 & 5.7e-13 \\
18 & NGC\,6318 & 2.6e-11 & 2.4e-11 & 4.7e-12 & 8.8e-11 & 1.2e-13 & -- & 9.2e-13 \\
19 & King\,10 & 2.4e-11 & -- & 3.5e-12 & -- & 1.1e-13 & 8.3e-15 & 6e-13 \\
20 & NGC\,3293 & 2.3e-11 & 4.7e-12 & 4.2e-12 & 1.8e-11 & 1.1e-13 & -- & 8.1e-13 \\
21 & Berkeley\,87 & 2.3e-11 & -- & 5.4e-12 & -- & 1.1e-13 & 1.2e-14 & 1.2e-12 \\
22 & NGC\,6396 & 2.3e-11 & 2.7e-12 & 3.4e-12 & 1.2e-11 & 1.1e-13 & -- & 6e-13 \\
23 & NGC\,6193 & 2e-11 & 4.7e-12 & 4.4e-12 & 1.5e-11 & 9.5e-14 & -- & 9.3e-13 \\
24 & Bica\,3 & 1.9e-11 & 2.2e-11 & 2.8e-12 & 1.1e-10 & 8.9e-14 & 1.3e-13 & 4.8e-13 \\
25 & Trumpler\,16 & 1.8e-11 & 6.8e-12 & 4.2e-12 & 2.1e-11 & 8.5e-14 & -- & 9.1e-13 \\
\bottomrule 
\end{tabular}
\end{center}
\end{rotatetable*}
\end{table*}

\section{Description of online data}
\label{sec:online}

\renewcommand{\thetable}{\thesection.\arabic{table}}
\renewcommand{\theHtable}{\thesection.\arabic{table}}
\setcounter{table}{0}

The full results from our model corresponding to the example tables \ref{tab:cluster_top}, \ref{tab:clusterflux_min} and \ref{tab:clusterflux_max} is available as supplementary material online in the form of two machine readable tables, corresponding to the \emph{baseline} and \emph{maximal} models respectively. 
The content and the format of these files is described in Tab.~\ref{tab:fitsfile}. Note that the upper limits are not included as these are not direct results of our model. 

\begin{table}
    \centering
    \begin{tabular}{c|c|c|l} \hline\hline
        Column & Data type & Unit & Description \\ \hline\hline
        cluster & str17 && Unified cluster name \\
        GLON & float64 & deg & Galactic Longitude \\
        GLON & float64 & deg & Galactic Latitude \\
        age & float64 & Myr & Cluster age \\
        distance & float64 & kpc & Distance to the cluster \\
        n$_{\rm ism}$ & float64 & cm$^{-3}$ & Local ISM density \\ 
        R$_c$ & float64 & pc & Cluster core radius \\
        R$_s$ & float64 & pc & Cluster shock radius \\
        R$_b$ & float64 & pc & Cluster bubble physical radius \\
        $\theta_b$ & float64 & deg & Cluster bubble angular radius \\
        $\dot{M}$ & float64 & M$_\odot$ / yr & Cluster mass loss rate \\  
        $v_w$ & float64 & cm / s & Cluster wind velocity \\          
        $L_w$ & float64 & erg / s & Cluster wind luminosity \\  
        EF1 & float64 & $\rm{TeV}$ / ($\rm{cm}^2$ s) & Integrated gamma-ray energy flux above 1 TeV\\
        EF10 & float64 & $\rm{TeV}$ / ($\rm{cm}^2$ s) & Integrated gamma-ray energy flux above 10 TeV\\
        DF7 & float64 & $\rm{TeV}^{-1}\rm{cm}^{-1}$s$^{-1}$ & Differential gamma-ray flux at 7 TeV \\
        EFnu10& float64 & $\rm{TeV}$ / ($\rm{cm}^2$ s) & Integrated muon neutrino energy flux above 10 TeV\\
        E$_{\rm max}$Kra & float64 & TeV & Maximum particle energy under Kraichnan diffusion \\
        E$_{\rm max}$Bohm & float64 & TeV & Maximum particle energy under Bohm diffusion \\
        E$_{\rm max}$SNe & float64 & TeV & Maximum particle energy due to SNe \\
    \hline\hline \end{tabular}
    \caption{Information on columns provided in the fits data product. This file is provided for both the \emph{baseline} and the \emph{maximal} scenarios. The integrated energy flux is always calculated over the cluster bubble. }
    \label{tab:fitsfile}
\end{table}


\bibliography{references}{}

@ARTICLE{Chen2020,
       author = {{Chen}, B. -Q. and {Li}, G. -X. and {Yuan}, H. -B. and {Huang}, Y. and {Tian}, Z. -J. and {Wang}, H. -F. and {Zhang}, H. -W. and {Wang}, C. and {Liu}, X. -W.},
        title = "{A large catalogue of molecular clouds with accurate distances within 4 kpc of the Galactic disc}",
      journal = {MNRAS},
         year = 2020,
        month = mar,
       volume = {493},
       number = {1},
        pages = {351-361},
          doi = {10.1093/mnras/staa235},
archivePrefix = {arXiv},
       eprint = {2001.11682},
 primaryClass = {astro-ph.GA},
       adsurl = {https://ui.adsabs.harvard.edu/abs/2020MNRAS.493..351C}
}

@ARTICLE{Larson1981,
       author = {{Larson}, R.~B.},
        title = "{Turbulence and star formation in molecular clouds.}",
      journal = {MNRAS},
         year = 1981,
        month = mar,
       volume = {194},
        pages = {809-826},
          doi = {10.1093/mnras/194.4.809},
       adsurl = {https://ui.adsabs.harvard.edu/abs/1981MNRAS.194..809L}
}

@ARTICLE{Miville2017,
       author = {{Miville-Desch{\^e}nes}, Marc-Antoine and {Murray}, Norman and {Lee}, Eve J.},
        title = "{Physical Properties of Molecular Clouds for the Entire Milky Way Disk}",
      journal = {ApJ},
         year = 2017,
        month = jan,
       volume = {834},
       number = {1},
          eid = {57},
        pages = {57},
          doi = {10.3847/1538-4357/834/1/57},
archivePrefix = {arXiv},
       eprint = {1610.05918},
 primaryClass = {astro-ph.GA},
       adsurl = {https://ui.adsabs.harvard.edu/abs/2017ApJ...834...57M}
}

@ARTICLE{Sun2024,
       author = {{Sun}, Yan and {Yang}, Ji and {Yan}, Qing-Zeng and {Zhang}, Shaobo and {Su}, Yang and {Chen}, Xuepeng and {Zhou}, Xin and {Ma}, Yuehui and {Yuan}, Lixia},
        title = "{Molecular Clouds in the Outer Milky Way Disk: Sample, Integrated Properties, and Radial Trends with Galactocentric Radius}",
      journal = {ApJS},
         year = 2024,
        month = dec,
       volume = {275},
       number = {2},
          eid = {35},
        pages = {35},
          doi = {10.3847/1538-4365/ad8237},
       adsurl = {https://ui.adsabs.harvard.edu/abs/2024ApJS..275...35S}
}

@ARTICLE{Kruijssen2019,
       author = {{Kruijssen}, J.~M. Diederik and {Schruba}, Andreas and {Chevance}, M{\'e}lanie and {Longmore}, Steven N. and {Hygate}, Alexander P.~S. and {Haydon}, Daniel T. and {McLeod}, Anna F. and {Dalcanton}, Julianne J. and {Tacconi}, Linda J. and {van Dishoeck}, Ewine F.},
        title = "{Fast and inefficient star formation due to short-lived molecular clouds and rapid feedback}",
      journal = {Nature},
         year = 2019,
        month = may,
       volume = {569},
       number = {7757},
        pages = {519-522},
          doi = {10.1038/s41586-019-1194-3},
archivePrefix = {arXiv},
       eprint = {1905.08801},
 primaryClass = {astro-ph.GA},
       adsurl = {https://ui.adsabs.harvard.edu/abs/2019Natur.569..519K}
}

@ARTICLE{tim2024,
       author = {{Unbehaun}, T. and {Mohrmann}, L. and {Funk}, S. and {CTA Consortium} and {KM3NeT Collaboration} and {Aiello}, S. and {Albert}, A. and {Garre}, S. Alves and {Aly}, Z. and {Ambrosone}, A. and {Ameli}, F. and {Andre}, M. and {Androutsou}, E. and {Anghinolfi}, M. and {Anguita}, M. and {Aphecetche}, L. and {Ardid}, M. and {Ardid}, S. and {Atmani}, H. and {Aublin}, J. and {Bagatelas}, C. and {Bailly-Salins}, L. and {Baret}, B. and {du Pree}, S. Basegmez and {Becherini}, Y. and {Bendahman}, M. and {Benfenati}, F. and {Benhassi}, M. and {Benoit}, D.~M. and {Berbee}, E. and {Bertin}, V. and {Biagi}, S. and {Boettcher}, M. and {Cabo}, M. Bou and {Boumaaza}, J. and {Bouta}, M. and {Bouwhuis}, M. and {Bozza}, C. and {Bozza}, R.~M. and {Br{\^a}nza{\c{s}}}, H. and {Bretaudeau}, F. and {Bruijn}, R. and {Brunner}, J. and {Bruno}, R. and {Buis}, E. and {Buompane}, R. and {Busto}, J. and {Caiffi}, B. and {Calvo}, D. and {Campion}, S. and {Capone}, A. and {Carenini}, F. and {Carretero}, V. and {Cartraud}, T. and {Castaldi}, P. and {Cecchini}, V. and {Celli}, S. and {Cerisy}, L. and {Chabab}, M. and {Chadolias}, M. and {Chen}, A. and {Cherubini}, S. and {Chiarusi}, T. and {Circella}, M. and {Cocimano}, R. and {Coelho}, J.~A.~B. and {Coleiro}, A. and {Coniglione}, R. and {Coyle}, P. and {Creusot}, A. and {Cruz}, A. and {Cuttone}, G. and {Dallier}, R. and {Darras}, Y. and {De Benedittis}, A. and {De Martino}, B. and {Decoene}, V. and {Del Burgo}, R. and {Di Mauro}, L.~S. and {Di Palma}, I. and {D{\'\i}az}, A.~F. and {Diego-Tortosa}, D. and {Distefano}, C. and {Domi}, A. and {Donzaud}, C. and {Dornic}, D. and {D{\"o}rr}, M. and {Drakopoulou}, E. and {Drouhin}, D. and {Dvornick{\'y}}, R. and {Eberl}, T. and {Eddymaoui}, A. and {van Eeden}, T. and {Eff}, M. and {van Eijk}, D. and {El Bojaddaini}, I. and {El Hedri}, S. and {Enzenh{\"o}fer}, A. and {Ferrara}, G. and {Filipovi{\'c}}, M.~D. and {Filippini}, F. and {Fusco}, L.~A. and {Gabriel}, J. and {Gal}, T. and {M{\'e}ndez}, J. Garc{\'\i}a and {Soto}, A. Garcia and {Oliver}, C. Gatius and {Gei{\ss}elbrecht}, N. and {Ghaddari}, H. and {Gialanella}, L. and {Gibson}, B.~K. and {Giorgio}, E. and {Girardi}, A. and {Goos}, I. and {Gozzini}, S.~R. and {Gracia}, R. and {Graf}, K. and {Guderian}, D. and {Guidi}, C. and {Guillon}, B. and {Guti{\'e}rrez}, M. and {van Haren}, H. and {Heijboer}, A. and {Hekalo}, A. and {Hennig}, L. and {Hern{\'a}ndez-Rey}, J.~J. and {Huang}, F. and {Ibnsalih}, W. Idrissi and {Illuminati}, G. and {James}, C.~W. and {de Jong}, M. and {de Jong}, P. and {Jung}, B.~J. and {Kalaczy{\'n}ski}, P. and {Kalekin}, O. and {Katz}, U.~F. and {Chowdhury}, N.~R. Khan and {Khatun}, A. and {Kistauri}, G. and {van der Knaap}, F. and {Kouchner}, A. and {Kulikovskiy}, V. and {Kvatadze}, R. and {Labalme}, M. and {Lahmann}, R. and {Larosa}, G. and {Lastoria}, C. and {Lazo}, A. and {Le Stum}, S. and {Lehaut}, G. and {Leonora}, E. and {Lessing}, N. and {Levi}, G. and {Clark}, M. Lindsey and {Longhitano}, F. and {Majumdar}, J. and {Malerba}, L. and {Ma{\'n}czak}, J. and {Manfreda}, A. and {Marconi}, M. and {Margiotta}, A. and {Marinelli}, A. and {Markou}, C. and {Martin}, L. and {Marzaioli}, F. and {Mastrodicasa}, M. and {Mastroianni}, S. and {Miccich{\`e}}, S. and {Miele}, G. and {Migliozzi}, P. and {Migneco}, E. and {Mijakowski}, P. and {Mitsou}, M.~L. and {Mollo}, C.~M. and {Morales-Gallegos}, L. and {Morley-Wong}, C. and {Mosbrugger}, A. and {Moussa}, A. and {Mateo}, I. Mozun and {Muller}, R. and {Musone}, M.~R. and {Musumeci}, M. and {Nauta}, L. and {Navas}, S. and {Nayerhoda}, A. and {Nicolau}, C.~A. and {Nkosi}, B. and {{\'O}'Fearraigh}, B. and {Oliviero}, V. and {Orlando}, A. and {Oukacha}, E. and {Gonz{\'a}lez}, J. Palacios and {Papalashvili}, G. and {Gomez}, E.~J. Pastor and {P{\v{a}}un}, A.~M. and {P{\v{a}}v{\v{a}}la{\c{s}}}, G.~E. and {Mart{\'\i}nez}, S. Pe{\~n}a and {Perrin-Terrin}, M. and {Perronnel}, J. and {Pestel}, V. and {Pestes}, R. and {Piattelli}, P. and {Poir{\`e}}, C. and {Popa}, V. and {Pradier}, T. and {Pulvirenti}, S. and {Qu{\'e}m{\'e}ner}, G. and {Quiroz}, C. and {Rahaman}, U. and {Randazzo}, N. and {Razzaque}, S. and {Rea}, I.~C. and {Real}, D. and {Reck}, S. and {Riccobene}, G. and {Robinson}, J. and {Romanov}, A. and {Roscilli}, L. and {Saina}, A. and {Greus}, F. Salesa and {Samtleben}, D.~F.~E. and {Losa}, A.~S. {\'a}nchez and {Sanguineti}, M. and {Santonastaso}, C. and {Santonocito}, D. and {Sapienza}, P. and {Schnabel}, J. and {Schneider}, M.~F. and {Schumann}, J. and {Schutte}, H.~M. and {Seneca}, J. and {Sennan}, N. and {Setter}, B. and {Sgura}, I. and {Shanidze}, R. and {Shitov}, Y. and {{\v{S}}imkovic}, F. and {Simonelli}, A. and {Sinopoulou}, A. and {Smirnov}, M.~V. and {Spisso}, B. and {Spurio}, M. and {Stavropoulos}, D. and {{\v{S}}tekl}, I. and {Taiuti}, M. and {Tayalati}, Y. and {Tedjditi}, H. and {Thiersen}, H. and {Tosta e Melo}, I. and {Trocme}, B. and {Tsagkli}, S. and {Tsourapis}, V. and {Tzamariudaki}, E. and {Vacheret}, A. and {Valsecchi}, V. and {Van Elewyck}, V. and {Vannoye}, G. and {Vasileiadis}, G. and {de Sola}, F. Vazquez and {Verilhac}, C. and {Veutro}, A. and {Viola}, S. and {Vivolo}, D. and {Warnhofer}, H. and {Wilms}, J. and {de Wolf}, E. and {Yousfi}, T. and {Zarpapis}, G. and {Zavatarelli}, S. and {Zegarelli}, A. and {Zito}, D. and {Zornoza}, J.~D. and {Z{\'u}{\~n}iga}, J. and {Zywucka}, N.},
        title = "{Prospects for combined analyses of hadronic emission from {\ensuremath{\gamma}} -ray sources in the Milky Way with CTA and KM3NeT}",
      journal = {European Physical Journal C},
         year = 2024,
        month = feb,
       volume = {84},
       number = {2},
          eid = {112},
        pages = {112},
          doi = {10.1140/epjc/s10052-023-12279-z},
archivePrefix = {arXiv},
       eprint = {2309.03007},
 primaryClass = {astro-ph.HE},
       adsurl = {https://ui.adsabs.harvard.edu/abs/2024EPJC...84..112U}
}

@ARTICLE{thijs,
       author = {{KM3NeT Collaboration} and {Aiello}, S. and {Albert}, A. and {Alshamsi}, M. and {Alves Garre}, S. and {Aly}, Z. and {Ambrosone}, A. and {Ameli}, F. and {Andre}, M. and {Androutsou}, E. and {Anguita}, M. and {Aphecetche}, L. and {Ardid}, M. and {Ardid}, S. and {Atmani}, H. and {Aublin}, J. and {Badaracco}, F. and {Bailly-Salins}, L. and {Barda{\v{c}}ov{\'a}}, Z. and {Baret}, B. and {Bariego-Quintana}, A. and {Baruzzi}, A. and {Basegmez du Pree}, S. and {Becherini}, Y. and {Bendahman}, M. and {Benfenati}, F. and {Benhassi}, M. and {Benoit}, D.~M. and {Berbee}, E. and {Bertin}, V. and {Biagi}, S. and {Boettcher}, M. and {Bonanno}, D. and {Boumaaza}, J. and {Bouta}, M. and {Bouwhuis}, M. and {Bozza}, C. and {Bozza}, R.~M. and {Br{\^a}nza{\c{s}}}, H. and {Bretaudeau}, F. and {Breuhaus}, M. and {Bruijn}, R. and {Brunner}, J. and {Bruno}, R. and {Buis}, E. and {Buompane}, R. and {Busto}, J. and {Caiffi}, B. and {Calvo}, D. and {Campion}, S. and {Capone}, A. and {Carenini}, F. and {Carretero}, V. and {Cartraud}, T. and {Castaldi}, P. and {Cecchini}, V. and {Celli}, S. and {Cerisy}, L. and {Chabab}, M. and {Chadolias}, M. and {Chen}, A. and {Cherubini}, S. and {Chiarusi}, T. and {Circella}, M. and {Cocimano}, R. and {Coelho}, J.~A.~B. and {Coleiro}, A. and {Coniglione}, R. and {Coyle}, P. and {Creusot}, A. and {Cuttone}, G. and {Dallier}, R. and {Darras}, Y. and {De Benedittis}, A. and {De Martino}, B. and {Decoene}, V. and {Del Burgo}, R. and {Del Rosso}, I. and {Di Mauro}, L.~S. and {Di Palma}, I. and {D{\'\i}az}, A.~F. and {Diaz}, C. and {Diego-Tortosa}, D. and {Distefano}, C. and {Domi}, A. and {Donzaud}, C. and {Dornic}, D. and {D{\"o}rr}, M. and {Drakopoulou}, E. and {Drouhin}, D. and {Ducoin}, J. -G. and {Dvornick{\'y}}, R. and {Eberl}, T. and {Eckerov{\'a}}, E. and {Eddymaoui}, A. and {van Eeden}, T. and {Eff}, M. and {van Eijk}, D. and {El Bojaddaini}, I. and {El Hedri}, S. and {Enzenh{\"o}fer}, A. and {Ferrara}, G. and {Filipovi{\'c}}, M.~D. and {Filippini}, F. and {Franciotti}, D. and {Fusco}, L.~A. and {Gabriel}, J. and {Gagliardini}, S. and {Gal}, T. and {Garc{\'\i}a M{\'e}ndez}, J. and {Garcia Soto}, A. and {Gatius Oliver}, C. and {Gei{\ss}elbrecht}, N. and {Ghaddari}, H. and {Gialanella}, L. and {Gibson}, B.~K. and {Giorgio}, E. and {Goos}, I. and {Goswami}, P. and {Goupilliere}, D. and {Gozzini}, S.~R. and {Gracia}, R. and {Graf}, K. and {Guidi}, C. and {Guillon}, B. and {Guti{\'e}rrez}, M. and {van Haren}, H. and {Heijboer}, A. and {Hekalo}, A. and {Hennig}, L. and {Hern{\'a}ndez-Rey}, J.~J. and {Idrissi Ibnsalih}, W. and {Illuminati}, G. and {de Jong}, M. and {de Jong}, P. and {Jung}, B.~J. and {Kalaczy{\'n}ski}, P. and {Kalekin}, O. and {Katz}, U.~F. and {Kistauri}, G. and {Kopper}, C. and {Kouchner}, A. and {Kueviakoe}, V. and {Kulikovskiy}, V. and {Kvatadze}, R. and {Labalme}, M. and {Lahmann}, R. and {Larosa}, G. and {Lastoria}, C. and {Lazo}, A. and {Le Stum}, S. and {Lehaut}, G. and {Leonora}, E. and {Lessing}, N. and {Levi}, G. and {Lindsey Clark}, M. and {Longhitano}, F. and {Magnani}, F. and {Majumdar}, J. and {Malerba}, L. and {Mamedov}, F. and {Ma{\'n}czak}, J. and {Manfreda}, A. and {Marconi}, M. and {Margiotta}, A. and {Marinelli}, A. and {Markou}, C. and {Martin}, L. and {Mart{\'\i}nez-Mora}, J.~A. and {Marzaioli}, F. and {Mastrodicasa}, M. and {Mastroianni}, S. and {Miccich{\`e}}, S. and {Miele}, G. and {Migliozzi}, P. and {Migneco}, E. and {Mitsou}, M.~L. and {Mollo}, C.~M. and {Morales-Gallegos}, L. and {Morga}, M. and {Moussa}, A. and {Mozun Mateo}, I. and {Muller}, R. and {Musone}, M.~R. and {Musumeci}, M. and {Navas}, S. and {Nayerhoda}, A. and {Nicolau}, C.~A. and {Nkosi}, B. and {{\'O} Fearraigh}, B. and {Oliviero}, V. and {Orlando}, A. and {Oukacha}, E. and {Paesani}, D. and {Palacios Gonz{\'a}lez}, J. and {Papalashvili}, G. and {Parisi}, V. and {Pastor Gomez}, E.~J. and {P{\u{a}}un}, A.~M. and {P{\u{a}}v{\u{a}}la{\c{s}}}, G.~E.},
        title = "{Astronomy potential of KM3NeT/ARCA}",
      journal = {European Physical Journal C},
         year = 2024,
        month = sep,
       volume = {84},
       number = {9},
          eid = {885},
        pages = {885},
          doi = {10.1140/epjc/s10052-024-13137-2},
archivePrefix = {arXiv},
       eprint = {2402.08363},
 primaryClass = {astro-ph.HE},
       adsurl = {https://ui.adsabs.harvard.edu/abs/2024EPJC...84..885K}
}

@ARTICLE{2022ApJ...935..167A,
       author = {{Astropy Collaboration} and {Price-Whelan}, Adrian M. and {Lim}, Pey Lian and {Earl}, Nicholas and {Starkman}, Nathaniel and {Bradley}, Larry and {Shupe}, David L. and {Patil}, Aarya A. and {Corrales}, Lia and {Brasseur}, C.~E. and {N{\"o}the}, Maximilian and {Donath}, Axel and {Tollerud}, Erik and {Morris}, Brett M. and {Ginsburg}, Adam and {Vaher}, Eero and {Weaver}, Benjamin A. and {Tocknell}, James and {Jamieson}, William and {van Kerkwijk}, Marten H. and {Robitaille}, Thomas P. and {Merry}, Bruce and {Bachetti}, Matteo and {G{\"u}nther}, H. Moritz and {Aldcroft}, Thomas L. and {Alvarado-Montes}, Jaime A. and {Archibald}, Anne M. and {B{\'o}di}, Attila and {Bapat}, Shreyas and {Barentsen}, Geert and {Baz{\'a}n}, Juanjo and {Biswas}, Manish and {Boquien}, M{\'e}d{\'e}ric and {Burke}, D.~J. and {Cara}, Daria and {Cara}, Mihai and {Conroy}, Kyle E. and {Conseil}, Simon and {Craig}, Matthew W. and {Cross}, Robert M. and {Cruz}, Kelle L. and {D'Eugenio}, Francesco and {Dencheva}, Nadia and {Devillepoix}, Hadrien A.~R. and {Dietrich}, J{\"o}rg P. and {Eigenbrot}, Arthur Davis and {Erben}, Thomas and {Ferreira}, Leonardo and {Foreman-Mackey}, Daniel and {Fox}, Ryan and {Freij}, Nabil and {Garg}, Suyog and {Geda}, Robel and {Glattly}, Lauren and {Gondhalekar}, Yash and {Gordon}, Karl D. and {Grant}, David and {Greenfield}, Perry and {Groener}, Austen M. and {Guest}, Steve and {Gurovich}, Sebastian and {Handberg}, Rasmus and {Hart}, Akeem and {Hatfield-Dodds}, Zac and {Homeier}, Derek and {Hosseinzadeh}, Griffin and {Jenness}, Tim and {Jones}, Craig K. and {Joseph}, Prajwel and {Kalmbach}, J. Bryce and {Karamehmetoglu}, Emir and {Ka{\l}uszy{\'n}ski}, Miko{\l}aj and {Kelley}, Michael S.~P. and {Kern}, Nicholas and {Kerzendorf}, Wolfgang E. and {Koch}, Eric W. and {Kulumani}, Shankar and {Lee}, Antony and {Ly}, Chun and {Ma}, Zhiyuan and {MacBride}, Conor and {Maljaars}, Jakob M. and {Muna}, Demitri and {Murphy}, N.~A. and {Norman}, Henrik and {O'Steen}, Richard and {Oman}, Kyle A. and {Pacifici}, Camilla and {Pascual}, Sergio and {Pascual-Granado}, J. and {Patil}, Rohit R. and {Perren}, Gabriel I. and {Pickering}, Timothy E. and {Rastogi}, Tanuj and {Roulston}, Benjamin R. and {Ryan}, Daniel F. and {Rykoff}, Eli S. and {Sabater}, Jose and {Sakurikar}, Parikshit and {Salgado}, Jes{\'u}s and {Sanghi}, Aniket and {Saunders}, Nicholas and {Savchenko}, Volodymyr and {Schwardt}, Ludwig and {Seifert-Eckert}, Michael and {Shih}, Albert Y. and {Jain}, Anany Shrey and {Shukla}, Gyanendra and {Sick}, Jonathan and {Simpson}, Chris and {Singanamalla}, Sudheesh and {Singer}, Leo P. and {Singhal}, Jaladh and {Sinha}, Manodeep and {Sip{\H{o}}cz}, Brigitta M. and {Spitler}, Lee R. and {Stansby}, David and {Streicher}, Ole and {{\v{S}}umak}, Jani and {Swinbank}, John D. and {Taranu}, Dan S. and {Tewary}, Nikita and {Tremblay}, Grant R. and {de Val-Borro}, Miguel and {Van Kooten}, Samuel J. and {Vasovi{\'c}}, Zlatan and {Verma}, Shresth and {de Miranda Cardoso}, Jos{\'e} Vin{\'\i}cius and {Williams}, Peter K.~G. and {Wilson}, Tom J. and {Winkel}, Benjamin and {Wood-Vasey}, W.~M. and {Xue}, Rui and {Yoachim}, Peter and {Zhang}, Chen and {Zonca}, Andrea and {Astropy Project Contributors}},
        title = "{The Astropy Project: Sustaining and Growing a Community-oriented Open-source Project and the Latest Major Release (v5.0) of the Core Package}",
      journal = {\apj},
         year = 2022,
        month = aug,
       volume = {935},
       number = {2},
          eid = {167},
        pages = {167},
          doi = {10.3847/1538-4357/ac7c74},
archivePrefix = {arXiv},
       eprint = {2206.14220},
 primaryClass = {astro-ph.IM},
       adsurl = {https://ui.adsabs.harvard.edu/abs/2022ApJ...935..167A}
}

@ARTICLE{2018AJ....156..123A,
       author = {{Astropy Collaboration} and {Price-Whelan}, A.~M. and {Sip{\H{o}}cz}, B.~M. and {G{\"u}nther}, H.~M. and {Lim}, P.~L. and {Crawford}, S.~M. and {Conseil}, S. and {Shupe}, D.~L. and {Craig}, M.~W. and {Dencheva}, N. and {Ginsburg}, A. and {VanderPlas}, J.~T. and {Bradley}, L.~D. and {P{\'e}rez-Su{\'a}rez}, D. and {de Val-Borro}, M. and {Aldcroft}, T.~L. and {Cruz}, K.~L. and {Robitaille}, T.~P. and {Tollerud}, E.~J. and {Ardelean}, C. and {Babej}, T. and {Bach}, Y.~P. and {Bachetti}, M. and {Bakanov}, A.~V. and {Bamford}, S.~P. and {Barentsen}, G. and {Barmby}, P. and {Baumbach}, A. and {Berry}, K.~L. and {Biscani}, F. and {Boquien}, M. and {Bostroem}, K.~A. and {Bouma}, L.~G. and {Brammer}, G.~B. and {Bray}, E.~M. and {Breytenbach}, H. and {Buddelmeijer}, H. and {Burke}, D.~J. and {Calderone}, G. and {Cano Rodr{\'\i}guez}, J.~L. and {Cara}, M. and {Cardoso}, J.~V.~M. and {Cheedella}, S. and {Copin}, Y. and {Corrales}, L. and {Crichton}, D. and {D'Avella}, D. and {Deil}, C. and {Depagne}, {\'E}. and {Dietrich}, J.~P. and {Donath}, A. and {Droettboom}, M. and {Earl}, N. and {Erben}, T. and {Fabbro}, S. and {Ferreira}, L.~A. and {Finethy}, T. and {Fox}, R.~T. and {Garrison}, L.~H. and {Gibbons}, S.~L.~J. and {Goldstein}, D.~A. and {Gommers}, R. and {Greco}, J.~P. and {Greenfield}, P. and {Groener}, A.~M. and {Grollier}, F. and {Hagen}, A. and {Hirst}, P. and {Homeier}, D. and {Horton}, A.~J. and {Hosseinzadeh}, G. and {Hu}, L. and {Hunkeler}, J.~S. and {Ivezi{\'c}}, {\v{Z}}. and {Jain}, A. and {Jenness}, T. and {Kanarek}, G. and {Kendrew}, S. and {Kern}, N.~S. and {Kerzendorf}, W.~E. and {Khvalko}, A. and {King}, J. and {Kirkby}, D. and {Kulkarni}, A.~M. and {Kumar}, A. and {Lee}, A. and {Lenz}, D. and {Littlefair}, S.~P. and {Ma}, Z. and {Macleod}, D.~M. and {Mastropietro}, M. and {McCully}, C. and {Montagnac}, S. and {Morris}, B.~M. and {Mueller}, M. and {Mumford}, S.~J. and {Muna}, D. and {Murphy}, N.~A. and {Nelson}, S. and {Nguyen}, G.~H. and {Ninan}, J.~P. and {N{\"o}the}, M. and {Ogaz}, S. and {Oh}, S. and {Parejko}, J.~K. and {Parley}, N. and {Pascual}, S. and {Patil}, R. and {Patil}, A.~A. and {Plunkett}, A.~L. and {Prochaska}, J.~X. and {Rastogi}, T. and {Reddy Janga}, V. and {Sabater}, J. and {Sakurikar}, P. and {Seifert}, M. and {Sherbert}, L.~E. and {Sherwood-Taylor}, H. and {Shih}, A.~Y. and {Sick}, J. and {Silbiger}, M.~T. and {Singanamalla}, S. and {Singer}, L.~P. and {Sladen}, P.~H. and {Sooley}, K.~A. and {Sornarajah}, S. and {Streicher}, O. and {Teuben}, P. and {Thomas}, S.~W. and {Tremblay}, G.~R. and {Turner}, J.~E.~H. and {Terr{\'o}n}, V. and {van Kerkwijk}, M.~H. and {de la Vega}, A. and {Watkins}, L.~L. and {Weaver}, B.~A. and {Whitmore}, J.~B. and {Woillez}, J. and {Zabalza}, V. and {Astropy Contributors}},
        title = "{The Astropy Project: Building an Open-science Project and Status of the v2.0 Core Package}",
      journal = {\aj},
         year = 2018,
        month = sep,
       volume = {156},
       number = {3},
          eid = {123},
        pages = {123},
          doi = {10.3847/1538-3881/aabc4f},
archivePrefix = {arXiv},
       eprint = {1801.02634},
 primaryClass = {astro-ph.IM},
       adsurl = {https://ui.adsabs.harvard.edu/abs/2018AJ....156..123A}
}

@ARTICLE{2013A&A...558A..33A,
       author = {{Astropy Collaboration} and {Robitaille}, Thomas P. and {Tollerud}, Erik J. and {Greenfield}, Perry and {Droettboom}, Michael and {Bray}, Erik and {Aldcroft}, Tom and {Davis}, Matt and {Ginsburg}, Adam and {Price-Whelan}, Adrian M. and {Kerzendorf}, Wolfgang E. and {Conley}, Alexander and {Crighton}, Neil and {Barbary}, Kyle and {Muna}, Demitri and {Ferguson}, Henry and {Grollier}, Fr{\'e}d{\'e}ric and {Parikh}, Madhura M. and {Nair}, Prasanth H. and {Unther}, Hans M. and {Deil}, Christoph and {Woillez}, Julien and {Conseil}, Simon and {Kramer}, Roban and {Turner}, James E.~H. and {Singer}, Leo and {Fox}, Ryan and {Weaver}, Benjamin A. and {Zabalza}, Victor and {Edwards}, Zachary I. and {Azalee Bostroem}, K. and {Burke}, D.~J. and {Casey}, Andrew R. and {Crawford}, Steven M. and {Dencheva}, Nadia and {Ely}, Justin and {Jenness}, Tim and {Labrie}, Kathleen and {Lim}, Pey Lian and {Pierfederici}, Francesco and {Pontzen}, Andrew and {Ptak}, Andy and {Refsdal}, Brian and {Servillat}, Mathieu and {Streicher}, Ole},
        title = "{Astropy: A community Python package for astronomy}",
      journal = {\aap},
         year = 2013,
        month = oct,
       volume = {558},
          eid = {A33},
        pages = {A33},
          doi = {10.1051/0004-6361/201322068},
archivePrefix = {arXiv},
       eprint = {1307.6212},
 primaryClass = {astro-ph.IM},
       adsurl = {https://ui.adsabs.harvard.edu/abs/2013A&A...558A..33A}
}

@Article{numpy_harris2020array,
 title         = {Array programming with {NumPy}},
 author        = {Charles R. Harris and K. Jarrod Millman and St{\'{e}}fan J.
                 van der Walt and Ralf Gommers and Pauli Virtanen and David
                 Cournapeau and Eric Wieser and Julian Taylor and Sebastian
                 Berg and Nathaniel J. Smith and Robert Kern and Matti Picus
                 and Stephan Hoyer and Marten H. van Kerkwijk and Matthew
                 Brett and Allan Haldane and Jaime Fern{\'{a}}ndez del
                 R{\'{i}}o and Mark Wiebe and Pearu Peterson and Pierre
                 G{\'{e}}rard-Marchant and Kevin Sheppard and Tyler Reddy and
                 Warren Weckesser and Hameer Abbasi and Christoph Gohlke and
                 Travis E. Oliphant},
 year          = {2020},
 month         = sep,
 journal       = {Nature},
 volume        = {585},
 number        = {7825},
 pages         = {357--362},
 doi           = {10.1038/s41586-020-2649-2},
 publisher     = {Springer Science and Business Media {LLC}},
 url           = {https://doi.org/10.1038/s41586-020-2649-2}
}

@ARTICLE{2020SciPy-NMeth,
  author  = {Virtanen, Pauli and Gommers, Ralf and Oliphant, Travis E. and
            Haberland, Matt and Reddy, Tyler and Cournapeau, David and
            Burovski, Evgeni and Peterson, Pearu and Weckesser, Warren and
            Bright, Jonathan and {van der Walt}, St{\'e}fan J. and
            Brett, Matthew and Wilson, Joshua and Millman, K. Jarrod and
            Mayorov, Nikolay and Nelson, Andrew R. J. and Jones, Eric and
            Kern, Robert and Larson, Eric and Carey, C J and
            Polat, {\.I}lhan and Feng, Yu and Moore, Eric W. and
            {VanderPlas}, Jake and Laxalde, Denis and Perktold, Josef and
            Cimrman, Robert and Henriksen, Ian and Quintero, E. A. and
            Harris, Charles R. and Archibald, Anne M. and
            Ribeiro, Ant{\^o}nio H. and Pedregosa, Fabian and
            {van Mulbregt}, Paul and {SciPy 1.0 Contributors}},
  title   = {{{SciPy} 1.0: Fundamental Algorithms for Scientific
            Computing in Python}},
  journal = {Nature Methods},
  year    = {2020},
  volume  = {17},
  pages   = {261--272},
  adsurl  = {https://rdcu.be/b08Wh},
  doi     = {10.1038/s41592-019-0686-2},
}

@ARTICLE{2018APh...100...69A,
       author = {{Ambrogi}, L. and {Celli}, S. and {Aharonian}, F.},
        title = "{On the potential of Cherenkov Telescope Arrays and KM3 Neutrino Telescopes for the detection of extended sources}",
      journal = {Astroparticle Physics},
         year = 2018,
        month = jul,
       volume = {100},
        pages = {69-79},
          doi = {10.1016/j.astropartphys.2018.03.001},
archivePrefix = {arXiv},
       eprint = {1803.03565},
 primaryClass = {astro-ph.HE},
       adsurl = {https://ui.adsabs.harvard.edu/abs/2018APh...100...69A}
}

@ARTICLE{2017Y,
       author = {{Yang}, Rui-zhi and {Aharonian}, Felix},
        title = "{Diffuse {\ensuremath{\gamma}}-ray emission near the young massive cluster NGC 3603}",
      journal = {A\&A},
         year = 2017,
        month = apr,
       volume = {600},
          eid = {A107},
        pages = {A107},
          doi = {10.1051/0004-6361/201630213},
archivePrefix = {arXiv},
       eprint = {1612.02250},
 primaryClass = {astro-ph.HE},
       adsurl = {https://ui.adsabs.harvard.edu/abs/2017A&A...600A.107Y}
}

@ARTICLE{celli2020,
       author = {{Celli}, Silvia and {Aharonian}, Felix and {Gabici}, Stefano},
        title = "{Spectral Signatures of PeVatrons}",
      journal = {ApJ},
         year = 2020,
        month = nov,
       volume = {903},
       number = {1},
          eid = {61},
        pages = {61},
          doi = {10.3847/1538-4357/abb805},
archivePrefix = {arXiv},
       eprint = {2009.05999},
 primaryClass = {astro-ph.HE},
       adsurl = {https://ui.adsabs.harvard.edu/abs/2020ApJ...903...61C}
}

@ARTICLE{LHAASOcygnus:2024,
       author = {{LHAASO Collaboration}},
        title = "{An ultrahigh-energy {\ensuremath{\gamma}} -ray bubble powered by a super PeVatron}",
      journal = {Science Bulletin},
         year = 2024,
        month = feb,
       volume = {69},
       number = {4},
        pages = {449-457},
          doi = {10.1016/j.scib.2023.12.040},
archivePrefix = {arXiv},
       eprint = {2310.10100},
 primaryClass = {astro-ph.HE},
       adsurl = {https://ui.adsabs.harvard.edu/abs/2024SciBu..69..449L}
}

@ARTICLE{celli2023,
       author = {{Celli}, S. and {Specovius}, A. and {Menchiari}, S. and {Mitchell}, A. and {Morlino}, G.},
        title = "{Mass and wind luminosity of young Galactic open clusters in Gaia DR2}",
      journal = {A\&A},
         year = 2024,
        month = jun,
       volume = {686},
          eid = {A118},
        pages = {A118},
          doi = {10.1051/0004-6361/202348541},
archivePrefix = {arXiv},
       eprint = {2311.09089},
 primaryClass = {astro-ph.GA},
       adsurl = {https://ui.adsabs.harvard.edu/abs/2024A&A...686A.118C}
}

@BOOK{Menchiari_PhD:2023,
       author = {{Menchiari}, S.},
        title = "{Probing star clusters as cosmic ray factories}",
         year = 2023,
       adsurl = {https://hdl.handle.net/11365/1235294}
}

@ARTICLE{Menchiari+2024,
       author = {{Menchiari}, S. and {Morlino}, G. and {Amato}, E. and {Bucciantini}, N. and {Beltr{\'a}n}, M.~T.},
        title = "{Cygnus OB2 as a test case for particle acceleration in young massive star clusters}",
      journal = {A\&A},
         year = 2024,
        month = jun,
       volume = {686},
          eid = {A242},
        pages = {A242},
          doi = {10.1051/0004-6361/202348817},
archivePrefix = {arXiv},
       eprint = {2402.07784},
 primaryClass = {astro-ph.HE},
       adsurl = {https://ui.adsabs.harvard.edu/abs/2024A&A...686A.242M}
}

@misc{zenodoCTA,
doi = {10.5281/ZENODO.5499840},
url = {https://zenodo.org/record/5499840},
author = {{Cherenkov Telescope Array Observatory} and {Cherenkov Telescope Array Consortium}},
title = {CTAO Instrument Response Functions - prod5 version v0.1},
publisher = {Zenodo},
year = {2021},
copyright = {Creative Commons Attribution 4.0 International}
}

@article{AMS2015,
  title = {Precision Measurement of the Proton Flux in Primary Cosmic Rays from Rigidity 1 GV to 1.8 TV with the Alpha Magnetic Spectrometer on the International Space Station},
  author = {Aguilar, M. and Aisa, D. and Alpat, B. and Alvino, A. and Ambrosi, G. and Andeen, K. and Arruda, L. and Attig, N. and Azzarello, P. and Bachlechner, A. and Barao, F. and Barrau, A. and Barrin, L. and Bartoloni, A. and Basara, L. and Battarbee, M. and Battiston, R. and Bazo, J. and Becker, U. and Behlmann, M. and Beischer, B. and Berdugo, J. and Bertucci, B. and Bigongiari, G. and Bindi, V. and Bizzaglia, S. and Bizzarri, M. and Boella, G. and de Boer, W. and Bollweg, K. and Bonnivard, V. and Borgia, B. and Borsini, S. and Boschini, M. J. and Bourquin, M. and Burger, J. and Cadoux, F. and Cai, X. D. and Capell, M. and Caroff, S. and Casaus, J. and Cascioli, V. and Castellini, G. and Cernuda, I. and Cerreta, D. and Cervelli, F. and Chae, M. J. and Chang, Y. H. and Chen, A. I. and Chen, H. and Cheng, G. M. and Chen, H. S. and Cheng, L. and Chou, H. Y. and Choumilov, E. and Choutko, V. and Chung, C. H. and Clark, C. and Clavero, R. and Coignet, G. and Consolandi, C. and Contin, A. and Corti, C. and Gil, E. Cortina and Coste, B. and Creus, W. and Crispoltoni, M. and Cui, Z. and Dai, Y. M. and Delgado, C. and Della Torre, S. and Demirk\"oz, M. B. and Derome, L. and Di Falco, S. and Di Masso, L. and Dimiccoli, F. and D\'{\i}az, C. and von Doetinchem, P. and Donnini, F. and Du, W. J. and Duranti, M. and D'Urso, D. and Eline, A. and Eppling, F. J. and Eronen, T. and Fan, Y. Y. and Farnesini, L. and Feng, J. and Fiandrini, E. and Fiasson, A. and Finch, E. and Fisher, P. and Galaktionov, Y. and Gallucci, G. and Garc\'{\i}a, B. and Garc\'{\i}a-L\'opez, R. and Gargiulo, C. and Gast, H. and Gebauer, I. and Gervasi, M. and Ghelfi, A. and Gillard, W. and Giovacchini, F. and Goglov, P. and Gong, J. and Goy, C. and Grabski, V. and Grandi, D. and Graziani, M. and Guandalini, C. and Guerri, I. and Guo, K. H. and Haas, D. and Habiby, M. and Haino, S. and Han, K. C. and He, Z. H. and Heil, M. and Hoffman, J. and Hsieh, T. H. and Huang, Z. C. and Huh, C. and Incagli, M. and Ionica, M. and Jang, W. Y. and Jinchi, H. and Kanishev, K. and Kim, G. N. and Kim, K. S. and Kirn, Th. and Kossakowski, R. and Kounina, O. and Kounine, A. and Koutsenko, V. and Krafczyk, M. S. and La Vacca, G. and Laudi, E. and Laurenti, G. and Lazzizzera, I. and Lebedev, A. and Lee, H. T. and Lee, S. C. and Leluc, C. and Levi, G. and Li, H. L. and Li, J. Q. and Li, Q. and Li, Q. and Li, T. X. and Li, W. and Li, Y. and Li, Z. H. and Li, Z. Y. and Lim, S. and Lin, C. H. and Lipari, P. and Lippert, T. and Liu, D. and Liu, H. and Lolli, M. and Lomtadze, T. and Lu, M. J. and Lu, S. Q. and Lu, Y. S. and Luebelsmeyer, K. and Luo, J. Z. and Lv, S. S. and Majka, R. and Ma\~n\'a, C. and Mar\'{\i}n, J. and Martin, T. and Mart\'{\i}nez, G. and Masi, N. and Maurin, D. and Menchaca-Rocha, A. and Meng, Q. and Mo, D. C. and Morescalchi, L. and Mott, P. and M\"uller, M. and Ni, J. Q. and Nikonov, N. and Nozzoli, F. and Nunes, P. and Obermeier, A. and Oliva, A. and Orcinha, M. and Palmonari, F. and Palomares, C. and Paniccia, M. and Papi, A. and Pauluzzi, M. and Pedreschi, E. and Pensotti, S. and Pereira, R. and Picot-Clemente, N. and Pilo, F. and Piluso, A. and Pizzolotto, C. and Plyaskin, V. and Pohl, M. and Poireau, V. and Postaci, E. and Putze, A. and Quadrani, L. and Qi, X. M. and Qin, X. and Qu, Z. Y. and R\"aih\"a, T. and Rancoita, P. G. and Rapin, D. and Ricol, J. S. and Rodr\'{\i}guez, I. and Rosier-Lees, S. and Rozhkov, A. and Rozza, D. and Sagdeev, R. and Sandweiss, J. and Saouter, P. and Sbarra, C. and Schael, S. and Schmidt, S. M. and von Dratzig, A. Schulz and Schwering, G. and Scolieri, G. and Seo, E. S. and Shan, B. S. and Shan, Y. H. and Shi, J. Y. and Shi, X. Y. and Shi, Y. M. and Siedenburg, T. and Son, D. and Spada, F. and Spinella, F. and Sun, W. and Sun, W. H. and Tacconi, M. and Tang, C. P. and Tang, X. W. and Tang, Z. C. and Tao, L. and Tescaro, D. and Ting, Samuel C. C. and Ting, S. M. and Tomassetti, N. and Torsti, J. and T\"urko\ifmmode \breve{g}\else \u{g}\fi{}lu, C. and Urban, T. and Vagelli, V. and Valente, E. and Vannini, C. and Valtonen, E. and Vaurynovich, S. and Vecchi, M. and Velasco, M. and Vialle, J. P. and Vitale, V. and Vitillo, S. and Wang, L. Q. and Wang, N. H. and Wang, Q. L. and Wang, R. S. and Wang, X. and Wang, Z. X. and Weng, Z. L. and Whitman, K. and Wienkenh\"over, J. and Wu, H. and Wu, X. and Xia, X. and Xie, M. and Xie, S. and Xiong, R. Q. and Xin, G. M. and Xu, N. S. and Xu, W. and Yan, Q. and Yang, J. and Yang, M. and Ye, Q. H. and Yi, H. and Yu, Y. J. and Yu, Z. Q. and Zeissler, S. and Zhang, J. H. and Zhang, M. T. and Zhang, X. B. and Zhang, Z. and Zheng, Z. M. and Zhuang, H. L. and Zhukov, V. and Zichichi, A. and Zimmermann, N. and Zuccon, P. and Zurbach, C.},
  collaboration = {AMS},
  journal = {Phys. Rev. Lett.},
  volume = {114},
  issue = {17},
  pages = {171103},
  numpages = {9},
  year = {2015},
  month = {Apr},
  publisher = {American Physical Society},
  doi = {10.1103/PhysRevLett.114.171103},
  url = {https://link.aps.org/doi/10.1103/PhysRevLett.114.171103}
}

@ARTICLE{Caprioli-Spitkovsky:2014,
       author = {{Caprioli}, D. and {Spitkovsky}, A.},
        title = "{Simulations of Ion Acceleration at Non-relativistic Shocks. I. Acceleration Efficiency}",
      journal = {ApJ},
         year = 2014,
        month = mar,
       volume = {783},
       number = {2},
          eid = {91},
        pages = {91},
          doi = {10.1088/0004-637X/783/2/91},
archivePrefix = {arXiv},
       eprint = {1310.2943},
 primaryClass = {astro-ph.HE},
       adsurl = {https://ui.adsabs.harvard.edu/abs/2014ApJ...783...91C}
}

@ARTICLE{Haggerty-Caprioli:2020,
       author = {{Haggerty}, Colby C. and {Caprioli}, Damiano},
        title = "{Kinetic Simulations of Cosmic-Ray-modified Shocks. I. Hydrodynamics}",
      journal = {\apj},
         year = 2020,
        month = dec,
       volume = {905},
       number = {1},
          eid = {1},
        pages = {1},
          doi = {10.3847/1538-4357/abbe06},
archivePrefix = {arXiv},
       eprint = {2008.12308},
 primaryClass = {astro-ph.HE},
       adsurl = {https://ui.adsabs.harvard.edu/abs/2020ApJ...905....1H}
}

@ARTICLE{2025RocamoraNGC3603,
       author = {{Rocamora}, M. and {Reimer}, A. and {Mart{\'\i}-Devesa}, G. and {Kissmann}, R.},
        title = "{Exploring non-thermal emission from the star-forming region NGC 3603 through a realistic modelling of its environment}",
      journal = {\aap},
         year = 2025,
        month = jul,
       volume = {699},
          eid = {A136},
        pages = {A136},
          doi = {10.1051/0004-6361/202452883},
archivePrefix = {arXiv},
       eprint = {2411.05206},
 primaryClass = {astro-ph.HE},
       adsurl = {https://ui.adsabs.harvard.edu/abs/2025A&A...699A.136R}
}

@ARTICLE{Morlino-Caprioli:2012,
       author = {{Morlino}, G. and {Caprioli}, D.},
        title = "{Strong evidence for hadron acceleration in Tycho's supernova remnant}",
      journal = {\aap},
         year = 2012,
        month = feb,
       volume = {538},
          eid = {A81},
        pages = {A81},
          doi = {10.1051/0004-6361/201117855},
archivePrefix = {arXiv},
       eprint = {1105.6342},
 primaryClass = {astro-ph.HE},
       adsurl = {https://ui.adsabs.harvard.edu/abs/2012A&A...538A..81M}
}

@ARTICLE{Caprioli:2012,
       author = {{Caprioli}, Damiano},
        title = "{Cosmic-ray acceleration in supernova remnants: non-linear theory revised}",
      journal = {\jcap},
         year = 2012,
        month = jul,
       volume = {2012},
       number = {7},
          eid = {038},
        pages = {038},
          doi = {10.1088/1475-7516/2012/07/038},
archivePrefix = {arXiv},
       eprint = {1206.1360},
 primaryClass = {astro-ph.HE},
       adsurl = {https://ui.adsabs.harvard.edu/abs/2012JCAP...07..038C}
}

@ARTICLE{El-Badry+2019MNRAS,
       author = {{El-Badry}, Kareem and {Ostriker}, Eve C. and {Kim}, Chang-Goo and {Quataert}, Eliot and {Weisz}, Daniel R.},
        title = "{Evolution of supernovae-driven superbubbles with conduction and cooling}",
      journal = {\mnras},
         year = 2019,
        month = dec,
       volume = {490},
       number = {2},
        pages = {1961-1990},
          doi = {10.1093/mnras/stz2773},
archivePrefix = {arXiv},
       eprint = {1902.09547},
 primaryClass = {astro-ph.GA},
       adsurl = {https://ui.adsabs.harvard.edu/abs/2019MNRAS.490.1961E}
}

@ARTICLE{Rodriguez+2025,
       author = {{Rodriguez}, Jennifer A. and {Lopez}, Laura A. and {Lancaster}, Lachlan and {Rosen}, Anna L. and {Nayak}, Omnarayani and {Lopez}, Sebastian and {Holland-Ashford}, Tyler and {Webb}, Trinity L.},
        title = "{Taming the Tarantula: How Stellar Wind Feedback Shapes Gas and Dust in 30 Doradus}",
      journal = {arXiv e-prints},
         year = 2025,
        month = dec,
          eid = {arXiv:2512.03129},
        pages = {arXiv:2512.03129},
          doi = {10.48550/arXiv.2512.03129},
archivePrefix = {arXiv},
       eprint = {2512.03129},
 primaryClass = {astro-ph.HE},
       adsurl = {https://ui.adsabs.harvard.edu/abs/2025arXiv251203129R}
}

@ARTICLE{Vieu-Reville:2023,
       author = {{Vieu}, T. and {Reville}, B.},
        title = "{Massive star cluster origin for the galactic cosmic ray population at very-high energies}",
      journal = {MNRAS},
         year = 2023,
        month = feb,
       volume = {519},
       number = {1},
        pages = {136-147},
          doi = {10.1093/mnras/stac3469},
archivePrefix = {arXiv},
       eprint = {2211.11625},
 primaryClass = {astro-ph.HE},
       adsurl = {https://ui.adsabs.harvard.edu/abs/2023MNRAS.519..136V}
}

@ARTICLE{vieu2022,
       author = {{Vieu}, T. and {Reville}, B. and {Aharonian}, F.},
        title = "{Can superbubbles accelerate ultrahigh energy protons?}",
      journal = {MNRAS},
         year = 2022,
        month = sep,
       volume = {515},
       number = {2},
        pages = {2256-2265},
          doi = {10.1093/mnras/stac1901},
archivePrefix = {arXiv},
       eprint = {2207.01432},
 primaryClass = {astro-ph.HE},
       adsurl = {https://ui.adsabs.harvard.edu/abs/2022MNRAS.515.2256V}
}

@ARTICLE{vieu2020,
       author = {{Vieu}, T. and {Gabici}, S. and {Tatischeff}, V.},
        title = "{Particle acceleration at colliding shock waves}",
      journal = {MNRAS},
         year = 2020,
        month = may,
       volume = {494},
       number = {3},
        pages = {3166-3176},
          doi = {10.1093/mnras/staa799},
archivePrefix = {arXiv},
       eprint = {2003.03411},
 primaryClass = {astro-ph.HE},
       adsurl = {https://ui.adsabs.harvard.edu/abs/2020MNRAS.494.3166V}
}

@ARTICLE{Harer+2023,
       author = {{H{\"a}rer}, Lucia K. and {Reville}, Brian and {Hinton}, Jim and {Mohrmann}, Lars and {Vieu}, Thibault},
        title = "{Understanding the TeV {\ensuremath{\gamma}}-ray emission surrounding the young massive star cluster Westerlund 1}",
      journal = {A\&A},
         year = 2023,
        month = mar,
       volume = {671},
          eid = {A4},
        pages = {A4},
          doi = {10.1051/0004-6361/202245444},
archivePrefix = {arXiv},
       eprint = {2301.10496},
 primaryClass = {astro-ph.HE},
       adsurl = {https://ui.adsabs.harvard.edu/abs/2023A&A...671A...4H}
}

@ARTICLE{Lemoine-Goumard+2025,
       author = {{Lemoine-Goumard}, Marianne and {H{\"a}rer}, Lucia and {Mohrmann}, Lars and {Bernet}, Romain and {Hinton}, Jim and {Peron}, Giada and {Reville}, Brian and {Tibaldo}, Luigi and {Vieu}, Thibault},
        title = "{A cosmic-ray loaded nascent outflow driven by a massive star cluster}",
      journal = {Nature Communications},
         year = 2025,
        month = dec,
       volume = {16},
       number = {1},
          eid = {10820},
        pages = {10820},
          doi = {10.1038/s41467-025-65592-4},
       adsurl = {https://ui.adsabs.harvard.edu/abs/2025NatCo..1610820L}
}

@ARTICLE{gupta2016,
       author = {{Gupta}, Siddhartha and {Nath}, Biman B. and {Sharma}, Prateek and {Shchekinov}, Yuri},
        title = "{How radiation affects superbubbles: through momentum injection in early phase and photo-heating thereafter}",
      journal = {\mnras},
         year = 2016,
        month = nov,
       volume = {462},
       number = {4},
        pages = {4532-4548},
          doi = {10.1093/mnras/stw1920},
archivePrefix = {arXiv},
       eprint = {1606.09127},
 primaryClass = {astro-ph.GA},
       adsurl = {https://ui.adsabs.harvard.edu/abs/2016MNRAS.462.4532G}
}

@ARTICLE{Vink2024,
       author = {{Vink}, Jacco},
        title = "{Superbubbles as Galactic PeVatrons: The Potential Role of Rapid Second-Order Fermi Acceleration}",
      journal = {arXiv e-prints},
         year = 2024,
        month = jun,
          eid = {arXiv:2406.03555},
        pages = {arXiv:2406.03555},
          doi = {10.48550/arXiv.2406.03555},
archivePrefix = {arXiv},
       eprint = {2406.03555},
 primaryClass = {astro-ph.HE},
       adsurl = {https://ui.adsabs.harvard.edu/abs/2024arXiv240603555V}
}

@ARTICLE{gaia2020,
       author = {{Cantat-Gaudin}, T. and {Anders}, F. and {Castro-Ginard}, A. and {Jordi}, C. and {Romero-G{\'o}mez}, M. and {Soubiran}, C. and {Casamiquela}, L. and {Tarricq}, Y. and {Moitinho}, A. and {Vallenari}, A. and {Bragaglia}, A. and {Krone-Martins}, A. and {Kounkel}, M.},
        title = "{Painting a portrait of the Galactic disc with its stellar clusters}",
      journal = {A\&A},
         year = 2020,
        month = aug,
       volume = {640},
          eid = {A1},
        pages = {A1},
          doi = {10.1051/0004-6361/202038192},
archivePrefix = {arXiv},
       eprint = {2004.07274},
 primaryClass = {astro-ph.GA},
       adsurl = {https://ui.adsabs.harvard.edu/abs/2020A&A...640A...1C}
}

@ARTICLE{gaiadr2_2018,
       author = {{Gaia Collaboration}},
        title = "{Gaia Data Release 2. Summary of the contents and survey properties}",
      journal = {A\&A},
         year = 2018,
        month = aug,
       volume = {616},
          eid = {A1},
        pages = {A1},
          doi = {10.1051/0004-6361/201833051},
archivePrefix = {arXiv},
       eprint = {1804.09365},
 primaryClass = {astro-ph.GA},
       adsurl = {https://ui.adsabs.harvard.edu/abs/2018A&A...616A...1G}
}

@ARTICLE{Weaver+1977,
       author = {{Weaver}, R. and {McCray}, R. and {Castor}, J. and {Shapiro}, P. and {Moore}, R.},
        title = "{Interstellar bubbles. II. Structure and evolution.}",
      journal = {ApJ},
         year = 1977,
        month = dec,
       volume = {218},
        pages = {377-395},
          doi = {10.1086/155692},
       adsurl = {https://ui.adsabs.harvard.edu/abs/1977ApJ...218..377W}
}

@article{Morlino2021,
    author={Morlino, G. and Blasi, P. and Peretti, E. and Cristofari, P.},
    title = {Particle acceleration in winds of star clusters},
    journal = {MNRAS},
    fjournal = {Monthly Notices of the Royal Astronomical Society},
    volume=504,
    pages={6096-6105},
    year=2021,
}

@ARTICLE{Lancaster+2021a,
       author = {{Lancaster}, Lachlan and {Ostriker}, Eve C. and {Kim}, Jeong-Gyu and {Kim}, Chang-Goo},
        title = "{Efficiently Cooled Stellar Wind Bubbles in Turbulent Clouds. I. Fractal Theory and Application to Star-forming Clouds}",
      journal = {ApJ},
         year = 2021,
        month = jun,
       volume = {914},
       number = {2},
          eid = {89},
        pages = {89},
          doi = {10.3847/1538-4357/abf8ab},
archivePrefix = {arXiv},
       eprint = {2104.07691},
 primaryClass = {astro-ph.GA},
       adsurl = {https://ui.adsabs.harvard.edu/abs/2021ApJ...914...89L}
}

@ARTICLE{Yadav+2017,
       author = {{Yadav}, Naveen and {Mukherjee}, Dipanjan and {Sharma}, Prateek and {Nath}, Biman B.},
        title = "{How multiple supernovae overlap to form superbubbles}",
      journal = {MNRAS},
         year = 2017,
        month = feb,
       volume = {465},
       number = {2},
        pages = {1720-1740},
          doi = {10.1093/mnras/stw2522},
archivePrefix = {arXiv},
       eprint = {1603.00815},
 primaryClass = {astro-ph.HE},
       adsurl = {https://ui.adsabs.harvard.edu/abs/2017MNRAS.465.1720Y}
}

@ARTICLE{Parizot+2004,
       author = {{Parizot}, E. and {Marcowith}, A. and {van der Swaluw}, E. and {Bykov}, A.~M. and {Tatischeff}, V.},
        title = "{Superbubbles and energetic particles in the Galaxy. I. Collective effects of particle acceleration}",
      journal = {A\&A},
         year = 2004,
        month = sep,
       volume = {424},
        pages = {747-760},
          doi = {10.1051/0004-6361:20041269},
archivePrefix = {arXiv},
       eprint = {astro-ph/0405531},
 primaryClass = {astro-ph},
       adsurl = {https://ui.adsabs.harvard.edu/abs/2004A&A...424..747P}
}

@ARTICLE{Dwarkadas:2005,
       author = {{Dwarkadas}, Vikram V.},
        title = "{The Evolution of Supernovae in Circumstellar Wind-Blown Bubbles. I. Introduction and One-Dimensional Calculations}",
      journal = {ApJ},
         year = 2005,
        month = sep,
       volume = {630},
       number = {2},
        pages = {892-910},
          doi = {10.1086/432109},
archivePrefix = {arXiv},
       eprint = {astro-ph/0410464},
 primaryClass = {astro-ph},
       adsurl = {https://ui.adsabs.harvard.edu/abs/2005ApJ...630..892D}
}

@ARTICLE{Dwarkadas:2007,
       author = {{Dwarkadas}, Vikram V.},
        title = "{The Evolution of Supernovae in Circumstellar Wind Bubbles. II. Case of a Wolf-Rayet Star}",
      journal = {ApJ},
         year = 2007,
        month = sep,
       volume = {667},
       number = {1},
        pages = {226-247},
          doi = {10.1086/520670},
archivePrefix = {arXiv},
       eprint = {0706.1049},
 primaryClass = {astro-ph},
       adsurl = {https://ui.adsabs.harvard.edu/abs/2007ApJ...667..226D}
}

@ARTICLE{Blasi:2013,
       author = {{Blasi}, Pasquale},
        title = "{The origin of galactic cosmic rays}",
      journal = {A\&Ar},
         year = 2013,
        month = nov,
       volume = {21},
          eid = {70},
        pages = {70},
          doi = {10.1007/s00159-013-0070-7},
archivePrefix = {arXiv},
       eprint = {1311.7346},
 primaryClass = {astro-ph.HE},
       adsurl = {https://ui.adsabs.harvard.edu/abs/2013A&ARv..21...70B}
}

@ARTICLE{Cristofari+2020,
       author = {{Cristofari}, Pierre and {Blasi}, Pasquale and {Amato}, Elena},
        title = "{The low rate of Galactic pevatrons}",
      journal = {Astroparticle Physics},
         year = 2020,
        month = dec,
       volume = {123},
          eid = {102492},
        pages = {102492},
          doi = {10.1016/j.astropartphys.2020.102492},
archivePrefix = {arXiv},
       eprint = {2007.04294},
 primaryClass = {astro-ph.HE},
       adsurl = {https://ui.adsabs.harvard.edu/abs/2020APh...12302492C}
}

@ARTICLE{Bell+2013,
       author = {{Bell}, A.~R. and {Schure}, K.~M. and {Reville}, B. and {Giacinti}, G.},
        title = "{Cosmic-ray acceleration and escape from supernova remnants}",
      journal = {MNRAS},
         year = 2013,
        month = may,
       volume = {431},
       number = {1},
        pages = {415-429},
          doi = {10.1093/mnras/stt179},
archivePrefix = {arXiv},
       eprint = {1301.7264},
 primaryClass = {astro-ph.HE},
       adsurl = {https://ui.adsabs.harvard.edu/abs/2013MNRAS.431..415B}
}

@ARTICLE{Kelner06,
       author = {{Kelner}, S.~R. and {Aharonian}, F.~A. and {Bugayov}, V.~V.},
        title = "{Energy spectra of gamma rays, electrons, and neutrinos produced at proton-proton interactions in the very high energy regime}",
      journal = {PRD},
         year = "2006",
        month = "Aug",
       volume = {74},
       number = {3},
          eid = {034018},
        pages = {034018},
          doi = {10.1103/PhysRevD.74.034018},
archivePrefix = {arXiv},
       eprint = {astro-ph/0606058},
 primaryClass = {astro-ph},
       adsurl = {https://ui.adsabs.harvard.edu/abs/2006PhRvD..74c4018K}
}

@ARTICLE{Rice16,
       author = {{Rice}, Thomas S. and {Goodman}, Alyssa A. and {Bergin}, Edwin A. and
         {Beaumont}, Christopher and {Dame}, T.~M.},
        title = "{A Uniform Catalog of Molecular Clouds in the Milky Way}",
      journal = {ApJ},
         year = "2016",
        month = "May",
       volume = {822},
       number = {1},
          eid = {52},
        pages = {52},
          doi = {10.3847/0004-637X/822/1/52},
archivePrefix = {arXiv},
       eprint = {1602.02791},
 primaryClass = {astro-ph.GA},
       adsurl = {https://ui.adsabs.harvard.edu/abs/2016ApJ...822...52R}
}

@ARTICLE{Lhaaso2021_pevatrons,
       author = {{Cao}, Zhen and {Aharonian}, F.~A. and {An}, Q. and {Axikegu}, Bai, L.~X. and {Bai}, Y.~X. and {Bao}, Y.~W. and {Bastieri}, D. and {Bi}, X.~J. and {Bi}, Y.~J. and {Cai}, H. and {Cai}, J.~T. and {Cao}, Zhe and {Chang}, J. and {Chang}, J.~F. and {Chang}, X.~C. and {Chen}, B.~M. and {Chen}, J. and {Chen}, L. and {Chen}, Liang and {Chen}, Long and {Chen}, M.~J. and {Chen}, M.~L. and {Chen}, Q.~H. and {Chen}, S.~H. and {Chen}, S.~Z. and {Chen}, T.~L. and {Chen}, X.~L. and {Chen}, Y. and {Cheng}, N. and {Cheng}, Y.~D. and {Cui}, S.~W. and {Cui}, X.~H. and {Cui}, Y.~D. and {Dai}, B.~Z. and {Dai}, H.~L. and {Dai}, Z.~G. and {Danzengluobu} and {della Volpe}, D. and {D'Ettorre Piazzoli}, B. and {Dong}, X.~J. and {Fan}, J.~H. and {Fan}, Y.~Z. and {Fan}, Z.~X. and {Fang}, J. and {Fang}, K. and {Feng}, C.~F. and {Feng}, L. and {Feng}, S.~H. and {Feng}, Y.~L. and {Gao}, B. and {Gao}, C.~D. and {Gao}, Q. and {Gao}, W. and {Ge}, M.~M. and {Geng}, L.~S. and {Gong}, G.~H. and {Gou}, Q.~B. and {Gu}, M.~H. and {Guo}, J.~G. and {Guo}, X.~L. and {Guo}, Y.~Q. and {Guo}, Y.~Y. and {Han}, Y.~A. and {He}, H.~H. and {He}, H.~N. and {He}, J.~C. and {He}, S.~L. and {He}, X.~B. and {He}, Y. and {Heller}, M. and {Hor}, Y.~K. and {Hou}, C. and {Hou}, X. and {Hu}, H.~B. and {Hu}, S. and {Hu}, S.~C. and {Hu}, X.~J. and {Huang}, D.~H. and {Huang}, Q.~L. and {Huang}, W.~H. and {Huang}, X.~T. and {Huang}, Z.~C. and {Ji}, F. and {Ji}, X.~L. and {Jia}, H.~Y. and {Jiang}, K. and {Jiang}, Z.~J. and {Jin}, C. and {Kuleshov}, D. and {Levochkin}, K. and {Li}, B.~B. and {Li}, Cong and {Li}, Cheng and {Li}, F. and {Li}, H.~B. and {Li}, H.~C. and {Li}, H.~Y. and {Li}, J. and {Li}, K. and {Li}, W.~L. and {Li}, X. and {Li}, Xin and {Li}, X.~R. and {Li}, Y. and {Li}, Y.~Z. and {Li}, Zhe and {Li}, Zhuo and {Liang}, E.~W. and {Liang}, Y.~F. and {Lin}, S.~J. and {Liu}, B. and {Liu}, C. and {Liu}, D. and {Liu}, H. and {Liu}, H.~D. and {Liu}, J. and {Liu}, J.~L. and {Liu}, J.~S. and {Liu}, J.~Y. and {Liu}, M.~Y. and {Liu}, R.~Y. and {Liu}, S.~M. and {Liu}, W. and {Liu}, Y.~N. and {Liu}, Z.~X. and {Long}, W.~J. and {Lu}, R. and {Lv}, H.~K. and {Ma}, B.~Q. and {Ma}, L.~L. and {Ma}, X.~H. and {Mao}, J.~R. and {Masood}, A. and {Mitthumsiri}, W. and {Montaruli}, T. and {Nan}, Y.~C. and {Pang}, B.~Y. and {Pattarakijwanich}, P. and {Pei}, Z.~Y. and {Qi}, M.~Y. and {Ruffolo}, D. and {Rulev}, V. and {S{\'a}iz}, A. and {Shao}, L. and {Shchegolev}, O. and {Sheng}, X.~D. and {Shi}, J.~R. and {Song}, H.~C. and {Stenkin}, Yu. V. and {Stepanov}, V. and {Sun}, Q.~N. and {Sun}, X.~N. and {Sun}, Z.~B. and {Tam}, P.~H.~T. and {Tang}, Z.~B. and {Tian}, W.~W. and {Wang}, B.~D. and {Wang}, C. and {Wang}, H. and {Wang}, H.~G. and {Wang}, J.~C. and {Wang}, J.~S. and {Wang}, L.~P. and {Wang}, L.~Y. and {Wang}, R.~N. and {Wang}, W. and {Wang}, W. and {Wang}, X.~G. and {Wang}, X.~J. and {Wang}, X.~Y. and {Wang}, Y.~D. and {Wang}, Y.~J. and {Wang}, Y.~P. and {Wang}, Zheng and {Wang}, Zhen and {Wang}, Z.~H. and {Wang}, Z.~X. and {Wei}, D.~M. and {Wei}, J.~J. and {Wei}, Y.~J. and {Wen}, T. and {Wu}, C.~Y. and {Wu}, H.~R. and {Wu}, S. and {Wu}, W.~X. and {Wu}, X.~F. and {Xi}, S.~Q. and {Xia}, J. and {Xia}, J.~J. and {Xiang}, G.~M. and {Xiao}, G. and {Xiao}, H.~B. and {Xin}, G.~G. and {Xin}, Y.~L. and {Xing}, Y. and {Xu}, D.~L. and {Xu}, R.~X. and {Xue}, L. and {Yan}, D.~H. and {Yang}, C.~W. and {Yang}, F.~F. and {Yang}, J.~Y. and {Yang}, L.~L. and {Yang}, M.~J. and {Yang}, R.~Z. and {Yang}, S.~B. and {Yao}, Y.~H. and {Yao}, Z.~G. and {Ye}, Y.~M. and {Yin}, L.~Q. and {Yin}, N. and {You}, X.~H. and {You}, Z.~Y. and {Yu}, Y.~H. and {Yuan}, Q. and {Zeng}, H.~D. and {Zeng}, T.~X. and {Zeng}, W. and {Zeng}, Z.~K. and {Zha}, M. and {Zhai}, X.~X. and {Zhang}, B.~B. and {Zhang}, H.~M. and {Zhang}, H.~Y. and {Zhang}, J.~L. and {Zhang}, J.~W. and {Zhang}, L. and {Zhang}, Li and {Zhang}, L.~X. and {Zhang}, P.~F. and {Zhang}, P.~P. and {Zhang}, R. and {Zhang}, S.~R. and {Zhang}, S.~S. and {Zhang}, X. and {Zhang}, X.~P. and {Zhang}, Yong and {Zhang}, Yi and {Zhang}, Y.~F. and {Zhang}, Y.~L. and {Zhao}, B. and {Zhao}, J. and {Zhao}, L. and {Zhao}, L.~Z. and {Zhao}, S.~P. and {Zheng}, F. and {Zheng}, Y. and {Zhou}, B. and {Zhou}, H. and {Zhou}, J.~N. and {Zhou}, P. and {Zhou}, R. and {Zhou}, X.~X. and {Zhu}, C.~G. and {Zhu}, F.~R. and {Zhu}, H. and {Zhu}, K.~J. and {Zuo}, X.},
        title = "{Ultrahigh-energy photons up to 1.4 petaelectronvolts from 12 {\ensuremath{\gamma}}-ray Galactic sources}",
      journal = {Nature},
         year = 2021,
        month = jun,
       volume = {594},
       number = {7861},
        pages = {33-36},
          doi = {10.1038/s41586-021-03498-z},
       adsurl = {https://ui.adsabs.harvard.edu/abs/2021Natur.594...33C}
}

@ARTICLE{hawc2020_56tev,
       author = {{Abeysekara}, A.~U. and {Albert}, A. and {Alfaro}, R. and {Angeles Camacho}, J.~R. and {Arteaga-Vel{\'a}zquez}, J.~C. and {Arunbabu}, K.~P. and {Avila Rojas}, D. and {Ayala Solares}, H.~A. and {Baghmanyan}, V. and {Belmont-Moreno}, E. and {BenZvi}, S.~Y. and {Brisbois}, C. and {Caballero-Mora}, K.~S. and {Capistr{\'a}n}, T. and {Carrami{\~n}ana}, A. and {Casanova}, S. and {Cotti}, U. and {Cotzomi}, J. and {Couti{\~n}o de Le{\'o}n}, S. and {De la Fuente}, E. and {de Le{\'o}n}, C. and {Dichiara}, S. and {Dingus}, B.~L. and {DuVernois}, M.~A. and {D{\'\i}az-V{\'e}lez}, J.~C. and {Ellsworth}, R.~W. and {Engel}, K. and {Espinoza}, C. and {Fleischhack}, H. and {Fraija}, N. and {Galv{\'a}n-G{\'a}mez}, A. and {Garcia}, D. and {Garc{\'\i}a-Gonz{\'a}lez}, J.~A. and {Garfias}, F. and {Gonz{\'a}lez}, M.~M. and {Goodman}, J.~A. and {Harding}, J.~P. and {Hernandez}, S. and {Hinton}, J. and {Hona}, B. and {Huang}, D. and {Hueyotl-Zahuantitla}, F. and {H{\"u}ntemeyer}, P. and {Iriarte}, A. and {Jardin-Blicq}, A. and {Joshi}, V. and {Kaufmann}, S. and {Kieda}, D. and {Lara}, A. and {Lee}, W.~H. and {Le{\'o}n Vargas}, H. and {Linnemann}, J.~T. and {Longinotti}, A.~L. and {Luis-Raya}, G. and {Lundeen}, J. and {L{\'o}pez-Coto}, R. and {Malone}, K. and {Marinelli}, S.~S. and {Martinez}, O. and {Martinez-Castellanos}, I. and {Mart{\'\i}nez-Castro}, J. and {Mart{\'\i}nez-Huerta}, H. and {Matthews}, J.~A. and {Miranda-Romagnoli}, P. and {Morales-Soto}, J.~A. and {Moreno}, E. and {Mostaf{\'a}}, M. and {Nayerhoda}, A. and {Nellen}, L. and {Newbold}, M. and {Nisa}, M.~U. and {Noriega-Papaqui}, R. and {Peisker}, A. and {P{\'e}rez-P{\'e}rez}, E.~G. and {Pretz}, J. and {Ren}, Z. and {Rho}, C.~D. and {Rivi{\`e}re}, C. and {Rosa-Gonz{\'a}lez}, D. and {Rosenberg}, M. and {Ruiz-Velasco}, E. and {Salesa Greus}, F. and {Sandoval}, A. and {Schneider}, M. and {Schoorlemmer}, H. and {Sinnis}, G. and {Smith}, A.~J. and {Springer}, R.~W. and {Surajbali}, P. and {Tabachnick}, E. and {Tanner}, M. and {Tibolla}, O. and {Tollefson}, K. and {Torres}, I. and {Torres-Escobedo}, R. and {Villase{\~n}or}, L. and {Weisgarber}, T. and {Wood}, J. and {Yapici}, T. and {Zhang}, H. and {Zhou}, H. and {HAWC Collaboration}},
        title = "{Multiple Galactic Sources with Emission Above 56 TeV Detected by HAWC}",
      journal = {PRL},
         year = 2020,
        month = jan,
       volume = {124},
       number = {2},
          eid = {021102},
        pages = {021102},
          doi = {10.1103/PhysRevLett.124.021102},
archivePrefix = {arXiv},
       eprint = {1909.08609},
 primaryClass = {astro-ph.HE},
       adsurl = {https://ui.adsabs.harvard.edu/abs/2020PhRvL.124b1102A}
}

@ARTICLE{Westerlund1_2022,
       author = {{Aharonian}, F. and {Ashkar}, H. and {Backes}, M. and {Barbosa Martins}, V. and {Becherini}, Y. and {Berge}, D. and {Bi}, B. and {B{\"o}ttcher}, M. and {de Bony de Lavergne}, M. and {Bradascio}, F. and {Brose}, R. and {Brun}, F. and {Bulik}, T. and {Burger-Scheidlin}, C. and {Cangemi}, F. and {Caroff}, S. and {Casanova}, S. and {Cerruti}, M. and {Chand}, T. and {Chandra}, S. and {Chen}, A. and {Chibueze}, O. and {Cristofari}, P. and {Damascene Mbarubucyeye}, J. and {Djannati-Ata{\"\i}}, A. and {Ernenwein}, J. -P. and {Feijen}, K. and {Fichet de Clairfontaine}, G. and {Fontaine}, G. and {Funk}, S. and {Gabici}, S. and {Gallant}, Y.~A. and {Ghafourizadeh}, S. and {Giavitto}, G. and {Giunti}, L. and {Glawion}, D. and {Glicenstein}, J.~F. and {Goswami}, P. and {Grondin}, M. -H. and {H{\"a}rer}, L.~K. and {Haupt}, M. and {Hinton}, J.~A. and {H{\"o}rbe}, M. and {Hofmann}, W. and {Holch}, T.~L. and {Holler}, M. and {Horns}, D. and {Jamrozy}, M. and {Joshi}, V. and {Jung-Richardt}, I. and {Kasai}, E. and {Katarzy{\'n}ski}, K. and {Katz}, U. and {Kh{\'e}lifi}, B. and {Klu{\'z}niak}, W. and {Komin}, Nu. and {Kosack}, K. and {Kostunin}, D. and {Kukec Mezek}, G. and {Lang}, R.~G. and {Le Stum}, S. and {Lemi{\`e}re}, A. and {Lemoine-Goumard}, M. and {Lenain}, J. -P. and {Leuschner}, F. and {Lohse}, T. and {Luashvili}, A. and {Lypova}, I. and {Mackey}, J. and {Majumdar}, J. and {Malyshev}, D. and {Marandon}, V. and {Marchegiani}, P. and {Marcowith}, A. and {Mart{\'\i}-Devesa}, G. and {Marx}, R. and {Maurin}, G. and {Meyer}, M. and {Mitchell}, A. and {Moderski}, R. and {Mohrmann}, L. and {Montanari}, A. and {Moulin}, E. and {Muller}, J. and {Murach}, T. and {Nakashima}, K. and {de Naurois}, M. and {Nayerhoda}, A. and {Niemiec}, J. and {Ohm}, S. and {Olivera-Nieto}, L. and {de Ona Wilhelmi}, E. and {Ostrowski}, M. and {Panny}, S. and {Panter}, M. and {Parsons}, R.~D. and {Peron}, G. and {Prokhorov}, D.~A. and {P{\"u}hlhofer}, G. and {Punch}, M. and {Quirrenbach}, A. and {Rauth}, R. and {Reichherzer}, P. and {Reimer}, A. and {Reimer}, O. and {Renaud}, M. and {Reville}, B. and {Rieger}, F. and {Rowell}, G. and {Rudak}, B. and {Ruiz-Velasco}, E. and {Sahakian}, V. and {Salzmann}, H. and {Sanchez}, D.~A. and {Santangelo}, A. and {Sasaki}, M. and {Sch{\"u}ssler}, F. and {Schutte}, H.~M. and {Schwanke}, U. and {Shapopi}, J.~N.~S. and {Specovius}, A. and {Spencer}, S. and {Stawarz}, {\L}. and {Steenkamp}, R. and {Steinmassl}, S. and {Steppa}, C. and {Sushch}, I. and {Suzuki}, H. and {Takahashi}, T. and {Tanaka}, T. and {Terrier}, R. and {Thorpe-Morgan}, C. and {Tsirou}, M. and {Tsuji}, N. and {Tuffs}, R. and {Unbehaun}, T. and {van Eldik}, C. and {van Soelen}, B. and {Vecchi}, M. and {Veh}, J. and {Venter}, C. and {Vink}, J. and {Wagner}, S.~J. and {White}, R. and {Wierzcholska}, A. and {Wong}, Y. Wun and {Zacharias}, M. and {Zargaryan}, D. and {Zdziarski}, A.~A. and {Zhu}, S.~J. and {Zouari}, S. and {{\.Z}ywucka}, N. and {Blackwell}, R. and {Braiding}, C. and {Burton}, M. and {Cubuk}, K. and {Filipovi{\'c}}, M. and {Tothill}, N. and {Wong}, G.},
        title = "{A deep spectromorphological study of the {\ensuremath{\gamma}}-ray emission surrounding the young massive stellar cluster Westerlund 1}",
      journal = {A\&A},
         year = 2022,
        month = oct,
       volume = {666},
          eid = {A124},
        pages = {A124},
          doi = {10.1051/0004-6361/202244323},
archivePrefix = {arXiv},
       eprint = {2207.10921},
 primaryClass = {astro-ph.HE},
       adsurl = {https://ui.adsabs.harvard.edu/abs/2022A&A...666A.124A}
}

@ARTICLE{Aharonian2019_natast,
       author = {{Aharonian}, Felix and {Yang}, Ruizhi and {de O{\~n}a Wilhelmi}, Emma},
        title = "{Massive stars as major factories of Galactic cosmic rays}",
      journal = {Nature Astronomy},
         year = 2019,
        month = mar,
       volume = {3},
        pages = {561-567},
          doi = {10.1038/s41550-019-0724-0},
archivePrefix = {arXiv},
       eprint = {1804.02331},
 primaryClass = {astro-ph.HE},
       adsurl = {https://ui.adsabs.harvard.edu/abs/2019NatAs...3..561A}
}

@ARTICLE{2018HGPS,
       author = {{H.\,E.\,S.\,S. Collaboration} and {Abdalla}, H. and {Abramowski}, A. and {Aharonian}, F. and {Ait Benkhali}, F. and {Ang{\"u}ner}, E.~O. and {Arakawa}, M. and {Arrieta}, M. and {Aubert}, P. and {Backes}, M. and {Balzer}, A. and {Barnard}, M. and {Becherini}, Y. and {Becker Tjus}, J. and {Berge}, D. and {Bernhard}, S. and {Bernl{\"o}hr}, K. and {Blackwell}, R. and {B{\"o}ttcher}, M. and {Boisson}, C. and {Bolmont}, J. and {Bonnefoy}, S. and {Bordas}, P. and {Bregeon}, J. and {Brun}, F. and {Brun}, P. and {Bryan}, M. and {B{\"u}chele}, M. and {Bulik}, T. and {Capasso}, M. and {Carrigan}, S. and {Caroff}, S. and {Carosi}, A. and {Casanova}, S. and {Cerruti}, M. and {Chakraborty}, N. and {Chaves}, R.~C.~G. and {Chen}, A. and {Chevalier}, J. and {Colafrancesco}, S. and {Condon}, B. and {Conrad}, J. and {Davids}, I.~D. and {Decock}, J. and {Deil}, C. and {Devin}, J. and {deWilt}, P. and {Dirson}, L. and {Djannati-Ata{\"\i}}, A. and {Domainko}, W. and {Donath}, A. and {Drury}, L.~O. 'C. and {Dutson}, K. and {Dyks}, J. and {Edwards}, T. and {Egberts}, K. and {Eger}, P. and {Emery}, G. and {Ernenwein}, J. -P. and {Eschbach}, S. and {Farnier}, C. and {Fegan}, S. and {Fernandes}, M.~V. and {Fiasson}, A. and {Fontaine}, G. and {F{\"o}rster}, A. and {Funk}, S. and {F{\"u}{\ss}ling}, M. and {Gabici}, S. and {Gallant}, Y.~A. and {Garrigoux}, T. and {Gast}, H. and {Gat{\'e}}, F. and {Giavitto}, G. and {Giebels}, B. and {Glawion}, D. and {Glicenstein}, J.~F. and {Gottschall}, D. and {Grondin}, M. -H. and {Hahn}, J. and {Haupt}, M. and {Hawkes}, J. and {Heinzelmann}, G. and {Henri}, G. and {Hermann}, G. and {Hinton}, J.~A. and {Hofmann}, W. and {Hoischen}, C. and {Holch}, T.~L. and {Holler}, M. and {Horns}, D. and {Ivascenko}, A. and {Iwasaki}, H. and {Jacholkowska}, A. and {Jamrozy}, M. and {Jankowsky}, D. and {Jankowsky}, F. and {Jingo}, M. and {Jouvin}, L. and {Jung-Richardt}, I. and {Kastendieck}, M.~A. and {Katarzy{\'n}ski}, K. and {Katsuragawa}, M. and {Katz}, U. and {Kerszberg}, D. and {Khangulyan}, D. and {Kh{\'e}lifi}, B. and {King}, J. and {Klepser}, S. and {Klochkov}, D. and {Klu{\'z}niak}, W. and {Komin}, Nu. and {Kosack}, K. and {Krakau}, S. and {Kraus}, M. and {Kr{\"u}ger}, P.~P. and {Laffon}, H. and {Lamanna}, G. and {Lau}, J. and {Lees}, J. -P. and {Lefaucheur}, J. and {Lemi{\`e}re}, A. and {Lemoine-Goumard}, M. and {Lenain}, J. -P. and {Leser}, E. and {Lohse}, T. and {Lorentz}, M. and {Liu}, R. and {L{\'o}pez-Coto}, R. and {Lypova}, I. and {Marandon}, V. and {Malyshev}, D. and {Marcowith}, A. and {Mariaud}, C. and {Marx}, R. and {Maurin}, G. and {Maxted}, N. and {Mayer}, M. and {Meintjes}, P.~J. and {Meyer}, M. and {Mitchell}, A.~M.~W. and {Moderski}, R. and {Mohamed}, M. and {Mohrmann}, L. and {Mor{\r{a}}}, K. and {Moulin}, E. and {Murach}, T. and {Nakashima}, S. and {de Naurois}, M. and {Ndiyavala}, H. and {Niederwanger}, F. and {Niemiec}, J. and {Oakes}, L. and {O'Brien}, P. and {Odaka}, H. and {Ohm}, S. and {Ostrowski}, M. and {Oya}, I. and {Padovani}, M. and {Panter}, M. and {Parsons}, R.~D. and {Paz Arribas}, M. and {Pekeur}, N.~W. and {Pelletier}, G. and {Perennes}, C. and {Petrucci}, P. -O. and {Peyaud}, B. and {Piel}, Q. and {Pita}, S. and {Poireau}, V. and {Poon}, H. and {Prokhorov}, D. and {Prokoph}, H. and {P{\"u}hlhofer}, G. and {Punch}, M. and {Quirrenbach}, A. and {Raab}, S. and {Rauth}, R. and {Reimer}, A. and {Reimer}, O. and {Renaud}, M. and {de los Reyes}, R. and {Rieger}, F. and {Rinchiuso}, L. and {Romoli}, C. and {Rowell}, G. and {Rudak}, B. and {Rulten}, C.~B. and {Safi-Harb}, S. and {Sahakian}, V. and {Saito}, S. and {Sanchez}, D.~A. and {Santangelo}, A. and {Sasaki}, M. and {Schandri}, M. and {Schlickeiser}, R. and {Sch{\"u}ssler}, F. and {Schulz}, A. and {Schwanke}, U. and {Schwemmer}, S. and {Seglar-Arroyo}, M. and {Settimo}, M. and {Seyffert}, A.~S. and {Shafi}, N. and {Shilon}, I. and {Shiningayamwe}, K. and {Simoni}, R. and {Sol}, H. and {Spanier}, F. and {Spir-Jacob}, M. and {Stawarz}, {\L}. and {Steenkamp}, R. and {Stegmann}, C. and {Steppa}, C. and {Sushch}, I. and {Takahashi}, T. and {Tavernet}, J. -P. and {Tavernier}, T. and {Taylor}, A.~M. and {Terrier}, R. and {Tibaldo}, L. and {Tiziani}, D. and {Tluczykont}, M. and {Trichard}, C. and {Tsirou}, M. and {Tsuji}, N. and {Tuffs}, R. and {Uchiyama}, Y. and {van der Walt}, D.~J. and {van Eldik}, C. and {van Rensburg}, C. and {van Soelen}, B. and {Vasileiadis}, G. and {Veh}, J. and {Venter}, C. and {Viana}, A. and {Vincent}, P. and {Vink}, J. and {Voisin}, F. and {V{\"o}lk}, H.~J. and {Vuillaume}, T. and {Wadiasingh}, Z. and {Wagner}, S.~J. and {Wagner}, P. and {Wagner}, R.~M. and {White}, R. and {Wierzcholska}, A. and {Willmann}, P. and {W{\"o}rnlein}, A. and {Wouters}, D. and {Yang}, R. and {Zaborov}, D. and {Zacharias}, M. and {Zanin}, R. and {Zdziarski}, A.~A. and {Zech}, A. and {Zefi}, F. and {Ziegler}, A. and {Zorn}, J. and {{\.Z}ywucka}, N.},
        title = "{The H.E.S.S. Galactic plane survey}",
      journal = {A\&A},
         year = 2018,
        month = apr,
       volume = {612},
          eid = {A1},
        pages = {A1},
          doi = {10.1051/0004-6361/201732098},
archivePrefix = {arXiv},
       eprint = {1804.02432},
 primaryClass = {astro-ph.HE},
       adsurl = {https://ui.adsabs.harvard.edu/abs/2018A&A...612A...1H}
}

@ARTICLE{3hwc_2020ApJ...905...76A,
       author = {{Albert}, A. and {Alfaro}, R. and {Alvarez}, C. and {Camacho}, J.~R. Angeles and {Arteaga-Vel{\'a}zquez}, J.~C. and {Arunbabu}, K.~P. and {Avila Rojas}, D. and {Ayala Solares}, H.~A. and {Baghmanyan}, V. and {Belmont-Moreno}, E. and {BenZvi}, S.~Y. and {Brisbois}, C. and {Caballero-Mora}, K.~S. and {Capistr{\'a}n}, T. and {Carrami{\~n}ana}, A. and {Casanova}, S. and {Cotti}, U. and {Couti{\~n}o de Le{\'o}n}, S. and {De la Fuente}, E. and {Diaz Hernandez}, R. and {Diaz-Cruz}, L. and {Dingus}, B.~L. and {DuVernois}, M.~A. and {Durocher}, M. and {D{\'\i}az-V{\'e}lez}, J.~C. and {Ellsworth}, R.~W. and {Engel}, K. and {Espinoza}, C. and {Fan}, K.~L. and {Fang}, K. and {Alonso}, M. Fern{\'a}ndez and {Fleischhack}, H. and {Fraija}, N. and {Galv{\'a}n-G{\'a}mez}, A. and {Garcia}, D. and {Garc{\'\i}a-Gonz{\'a}lez}, J.~A. and {Garfias}, F. and {Giacinti}, G. and {Gonz{\'a}lez}, M.~M. and {Goodman}, J.~A. and {Harding}, J.~P. and {Hernandez}, S. and {Hinton}, J. and {Hona}, B. and {Huang}, D. and {Hueyotl-Zahuantitla}, F. and {H{\"u}ntemeyer}, P. and {Iriarte}, A. and {Jardin-Blicq}, A. and {Joshi}, V. and {Kieda}, D. and {Lara}, A. and {Lee}, W.~H. and {Le{\'o}n Vargas}, H. and {Linnemann}, J.~T. and {Longinotti}, A.~L. and {Luis-Raya}, G. and {Lundeen}, J. and {L{\'o}pez-Coto}, R. and {Malone}, K. and {Marandon}, V. and {Martinez}, O. and {Martinez-Castellanos}, I. and {Mart{\'\i}nez-Castro}, J. and {Matthews}, J.~A. and {Miranda-Romagnoli}, P. and {Morales-Soto}, J.~A. and {Moreno}, E. and {Mostaf{\'a}}, M. and {Nayerhoda}, A. and {Nellen}, L. and {Newbold}, M. and {Nisa}, M.~U. and {Noriega-Papaqui}, R. and {Olivera-Nieto}, L. and {Omodei}, N. and {Peisker}, A. and {P{\'e}rez Araujo}, Y. and {P{\'e}rez-P{\'e}rez}, E.~G. and {Ren}, Z. and {Rho}, C.~D. and {Rivi{\`e}re}, C. and {Rosa-Gonz{\'a}lez}, D. and {Ruiz-Velasco}, E. and {Salazar}, H. and {Salesa Greus}, F. and {Sandoval}, A. and {Schneider}, M. and {Schoorlemmer}, H. and {Serna}, F. and {Sinnis}, G. and {Smith}, A.~J. and {Springer}, R.~W. and {Surajbali}, P. and {Tollefson}, K. and {Torres}, I. and {Torres-Escobedo}, R. and {Ukwatta}, T.~N. and {Ure{\~n}a-Mena}, F. and {Weisgarber}, T. and {Werner}, F. and {Willox}, E. and {Zepeda}, A. and {Zhou}, H. and {de Le{\'o}n}, C. and {{\'A}lvarez}, J.~D. and {HAWC Collaboration}},
        title = "{3HWC: The Third HAWC Catalog of Very-high-energy Gamma-Ray Sources}",
      journal = {ApJ},
         year = 2020,
        month = dec,
       volume = {905},
       number = {1},
          eid = {76},
        pages = {76},
          doi = {10.3847/1538-4357/abc2d8},
archivePrefix = {arXiv},
       eprint = {2007.08582},
 primaryClass = {astro-ph.HE},
       adsurl = {https://ui.adsabs.harvard.edu/abs/2020ApJ...905...76A}
}

@ARTICLE{WeidnerKroupa2004,
       author = {{Weidner}, C. and {Kroupa}, P.},
        title = "{Evidence for a fundamental stellar upper mass limit from clustered star formation}",
      journal = {MNRAS},
         year = 2004,
        month = feb,
       volume = {348},
       number = {1},
        pages = {187-191},
          doi = {10.1111/j.1365-2966.2004.07340.x},
archivePrefix = {arXiv},
       eprint = {astro-ph/0310860},
 primaryClass = {astro-ph},
       adsurl = {https://ui.adsabs.harvard.edu/abs/2004MNRAS.348..187W}
}

@ARTICLE{2021ApJ...913L..33Li_J1908,
       author = {{Li}, Jian and {Liu}, Ruo-Yu and {de O{\~n}a Wilhelmi}, Emma and {Torres}, Diego F. and {Liu}, Qian-Cheng and {Kerr}, Matthew and {B{\"u}hler}, Rolf and {Su}, Yang and {He}, Hao-Ning and {Xiao}, Meng-Yuan},
        title = "{Investigating the Nature of MGRO J1908+06 with Multiwavelength Observations}",
      journal = {ApJl},
         year = 2021,
        month = jun,
       volume = {913},
       number = {2},
          eid = {L33},
        pages = {L33},
          doi = {10.3847/2041-8213/abf925},
archivePrefix = {arXiv},
       eprint = {2102.05615},
 primaryClass = {astro-ph.HE},
       adsurl = {https://ui.adsabs.harvard.edu/abs/2021ApJ...913L..33L}
}

@ARTICLE{2021MNRAS.505.2309Crestan_J1908,
       author = {{Crestan}, S. and {Giuliani}, A. and {Mereghetti}, S. and {Sidoli}, L. and {Pintore}, F. and {La Palombara}, N.},
        title = "{Multiwavelength investigation of the candidate Galactic PeVatron MGRO J1908+06}",
      journal = {MNRAS},
         year = 2021,
        month = aug,
       volume = {505},
       number = {2},
        pages = {2309-2315},
          doi = {10.1093/mnras/stab1422},
archivePrefix = {arXiv},
       eprint = {2105.07001},
 primaryClass = {astro-ph.HE},
       adsurl = {https://ui.adsabs.harvard.edu/abs/2021MNRAS.505.2309C}
}

@ARTICLE{2020ApJ...897..131Saha,
       author = {{Saha}, L. and {Dom{\'\i}nguez}, A. and {Tibaldo}, L. and {Marchesi}, S. and {Ajello}, M. and {Lemoine-Goumard}, M. and {L{\'o}pez}, M.},
        title = "{Morphological and Spectral Study of 4FGL J1115.1-6118 in the Region of the Young Massive Stellar Cluster NGC 3603}",
      journal = {ApJ},
         year = 2020,
        month = jul,
       volume = {897},
       number = {2},
          eid = {131},
        pages = {131},
          doi = {10.3847/1538-4357/ab9ac2},
archivePrefix = {arXiv},
       eprint = {2006.00274},
 primaryClass = {astro-ph.HE},
       adsurl = {https://ui.adsabs.harvard.edu/abs/2020ApJ...897..131S}
}

@ARTICLE{Rocha+2022-Wd1,
       author = {{Rocha}, Danilo F. and {Almeida}, Leonardo A. and {Damineli}, Augusto and {Navarete}, Felipe and {Abdul-Masih}, Michael and {Mace}, Gregory N.},
        title = "{Distance and age of the massive stellar cluster Westerlund 1 - II. The eclipsing binary W36}",
      journal = {MNRAS},
         year = 2022,
        month = dec,
       volume = {517},
       number = {3},
        pages = {3749-3762},
          doi = {10.1093/mnras/stac2927},
archivePrefix = {arXiv},
       eprint = {2210.04985},
 primaryClass = {astro-ph.SR},
       adsurl = {https://ui.adsabs.harvard.edu/abs/2022MNRAS.517.3749R}
}

@ARTICLE{Navarete+2022-Wd1,
       author = {{Navarete}, Felipe and {Damineli}, Augusto and {Ramirez}, Aura E. and {Rocha}, Danilo F. and {Almeida}, Leonardo A.},
        title = "{Distance and age of the massive stellar cluster Westerlund 1. I. Parallax method using Gaia-EDR3}",
      journal = {MNRAS},
         year = 2022,
        month = oct,
       volume = {516},
       number = {1},
        pages = {1289-1301},
          doi = {10.1093/mnras/stac2374},
archivePrefix = {arXiv},
       eprint = {2204.09414},
 primaryClass = {astro-ph.SR},
       adsurl = {https://ui.adsabs.harvard.edu/abs/2022MNRAS.516.1289N}
}

@ARTICLE{Gennaro+2011,
       author = {{Gennaro}, M. and {Brandner}, W. and {Stolte}, A. and {Henning}, Th.},
        title = "{Mass segregation and elongation of the starburst cluster Westerlund 1}",
      journal = {MNRAS},
         year = 2011,
        month = apr,
       volume = {412},
       number = {4},
        pages = {2469-2488},
          doi = {10.1111/j.1365-2966.2010.18068.x},
archivePrefix = {arXiv},
       eprint = {1011.5223},
 primaryClass = {astro-ph.GA},
       adsurl = {https://ui.adsabs.harvard.edu/abs/2011MNRAS.412.2469G}
}

@ARTICLE{2010MNRAS.406.2633Schure,
       author = {{Schure}, K.~M. and {Achterberg}, A. and {Keppens}, R. and {Vink}, J.},
        title = "{Time-dependent particle acceleration in supernova remnants in different environments}",
      journal = {MNRAS},
         year = 2010,
        month = aug,
       volume = {406},
       number = {4},
        pages = {2633-2649},
          doi = {10.1111/j.1365-2966.2010.16857.x},
archivePrefix = {arXiv},
       eprint = {1004.2766},
 primaryClass = {astro-ph.HE},
       adsurl = {https://ui.adsabs.harvard.edu/abs/2010MNRAS.406.2633S}
}

@ARTICLE{2023A&A...673A.114Hunt,
       author = {{Hunt}, Emily L. and {Reffert}, Sabine},
        title = "{Improving the open cluster census. II. An all-sky cluster catalogue with Gaia DR3}",
      journal = {A\&A},
         year = 2023,
        month = may,
       volume = {673},
          eid = {A114},
        pages = {A114},
          doi = {10.1051/0004-6361/202346285},
archivePrefix = {arXiv},
       eprint = {2303.13424},
 primaryClass = {astro-ph.GA},
       adsurl = {https://ui.adsabs.harvard.edu/abs/2023A&A...673A.114H}
}

@ARTICLE{Blasi-Morlino:2023,
       author = {{Blasi}, Pasquale and {Morlino}, Giovanni},
        title = "{High-energy cosmic rays and gamma-rays from star clusters: the case of Cygnus OB2}",
      journal = {MNRAS},
         year = 2023,
        month = aug,
       volume = {523},
       number = {3},
        pages = {4015-4028},
          doi = {10.1093/mnras/stad1662},
archivePrefix = {arXiv},
       eprint = {2306.03762},
 primaryClass = {astro-ph.HE},
       adsurl = {https://ui.adsabs.harvard.edu/abs/2023MNRAS.523.4015B}
}

@ARTICLE{2013MNRAS.434.2289OhmWest1,
       author = {{Ohm}, S. and {Hinton}, J.~A. and {White}, R.},
      journal = {MNRAS},
        title = "{$\gamma$-ray emission from the Westerlund 1 region}",
         year = 2013,
        month = sep,
       volume = {434},
       number = {3},
        pages = {2289-2294},
          doi = {10.1093/mnras/stt1170},
archivePrefix = {arXiv},
       eprint = {1306.5642},
 primaryClass = {astro-ph.HE},
       adsurl = {https://ui.adsabs.harvard.edu/abs/2013MNRAS.434.2289O}
}

@ARTICLE{2015MNRAS.447.2322R_WRCat,
       author = {{Rosslowe}, C.~K. and {Crowther}, P.~A.},
        title = "{Spatial distribution of Galactic Wolf-Rayet stars and implications for the global population}",
      journal = {MNRAS},
         year = 2015,
        month = mar,
       volume = {447},
       number = {3},
        pages = {2322-2347},
          doi = {10.1093/mnras/stu2525},
archivePrefix = {arXiv},
       eprint = {1412.0699},
 primaryClass = {astro-ph.SR},
        url   = {https://pacrowther.staff.shef.ac.uk/WRcat/index.php},
       adsurl = {https://ui.adsabs.harvard.edu/abs/2015MNRAS.447.2322R}
}

@article{Cao_2024_1lhaaso,
doi = {10.3847/1538-4365/acfd29},
url = {https://dx.doi.org/10.3847/1538-4365/acfd29},
year = {2024},
month = {feb},
publisher = {The American Astronomical Society},
volume = {271},
number = {1},
pages = {25},
author = {Zhen Cao and F. Aharonian and Q. An and  Axikegu and Y. X. Bai and Y. W. Bao and D. Bastieri and X. J. Bi and Y. J. Bi and J. T. Cai and Q. Cao and W. Y. Cao and Zhe Cao and J. Chang and J. F. Chang and A. M. Chen and E. S. Chen and Liang Chen and Lin Chen and Long Chen and M. J. Chen and M. L. Chen and Q. H. Chen and S. H. Chen and S. Z. Chen and T. L. Chen and Y. Chen and N. Cheng and Y. D. Cheng and M. Y. Cui and S. W. Cui and X. H. Cui and Y. D. Cui and B. Z. Dai and H. L. Dai and Z. G. Dai and  Danzengluobu and D. della Volpe and X. Q. Dong and K. K. Duan and J. H. Fan and Y. Z. Fan and J. Fang and K. Fang and C. F. Feng and L. Feng and S. H. Feng and X. T. Feng and Y. L. Feng and S. Gabici and B. Gao and C. D. Gao and L. Q. Gao and Q. Gao and W. Gao and W. K. Gao and M. M. Ge and L. S. Geng and G. Giacinti and G. H. Gong and Q. B. Gou and M. H. Gu and F. L. Guo and X. L. Guo and Y. Q. Guo and Y. Y. Guo and Y. A. Han and H. H. He and H. N. He and J. Y. He and X. B. He and Y. He and M. Heller and Y. K. Hor and B. W. Hou and C. Hou and X. Hou and H. B. Hu and Q. Hu and S. C. Hu and D. H. Huang and T. Q. Huang and W. J. Huang and X. T. Huang and X. Y. Huang and Y. Huang and Z. C. Huang and X. L. Ji and H. Y. Jia and K. Jia and K. Jiang and X. W. Jiang and Z. J. Jiang and M. Jin and M. M. Kang and T. Ke and D. Kuleshov and K. Kurinov and B. B. Li and Cheng Li and Cong Li and D. Li and F. Li and H. B. Li and H. C. Li and H. Y. Li and J. Li and Jian Li and Jie Li and K. Li and W. L. Li and W. L. Li and X. R. Li and Xin Li and Y. Z. Li and Zhe Li and Zhuo Li and E. W. Liang and Y. F. Liang and S. J. Lin and B. Liu and C. Liu and D. Liu and H. Liu and H. D. Liu and J. Liu and J. L. Liu and J. Y. Liu and M. Y. Liu and R. Y. Liu and S. M. Liu and W. Liu and Y. Liu and Y. N. Liu and R. Lu and Q. Luo and H. K. Lv and B. Q. Ma and L. L. Ma and X. H. Ma and J. R. Mao and Z. Min and W. Mitthumsiri and H. J. Mu and Y. C. Nan and A. Neronov and Z. W. Ou and B. Y. Pang and P. Pattarakijwanich and Z. Y. Pei and M. Y. Qi and Y. Q. Qi and B. Q. Qiao and J. J. Qin and D. Ruffolo and A. SÃ¡iz and D. Semikoz and C. Y. Shao and L. Shao and O. Shchegolev and X. D. Sheng and F. W. Shu and H. C. Song and Yu. V. Stenkin and V. Stepanov and Y. Su and Q. N. Sun and X. N. Sun and Z. B. Sun and P. H. T. Tam and Q. W. Tang and Z. B. Tang and W. W. Tian and C. Wang and C. B. Wang and G. W. Wang and H. G. Wang and H. H. Wang and J. C. Wang and K. Wang and L. P. Wang and L. Y. Wang and P. H. Wang and R. Wang and W. Wang and X. G. Wang and X. Y. Wang and Y. Wang and Y. D. Wang and Y. J. Wang and Z. H. Wang and Z. X. Wang and Zhen Wang and Zheng Wang and D. M. Wei and J. J. Wei and Y. J. Wei and T. Wen and C. Y. Wu and H. R. Wu and S. Wu and X. F. Wu and Y. S. Wu and S. Q. Xi and J. Xia and J. J. Xia and G. M. Xiang and D. X. Xiao and G. Xiao and G. G. Xin and Y. L. Xin and Y. Xing and Z. Xiong and D. L. Xu and R. F. Xu and R. X. Xu and W. L. Xu and L. Xue and D. H. Yan and J. Z. Yan and T. Yan and C. W. Yang and F. Yang and F. F. Yang and H. W. Yang and J. Y. Yang and L. L. Yang and M. J. Yang and R. Z. Yang and S. B. Yang and Y. H. Yao and Z. G. Yao and Y. M. Ye and L. Q. Yin and N. Yin and X. H. You and Z. Y. You and Y. H. Yu and Q. Yuan and H. Yue and H. D. Zeng and T. X. Zeng and W. Zeng and M. Zha and B. B. Zhang and F. Zhang and H. M. Zhang and H. Y. Zhang and J. L. Zhang and L. X. Zhang and Li Zhang and P. F. Zhang and P. P. Zhang and R. Zhang and S. B. Zhang and S. R. Zhang and S. S. Zhang and X. Zhang and X. P. Zhang and Y. F. Zhang and Yi Zhang and Yong Zhang and B. Zhao and J. Zhao and L. Zhao and L. Z. Zhao and S. P. Zhao and F. Zheng and B. Zhou and H. Zhou and J. N. Zhou and M. Zhou and P. Zhou and R. Zhou and X. X. Zhou and C. G. Zhu and F. R. Zhu and H. Zhu and K. J. Zhu and X. Zuo and (The LHAASO Collaboration)},
title = {The First LHAASO Catalog of Gamma-Ray Sources},
journal = {The Astrophysical Journal Supplement Series}
}

@ARTICLE{2020ApJS..247...33A_4fgl,
       author = {{Abdollahi}, S. and {Acero}, F. and {Ackermann}, M. and {Ajello}, M. and {Atwood}, W.~B. and {Axelsson}, M. and {Baldini}, L. and {Ballet}, J. and {Barbiellini}, G. and {Bastieri}, D. and {Becerra Gonzalez}, J. and {Bellazzini}, R. and {Berretta}, A. and {Bissaldi}, E. and {Blandford}, R.~D. and {Bloom}, E.~D. and {Bonino}, R. and {Bottacini}, E. and {Brandt}, T.~J. and {Bregeon}, J. and {Bruel}, P. and {Buehler}, R. and {Burnett}, T.~H. and {Buson}, S. and {Cameron}, R.~A. and {Caputo}, R. and {Caraveo}, P.~A. and {Casandjian}, J.~M. and {Castro}, D. and {Cavazzuti}, E. and {Charles}, E. and {Chaty}, S. and {Chen}, S. and {Cheung}, C.~C. and {Chiaro}, G. and {Ciprini}, S. and {Cohen-Tanugi}, J. and {Cominsky}, L.~R. and {Coronado-Bl{\'a}zquez}, J. and {Costantin}, D. and {Cuoco}, A. and {Cutini}, S. and {D'Ammando}, F. and {DeKlotz}, M. and {de la Torre Luque}, P. and {de Palma}, F. and {Desai}, A. and {Digel}, S.~W. and {Di Lalla}, N. and {Di Mauro}, M. and {Di Venere}, L. and {Dom{\'\i}nguez}, A. and {Dumora}, D. and {Fana Dirirsa}, F. and {Fegan}, S.~J. and {Ferrara}, E.~C. and {Franckowiak}, A. and {Fukazawa}, Y. and {Funk}, S. and {Fusco}, P. and {Gargano}, F. and {Gasparrini}, D. and {Giglietto}, N. and {Giommi}, P. and {Giordano}, F. and {Giroletti}, M. and {Glanzman}, T. and {Green}, D. and {Grenier}, I.~A. and {Griffin}, S. and {Grondin}, M. -H. and {Grove}, J.~E. and {Guiriec}, S. and {Harding}, A.~K. and {Hayashi}, K. and {Hays}, E. and {Hewitt}, J.~W. and {Horan}, D. and {J{\'o}hannesson}, G. and {Johnson}, T.~J. and {Kamae}, T. and {Kerr}, M. and {Kocevski}, D. and {Kovac'evic'}, M. and {Kuss}, M. and {Landriu}, D. and {Larsson}, S. and {Latronico}, L. and {Lemoine-Goumard}, M. and {Li}, J. and {Liodakis}, I. and {Longo}, F. and {Loparco}, F. and {Lott}, B. and {Lovellette}, M.~N. and {Lubrano}, P. and {Madejski}, G.~M. and {Maldera}, S. and {Malyshev}, D. and {Manfreda}, A. and {Marchesini}, E.~J. and {Marcotulli}, L. and {Mart{\'\i}-Devesa}, G. and {Martin}, P. and {Massaro}, F. and {Mazziotta}, M.~N. and {McEnery}, J.~E. and {Mereu}, I. and {Meyer}, M. and {Michelson}, P.~F. and {Mirabal}, N. and {Mizuno}, T. and {Monzani}, M.~E. and {Morselli}, A. and {Moskalenko}, I.~V. and {Negro}, M. and {Nuss}, E. and {Ojha}, R. and {Omodei}, N. and {Orienti}, M. and {Orlando}, E. and {Ormes}, J.~F. and {Palatiello}, M. and {Paliya}, V.~S. and {Paneque}, D. and {Pei}, Z. and {Pe{\~n}a-Herazo}, H. and {Perkins}, J.~S. and {Persic}, M. and {Pesce-Rollins}, M. and {Petrosian}, V. and {Petrov}, L. and {Piron}, F. and {Poon}, H. and {Porter}, T.~A. and {Principe}, G. and {Rain{\`o}}, S. and {Rando}, R. and {Razzano}, M. and {Razzaque}, S. and {Reimer}, A. and {Reimer}, O. and {Remy}, Q. and {Reposeur}, T. and {Romani}, R.~W. and {Saz Parkinson}, P.~M. and {Schinzel}, F.~K. and {Serini}, D. and {Sgr{\`o}}, C. and {Siskind}, E.~J. and {Smith}, D.~A. and {Spandre}, G. and {Spinelli}, P. and {Strong}, A.~W. and {Suson}, D.~J. and {Tajima}, H. and {Takahashi}, M.~N. and {Tak}, D. and {Thayer}, J.~B. and {Thompson}, D.~J. and {Tibaldo}, L. and {Torres}, D.~F. and {Torresi}, E. and {Valverde}, J. and {Van Klaveren}, B. and {van Zyl}, P. and {Wood}, K. and {Yassine}, M. and {Zaharijas}, G.},
        title = "{Fermi Large Area Telescope Fourth Source Catalog}",
      journal = {ApJs},
         year = 2020,
        month = mar,
       volume = {247},
       number = {1},
          eid = {33},
        pages = {33},
          doi = {10.3847/1538-4365/ab6bcb},
archivePrefix = {arXiv},
       eprint = {1902.10045},
 primaryClass = {astro-ph.HE},
       adsurl = {https://ui.adsabs.harvard.edu/abs/2020ApJS..247...33A}
}

@ARTICLE{2011A&A...525A..46H_West2,
       author = {{H.~E.~S.~S. Collaboration} and {Abramowski}, A. and {Acero}, F. and {Aharonian}, F. and {Akhperjanian}, A.~G. and {Anton}, G. and {Barnacka}, A. and {Barres de Almeida}, U. and {Bazer-Bachi}, A.~R. and {Becherini}, Y. and {Becker}, J. and {Behera}, B. and {Bernl{\"o}hr}, K. and {Bochow}, A. and {Boisson}, C. and {Bolmont}, J. and {Bordas}, P. and {Borrel}, V. and {Brucker}, J. and {Brun}, F. and {Brun}, P. and {Bulik}, T. and {B{\"u}sching}, I. and {Boutelier}, T. and {Casanova}, S. and {Cerruti}, M. and {Chadwick}, P.~M. and {Charbonnier}, A. and {Chaves}, R.~C.~G. and {Cheesebrough}, A. and {Conrad}, J. and {Chounet}, L. -M. and {Clapson}, A.~C. and {Coignet}, G. and {Dalton}, M. and {Daniel}, M.~K. and {Davids}, I.~D. and {Degrange}, B. and {Deil}, C. and {Dickinson}, H.~J. and {Djannati-Ata{\"\i}}, A. and {Domainko}, W. and {Drury}, L. O'C. and {Dubois}, F. and {Dubus}, G. and {Dyks}, J. and {Dyrda}, M. and {Egberts}, K. and {Eger}, P. and {Espigat}, P. and {Fallon}, L. and {Farnier}, C. and {Fegan}, S. and {Feinstein}, F. and {Fernandes}, M.~V. and {Fiasson}, A. and {F{\"o}rster}, A. and {Fontaine}, G. and {F{\"u}{\ss}ling}, M. and {Gabici}, S. and {Gallant}, Y.~A. and {G{\'e}rard}, L. and {Gerbig}, D. and {Giebels}, B. and {Glicenstein}, J.~F. and {Gl{\"u}ck}, B. and {Goret}, P. and {G{\"o}ring}, D. and {Hague}, J.~D. and {Hampf}, D. and {Hauser}, M. and {Heinz}, S. and {Heinzelmann}, G. and {Henri}, G. and {Hermann}, G. and {Hinton}, J.~A. and {Hoffmann}, A. and {Hofmann}, W. and {Hofverberg}, P. and {Holleran}, M. and {Hoppe}, S. and {Horns}, D. and {Jacholkowska}, A. and {de Jager}, O.~C. and {Jahn}, C. and {Jung}, I. and {Katarzy{\'n}ski}, K. and {Katz}, U. and {Kaufmann}, S. and {Kerschhaggl}, M. and {Khangulyan}, D. and {Kh{\'e}lifi}, B. and {Keogh}, D. and {Klochkov}, D. and {Klu{\'z}niak}, W. and {Kneiske}, T. and {Komin}, Nu. and {Kosack}, K. and {Kossakowski}, R. and {Lamanna}, G. and {Lenain}, J. -P. and {Lennarz}, D. and {Lohse}, T. and {Lu}, C. -C. and {Marandon}, V. and {Marcowith}, A. and {Masbou}, J. and {Maurin}, D. and {McComb}, T.~J.~L. and {Medina}, M.~C. and {M{\'e}hault}, J. and {Moderski}, R. and {Moulin}, E. and {Naumann-Godo}, M. and {de Naurois}, M. and {Nedbal}, D. and {Nekrassov}, D. and {Nguyen}, N. and {Nicholas}, B. and {Niemiec}, J. and {Nolan}, S.~J. and {Ohm}, S. and {Olive}, J. -F. and {de O{\~n}a Wilhelmi}, E. and {Opitz}, B. and {Orford}, K.~J. and {Ostrowski}, M. and {Panter}, M. and {Paz Arribas}, M. and {Pedaletti}, G. and {Pelletier}, G. and {Petrucci}, P. -O. and {Pita}, S. and {P{\"u}hlhofer}, G. and {Punch}, M. and {Quirrenbach}, A. and {Raubenheimer}, B.~C. and {Raue}, M. and {Rayner}, S.~M. and {Reimer}, O. and {Reimer}, A. and {Renaud}, M. and {de los Reyes}, R. and {Rieger}, F. and {Ripken}, J. and {Rob}, L. and {Rosier-Lees}, S. and {Rowell}, G. and {Rudak}, B. and {Rulten}, C.~B. and {Ruppel}, J. and {Ryde}, F. and {Sahakian}, V. and {Santangelo}, A. and {Schlickeiser}, R. and {Sch{\"o}ck}, F.~M. and {Sch{\"o}nwald}, A. and {Schwanke}, U. and {Schwarzburg}, S. and {Schwemmer}, S. and {Shalchi}, A. and {Sushch}, I. and {Sikora}, M. and {Skilton}, J.~L. and {Sol}, H. and {Spengler}, G. and {Stawarz}, {\L}. and {Steenkamp}, R. and {Stegmann}, C. and {Stinzing}, F. and {Szostek}, A. and {Tam}, P.~H. and {Tavernet}, J. -P. and {Terrier}, R. and {Tibolla}, O. and {Tluczykont}, M. and {Valerius}, K. and {van Eldik}, C. and {Vasileiadis}, G. and {Venter}, C. and {Vialle}, J.~P. and {Vincent}, P. and {Vivier}, M. and {V{\"o}lk}, H.~J. and {Volpe}, F. and {Wagner}, S.~J. and {Ward}, M. and {Zdziarski}, A.~A. and {Zech}, A. and {Zechlin}, H. -S. and {Fukui}, Y. and {Furukawa}, N. and {Ohama}, A. and {Sano}, H. and {Dawson}, J. and {Kawamura}, A and {H.~E.~S.~S. Collaboration}},
        title = "{Revisiting the Westerlund 2 field with the HESS telescope array}",
      journal = {A\&A},
         year = 2011,
        month = jan,
       volume = {525},
          eid = {A46},
        pages = {A46},
          doi = {10.1051/0004-6361/201015290},
archivePrefix = {arXiv},
       eprint = {1009.3012},
 primaryClass = {astro-ph.HE},
       adsurl = {https://ui.adsabs.harvard.edu/abs/2011A&A...525A..46H}
}

@ARTICLE{Celli-Peron_Extended:2024,
       author = {{Celli}, S. and {Peron}, G.},
        title = "{Detection prospects of very and ultra high-energy gamma rays from extended sources with ASTRI, CTA, and LHAASO}",
      journal = {A\&A},
         year = 2024,
        month = sep,
       volume = {689},
          eid = {A258},
        pages = {A258},
          doi = {10.1051/0004-6361/202449837},
archivePrefix = {arXiv},
       eprint = {2403.03731},
 primaryClass = {astro-ph.HE},
       adsurl = {https://ui.adsabs.harvard.edu/abs/2024A&A...689A.258C}
}

@ARTICLE{2021MNRAS.505.2731Mestre,
       author = {{Mestre}, Enrique and {de O{\~n}a Wilhelmi}, Emma and {Torres}, Diego F. and {Holch}, Tim Lukas and {Schwanke}, Ullrich and {Aharonian}, Felix and {Parkinson}, Pablo Saz and {Yang}, Ruizhi and {Zanin}, Roberta},
        title = "{Probing the hadronic nature of the gamma-ray emission associated with Westerlund 2}",
      journal = {MNRAS},
         year = 2021,
        month = aug,
       volume = {505},
       number = {2},
        pages = {2731-2740},
          doi = {10.1093/mnras/stab1455},
archivePrefix = {arXiv},
       eprint = {2105.09155},
 primaryClass = {astro-ph.HE},
       adsurl = {https://ui.adsabs.harvard.edu/abs/2021MNRAS.505.2731M}
}
\bibliographystyle{aasjournalv7}

\end{document}